\documentclass{ws-ijmpe}
\usepackage{ws-rv-van}  
\usepackage{hyperref}
\usepackage{cite}
\usepackage{psfrag}
\usepackage{pdfcomment}
\usepackage{xcolor}

\usepackage[active]{srcltx}

\usepackage{amssymb}
\usepackage{amsmath}
\usepackage{graphicx}
\usepackage{bm} 

\def\npa#1{\newblock {\em Nucl.\ Phys.\ A}\ {\bf #1}}

\newcommand{\taup}{\tau_{\!+}}

\newcommand{\taux}{\tau_{\!\times}} 
\newcommand{\Del}{$\Delta(1232)$ }
\newcommand{\pppi}{\mbox{$pp\to \{pp\}_{\!s\,}\pi^0$}}

\newcommand{\pnpppi}{\mbox{$pn\to \{pp\}_s\pi^-$}}

\newcommand{\fmn}[2]{\mbox{${\textstyle \frac{#1}{#2}}$}}
\newcommand{\nppppi}{\mbox{$np\to \{pp\}_{\!s}\pi^-$}}
\newcommand{\vnvpppppi}{\mbox{$\pol{n}\,\pol{p}\to \{pp\}_{\!s}\pi^-$}}
\newcommand{\pol}[1]{\mathaccent"017E{#1}}

\newcommand{\NNLO}{N$^2$LO} 
\newcommand{\NNNLO}{N$^3$LO} 
\newcommand{\NNNNLO}{N$^4$LO}
\newcommand{\mpi}{\ensuremath{m_\pi}}   
\newcommand{\mN}{\ensuremath{m_{\!N}}}   
\newcommand{\gA}{\ensuremath{g_{\!A}}} 
\newcommand{\fpi}{\ensuremath{f_{\!\pi}}} 
\newcommand{\nolik}{\ensuremath{0}}    
\newcommand{\NNNNpi}{\ensuremath{NN \to NN\pi}}
\newcommand{\maM}{\mathcal{M}}
\newcommand{\maA}{\mathcal{A}}
\newcommand{\boldpi}{\bm{\pi}}
\newcommand{\boldtau}{\bm{\tau}}

\newcommand{\be}{\begin{eqnarray}}
\newcommand{\ee}{\end{eqnarray}}
\newcommand{\beq}{\begin{equation}}
\newcommand{\eeq}{\end{equation}}

\newcommand{\ga}{g_{A}}
\newcommand{\gpind}{g_{\pi N \Delta}}

\newcommand{\vdotq}{v \cdot q}

\newcommand{\ipipi}{{I_{\pi \pi}}}
\newcommand{\ipipifsi}{{I_{\pi \pi}^{\rm finite}}}

\newcommand{\jpid}{{J_{\pi  \Delta}}}
\newcommand{\jpidfsi}{{J_{\pi  \Delta}^{\rm finite}}}

\newcommand{\jpipid}{{J_{\pi \pi \Delta}}}
\newcommand{\jpipind}{{J_{\pi \pi N \Delta}}}

\newcommand{\intI}{\ipipi}
\newcommand{\intJ}{\frac{\jpid}{\delta}}
\newcommand{\intT}{\Delta\jpipid}
\newcommand{\intC}[1]{\frac{#1}{(4\pi)^2}}
\newcommand{\intL}{k_1^2 \jpipind}

\newcommand{\Sd}{\mathbb{S}}
\newcommand{\Sdd}{\mathbb{S}^{\dagger}}

\begin{document}

\markboth{Vadim Baru, Christoph Hanhart, Fred Myhrer}
             {Effective Field Theory calculations of $NN\to NN\pi$}


\title{Effective Field Theory  calculations of $NN\to NN\pi$. 
}

\author{\footnotesize Vadim Baru\footnote{vadimb@tp2.rub.de}}

\address{Institut f\"ur theoretische Physik II, Fakult\"at f\" ur Physik und Astronomie
Ruhr-Universit\"at Bochum, \\ 
44780 Bochum Germany, and \\
Institute for Theoretical and Experimental Physics,
 117218, B. Cheremushkinskaya 25, Moscow, Russia}

\author{Christoph Hanhart\footnote{c.hanhart@fz-juelich.de}}

\address{Institute for Advanced Simulation, 
Institute for Nuclear Physics, and J\"ulich Center for Hadron Physics,  \\
52425 J\"ulich, Germany}

\author{Fred Myhrer\footnote{myhrer@sc.edu}} 

\address{Department of Physics and Astronomy, 
University of South Carolina, Columbia, SC 29208, USA}

\maketitle

\begin{history}
\received{(received date)}
\revised{(revised date)}
\end{history}

\begin{abstract}
In this review we present the recent advances for calculations of
the reactions $NN\to NN\pi$ using chiral effective field  theory. 
Discussed are the next-to-next-to leading order loop contributions with 
nucleon and Delta-isobar for near threshold s-wave pion production.
Results of  recent experimental pion-production data for 
energies close to the threshold are analyzed. 
Several  particular applications  are discussed: 
(i)  it is shown how the measured charge symmetry violating 
pion-production reaction can be used to extract the 
strong-interaction contribution to the  proton-neutron mass difference; 
(ii)  the role  of  $NN\to NN\pi$  for the 
extraction of the pion-nucleon scattering lengths   from pionic atoms
data is illuminated. 

\end{abstract}

\newpage
	
\section{Introduction}
\label{sec:intro}
After the first high quality data for the reaction $NN\to NN\pi$ 
were published~\cite{IUCF1}, 
it became quickly clear that this first hadronic inelastic channel of
the $NN$ interaction is far more difficult to handle theoretically 
than what was initially expected. 
The best phenomenological 
model existing at the time~\cite{koltun} failed 
to explain the cross section data by factors of 2 and
10 for the reactions $pp\to d\pi^+$ and $pp\to pp\pi^0$, respectively.
Various attempts were made to identify the phenomenological mechanisms
responsible for this discrepancy --- the most popular ones were heavy meson
exchanges~\cite{Lee,HGM,Hpipl,jounicomment} and off--shell $\pi N$
rescattering~\cite{eulogio,unsers}. 
Both mechanisms were successful 
quantitively but no consensus on the underlying
dynamics was achieved in the literature. In addition, for the neutral channel
severe quantitative discrepancies appeared in the description of polarization
observables~\cite{ourpols}.

All phenomenolgical approaches have their drawbacks.
For example, they 
do not have a readily identifiable expansion scheme and 
there is no guiding principle 
that allows one to judge which diagrams 
are important and which ones can be neglected.  
In addition, the requirements of chiral symmetry can  easily be violated in 
phenomenological approaches.
On the other hand,
a quantitative understanding of the reactions $NN\to NN\pi$ at threshold 
is of particular importance. 
First of all the system is controlled by several
scales: the large initial momentum, the small final momenta, the
pion mass and the typical hadronic scale $\Lambda_\chi \simeq 1$ GeV. 
A systematic understanding of such systems is not only of
conceptual interest but is an important step towards the development of
effective field theories.  
Secondly, the
calculations of $\pi d$ scattering observables also need accurate input for
the pion absorption amplitudes and their corresponding 
dispersive corrections~\cite{disp,piddelta},
in order to reach a precision high enough to
extract the isoscalar $\pi N$ scattering amplitude~\cite{JOB,longJOB}. 

Since low-energy pion interactions are largely dictated by 
the requirements of chiral symmetry of the
strong interaction, chiral perturbation theory (ChPT) is the proper tool to
resolve the above mentioned discrepancy.  
ChPT allows for a systematic, model
independent expansion of the amplitudes in terms of momenta and pion masses 
measured in units of  $\Lambda_\chi$.  
However, the
use of the standard ChPT power counting, which is based on the assumption that
all relevant momenta are effectively of the order of the pion mass, was not
very successful.  
The first calculations of  \NNNNpi{} reactions 
in this framework were done at tree
level up to next-to-next-to leading order (\NNLO{}) for both 
$pp\to pp\pi^0 \ $~\cite{cohen,park,sato} as well as
for $pp\to d\pi^+$~\cite{rocha,unserd}.  
These early studies revealed, in
particular, that the discrepancy between theory and experiment increases for
both channels due to a destructive interference between the  
impulse approximation or leading order (LO) amplitude
and the isoscalar rescattering contributions that enter at 
next-to leading order (NLO) in standard counting.  
In addition, some loop contributions at \NNLO{} were found in
Refs.~\cite{DKMS,Ando} to be significantly larger than the NLO contribution, revealing a
problem regarding the convergence of the standard 
ChPT approach to the $pp\to pp\pi^0$ reaction \cite{novel}.

Based on the observation that  the  initial nucleon
three-momentum at threshold ($p =|\vec p \,| \simeq \sqrt{m_N\mpi}\sim 360$~MeV where
$\mpi$ and $m_N$  denote the pion and nucleon mass, respectively)  is significantly
larger than the pion mass,
 Refs.~\cite{cohen,rocha} proposed a 
 modification of the chiral counting scheme.
The expansion parameter in the new counting scheme is
\begin{equation}
\chi_{\rm MCS}= p/\Lambda_\chi \simeq \sqrt{m_\pi/m_N} \simeq \ 0.4,
\label{eq:expanpar}
\end{equation}
where  $\Lambda_\chi$ is here 
identified with the nucleon mass.
In what follows, this power counting will be referred to as the
momentum counting scheme (MCS)--- it will be discussed in detail in the next
section.  
This scale  was first implemented in  the  
actual calculations in Refs. \cite{ch3body,HanKai}, 
see Ref.~\cite{hanhart04} for an earlier review article.  
Ref.~\cite{hanhart04} demonstrated that the order of magnitude differences between 
the two-pion exchange diagrams of Ref.~\cite{DKMS} could 
easily be understood in the MCS, since the diagrams are of different orders 
in this new counting scheme 
as will be shown in    
Sec. \ref{sec:swave}. 
It  has  been  known for  years  that  the strength  of  the s-wave  
pion  production amplitude  
in the  charged  channels $pp\to d\pi^+$ and 
$pp\to pn\pi^+$   is dominated  by the  leading  order (LO) 
Weinberg-Tomozawa (WT) operator \cite{koltun}. 
However,   
the  cross  sections  for these two reactions were  underestimated  
by  a  factor   of  2\cite{hanhart04}. 
Meanwhile, the  application  of  the  MCS  to  s-wave  production  in  the  
$pp\to d\pi^+$ channel  at  next-to-leading order  (NLO) \cite{lensky2} 
revealed quite  good  agreement   with  experimental  data. 
In contrast, for  s-wave pion production in  the  neutral  channel  $pp\to pp\pi^0$ the 
situation  is completely  different.
In this channel  the  large isovector WT  rescattering vertex  does not  contribute while
the   other mechanisms  at LO  and NLO   are   either strongly  suppressed  or  vanish completely \cite{cohen,sato,park,HanKai,NNLOswave}.    
Therefore,  we  believe  that the experimentally measured  
$pp\to pp \pi^0$ reaction  
is unique in that it directly probes  the higher order MCS 
contributions which in the other channels 
are masked by the dominant lower order Weinberg--Tomozawa  term.   
This  was  the   motivation  for  the
derivation of  the  pion s-wave production operator   
at  N$^2$LO in MCS which was performed  in  Refs. \cite{NNLOswave,future}
and will be  discussed in Sec \ref{sec:swave}.

Pion production near threshold is also useful for studying isospin violation 
in, e.g. nucleon-nucleon reactions.
Within the standard model there are only two sources of isospin
violation, namely the electromagnetic interaction and the mass difference
of the two lightest quarks, $m_d - m_u \sim m_u$~\cite{SWeinberg77,Gasser:1982ap}.
In processes 
where we are able to disentangle these two
sources, the observation of isospin violation in hadronic reactions gives a
 window to quark mass effects~\cite{Gasser:1982ap,miller}\footnote{ 
Quark masses
themselves are not directly observable and additional information is
necessary to assign a scale to these fundamental parameters of the standard
model (see e.g.~\cite{Leutwyler:1996qg}).}.
The mass difference between the neutral and charged pions, 
which is almost completely due to electromagnetic interactions, is 
by far the largest isospin-violating effect.  
This mass difference drives, for example, the spectacular energy
dependence of the $\pi^0$--photoproduction amplitude near 
threshold (see, e.g.,  the review 
article~\cite{Bernard:2007zu} and references therein). 
Thus, in order to disentangle the two isospin-violating sources, 
it is important to use isospin violation observables where 
the pion mass difference does not contribute.
An example where the pion mass difference does not enter is  
  charge symmetry breaking (CSB) observables.

Within  effective field theory there is a close correlation  
between the leading CSB pion-nucleon scattering amplitude
and the proton--neutron mass difference.  
Already in 1977
Weinberg predicted a huge effect of up to 30\% difference in the 
$\pi^0 p$ and $\pi^0 n$ scattering
lengths due to CSB in $\pi^0 N$
scattering~\cite{SWeinberg77,Meissner:1997ii}.  
Since scattering experiments with neutral 
pions are not feasible,\footnote{
The $\pi^0 p$ scattering length
 might be measurable in polarized neutral pion photoproduction at
  threshold~\cite{AMBernstein98}.}
Refs.~\cite{jouni,filin,bolton} focused instead 
  on $NN$ induced 
pion-production reactions to study CSB.
There have been two 
successful CSB measurements close to the pion threshold, 
namely the forward-backward
asymmetry in the $pn\to{}d\pi^0$ reaction,  
$A_{fb}(pn\to d\pi^0)$~\cite{Opper:2003sb}, 
and the  $dd\to \alpha \pi^0$ reaction cross 
section~\cite{Stephenson:2003dv},  see an early review  article~\cite{Miller2006}.
We will concentrate our discussion on the first CSB reaction, see  Sec. \ref{sec:CSB},   
before we make some remarks regarding the $dd\to \alpha \pi^0$ reaction.
Following the arguments presented in Ref.~\cite{filin}, 
we will show that at leading CSB order within the MCS scheme, 
only the strong (quark-mass induced) contribution to  the  proton-neutron
mass difference enters   the CSB  pion-production operator in $pn\to d\pi^0$.  
Using data on $A_{fb}(pn\to d\pi^0)$~\cite{Opper:2003sb}, 
the strong   contribution to  the  proton-neutron
mass difference  was  extracted in Ref.~\cite{filin}  within  MCS.

\begin{figure*}[t]
\begin{center}
\includegraphics[height=7.0cm,keepaspectratio]{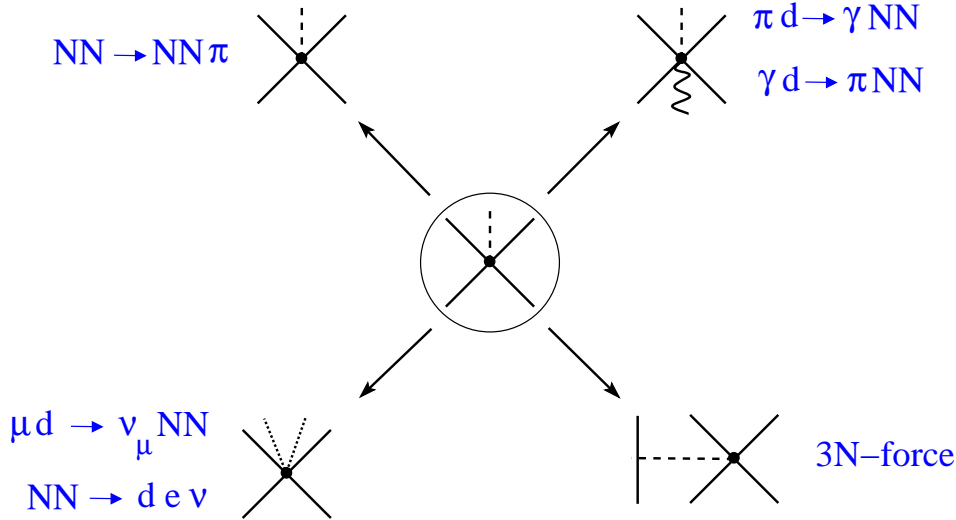}
\caption{
\label{ct}  An illustration of few-nucleon reactions where the LEC $d$ contributes.}
\end{center}
\end{figure*}

An effective field theory (EFT) is a low energy theory where 
the short distance physics is characterized by local  operators 
with their corresponding low energy constants (LECs), see Sec.~\ref{sec:Lagrangian}.  
In few-nucleon systems these LECs parametrize the 
short distance two-nucleon correlations 
not explicitly probed in low-energy reactions. 
One LEC appearing in \NNNNpi{} turns out to be important in a 
variety of other few-nucleon processes at relatively low orders: 
as was stressed in Ref.~\cite{ch3body}, 
the short range physics of the near threshold reactions $NN\to NN\pi$ 
is given by a $(N N)^2 \pi$  short ranged operator which is closely 
connected to the short ranged three--nucleon force. 
The latter is discussed in, 
e.g., Refs.~\cite{pdchiral,Epelbaum:2008ga,Machleidt:2011zz,gazit}.  
The strength of this short range two-nucleon operator is given  
by the LEC $d$, which can be thought of as the 
two-nucleon analog of the axial constant $g_A$. 
At the same time,
this LEC plays a pivotal role in various few-nucleon processes, 
like in the $pp$ fusion reaction~\cite{park2}, $ pp \to d e^+  \nu_e$, 
which is the primary process in stellar thermonuclear reactions, and  
the neutrino deuteron breakup reactions~\cite{Nakamura:2002jg,ando2003} 
which were experimentally measured at the 
Sudbury Neutrino Observatory~\cite{SNO} establishing the 
total neutrino flux from the sun.  In addition, the same short range 
operator is also of importance in $\gamma d\to\pi
NN$~\cite{gammad,gammad_nn} and $\pi d\to \gamma NN$~\cite{gardestig,phillips}
as well as in weak reactions like  tritium beta
decay~\cite{park2,gazit,phillips}. 
In these different reactions the
 operator appears in   
very different
kinematical regimes,  
ranging from very low energies for both incoming and
outgoing $NN$ pairs in weak reactions and proton-deuteron scattering up to
relatively high initial energies for the $NN$ induced pion
production.  
In Fig.~\ref{ct} we illustrate some few-nucleon processes 
where this short range operator 
(and the corresponding LEC $d$)  
parametrizes common short distance two-nucleon physics.  
In Ref.\cite{park2} a first determination of the LEC $d$ from data was done 
in a technically  involved  
EFT  calculation of  the tritium
$\beta$-decay.      
Ref.\cite{park2} then used this value for $d$ to
predict  the astrophysical  S-factors of the solar nuclear processes.
The results of Ref.\cite{park2}  for the rate of  
$pp\to d e^+\nu_e$   were later used  by Ref. ~\cite{phillips}  
to reduce  the cutoff dependence in the reaction 
$\pi^- d \to \gamma nn$,  which was proposed  
as a tool to extract  the $nn$-scattering
length \cite{Gardestig:2005pp,gardestigreview}.  
In a recent work using EFT~\cite{gazit}, 
it was shown that  both the $^3$H
and $^3$He binding energies and the tritium $\beta$-decay 
can be described with  the same contact term. 
Although there are good reasons to believe that the use of EFT for 
the three-nucleon system is reliable, 
it was realized early on that it is desirable to determine the 
LEC $d$ within the two-nucleon system, for this provides an independent
cross check for the formalism.    
The strength of this $(NN)^2\pi$ short range operator could be 
experimentally relatively well determined 
by the muon   capture rate on deuterons, $\mu^- d \to nn \nu_{\mu}$,  cf.
e.g., Refs.~\cite{Ando2000,MuSun,pisaPRL,Marcucci:2011tf,Adam}.  
The  ongoing measurement  of $\mu^- d \to nn \nu_{\mu}$ by the MuSun
Collaboration at PSI \cite{MuSun} aims at
the  determination of  the  rate  with a precision of 1.5 \%.  
Below we will argue that the direct access
to the strength of this $(NN)^2\pi$ short ranged operator  
comes also from the study of the $NN\to NN\pi$ reactions. 
Once this LEC is determined the EFT will have predictive power for 
all the reactions mentioned above.

How the short range operator connects these processes can easily be 
understood from the general structure of the chiral Lagrangian. 
The  Lagrangian is written in terms of the standard chiral field $U(x)$ 
which depends non-linearly on the pion field $\boldpi$. 
The LO Lagrangian reads~\cite{ulfbible}
 \begin{eqnarray}{\cal L}^{(1)}_{\pi N} &=&
 N^\dagger ( i v \cdot D + \mathring{g}_A\, S \cdot u ) N  + \cdots \, , 
\label{eq:lagU} 
\end{eqnarray} 
where the field $N(x)$ denotes the large component of the heavy-nucleon field, 
$\mathring{g}_A$  is the bare axial-vector coupling  of the nucleon, 
$D_\mu N$ is the covariant derivative, 
$D_\mu N= (\partial_\mu +\Gamma_\mu )N$ 
with $\Gamma_\mu = [\xi^\dagger , \partial_\mu \xi ]/2$ 
[we here ignore the external electroweak fields 
which  are, however, introduced in Eq.~(\ref{u})],  
and $\xi = \sqrt{U(x)}$. 
Furthermore, $ u_\mu\equiv i \,  (\xi^\dagger \partial_\mu 
\xi- \xi \partial_\mu \xi^\dagger)$, 
 the nucleon's four-velocity vector $v^\mu = (1,\vec{0})$ and 
the covariant nucleon spin operator~\cite{ulfbible} 
$S^\mu = (0, \vec{\sigma}/2)$. 
%
The following two-nucleon interaction  Lagrangian, which is part of 
the Lagrangians to be specified in Sec.~\ref{sec:Lagrangian}, describes the 
short-range contact interaction illustrated in Fig.~\ref{ct}.  
\be
{\cal L}_{NN}^{int}= -2 d\; \left(N^\dagger\; S\cdot u \;N\right) 
N^\dagger N      \, ,
\label{lag}
\ee 
where the strength of the short-ranged operator is $d$, 
and, including the external electromagnetic fields, 
we observe how $u^\mu$ connects  
pion production with 
the external vector $V_\mu$ and axial-vector $A_\mu$ currents via
\be
f_{\pi} { u_{\mu}}= - {\boldtau} 
\partial_{\mu} {\boldpi} 
- \varepsilon_{3ab}{ V_{\mu}} \pi_a \tau_b +f_{\pi}{ A_{\mu}}+ \cdots \, .  
\label{u}
\ee 
Here $\boldpi$ corresponds to the pion field 
and  $f_{\pi}$ is the pion decay constant ($f_{\pi}=92.4$~MeV). 
In Eq.~(\ref{u}) we have expanded $u^\mu$ in terms of the pion field, 
which as mentioned appears non-lineary in $u^\mu$, and where the ellipses 
correspond  to higher powers in pion field interaction terms.   
It is Eq.~(\ref{u}) that highlights the connection of pion production
and the weak currents.
As follows from Eqs.~(\ref{lag}) and (\ref{u}), the  
contact operator supplemented by the LEC $d$ is proportional to the pion derivative. 
Therefore it should contribute to  production of 
p-wave pions in $NN\to NN\pi$ while connecting to final state S-wave nucleons. 
There are  two    reaction channels  which  fulfil  this
condition.  
One  corresponds to  the  case   with
a spin-triplet  S-wave $NN$  final state  interaction  (FSI)  which is
realized  in the reactions $pp\to pn\pi^+$ and  
$pp\to d\pi^+$. 
The other  channel  is the spin-singlet S-wave $NN$ FSI 
which appears in the reaction $pn\to pp\pi^-$ as indicated in Table~\ref{table1}.  
As will be explained in Sec.~\ref{sec:pwave}, in order to 
experimentally ensure that the final two nucleons are in a relative S-wave, 
one restricts the final relative two-nucleon energy in \NNNNpi{} to be very small,
typically $< 3$~MeV. 
Note that this contact term does not contribute 
to the reaction $pp\to pp \pi^0$ where a p-wave pion can  only be produced in
combination with the final $NN$-pair in a relative P-wave.  
The initial and final $NN$ partial 
wave combinations  for production of  p-wave pions 
can be read off from   Table~\ref{table1}. 
They are
$^1S_0\to  {^3 S_1}p$  [and  $^1D_2\to  {^3 S_1}p$] for  $\pi^+$ production
and  $^3S_1\to  {^1 S_0} p$ [and  $^3D_1\to  {^1 S_0} p$]  for  $\pi^-$ production.
We will show in  Sec.~\ref{sec:pwave} that \NNNNpi{}  p-wave pion production may
be an excellent tool in order to extract  information  about this LEC.

\begin{table}
  \begin{tabular}[ht]{|c|c||r|}
  \hline
 Spin of the $NN$ state & Reaction channel &  Partial  waves   \\
  \hline
\multicolumn{1}{|c|}{}  & $pp\to d\pi^+$ & $ ^3P_1 \to { ^3\!S_1} s$  \\
\multicolumn{1}{|c|}{Spin-triplet S-wave $NN$ FSI}  & $pp\to pn\pi^+$ &  $^1S_0 \to {^3 S_1} p$  \\
\multicolumn{1}{|c|}{}     &$pp\to d\pi^+ $&  $^1D_2 \to {^3 S_1} p$   \\   
 \multicolumn{1}{|c|}{} & $pn\to d\pi^0 $ & \ {\rm (CSB)} $^1P_1 \to { ^3\!S_1} s $ \\
 \hline 
 \hline
\multicolumn{1}{|c|}{}  &     $pp\to pp\pi^0$ & $\!^3\!P_0 \to {^1 S_0} s $ \\
\multicolumn{1}{|c|}{} & & $ (^3\!P_2-^3\!\!F_2) \to  {^1 S_0} d $\\
\cline{2-3}
 \multicolumn{1}{|c|}{Spin-singlet S-wave  $NN$ FSI}  &  & $\!^3\!P_0 \to {^1 S_0} s $ \\
&   $pn\to pp\pi^-$& $ (^3\!P_2-^3\!\!F_2) \to  {^1 S_0} d $\\
\multicolumn{1}{|c|}{}   &    &  $ (^3\!S_1-^3\!\!D_1) \to  {^1 S_0} p$\\
\hline
 \end{tabular}
  \caption{
   \label{table1}  Partial waves and reaction
   channels restricted to the spin-triplet and spin-singlet S-wave $NN$  final
   state  interaction.  }
\end{table}

The paper  is  organizied as follows.  First,  
in Sec.~\ref{sec:swave} we will present the evaluations of the amplitudes for s-wave pion 
production where  
some pertinent details of the lengthy loop-evaluations of 
the NLO and \NNLO{} two-pion-exchange amplitudes will be given. 
In MCS, the pion p-wave production amplitude is 
given by tree level diagrams up to \NNLO{} while for
pion s-wave production, loop diagrams 
start to contribute individually already at NLO. 
The reason for this apparent asymmetry lies in the Goldstone nature of
the pion: since pions have to decouple from matter in the chiral limit for
vanishing momenta, direct production (where a nucleon emits a pion) in
the s-wave can only occur through a nucleon recoil process.  
However, these NLO loops
turn out to cancel completely both for the neutral~\cite{HanKai} and
charged~\cite{lensky2} pion production --- this cancellation, a 
necessary requirement of field theoretic consistency, 
is discussed in detail in
Sec.~\ref{sec:swave}.  
As a by-product of this systematic treatment of 
nucleon recoil effects in the $\pi N\to\pi N$ vertex in Ref.~\cite{lensky2}, 
the isovector rescattering one-pion
exchange amplitude at LO was found to be enhanced by a factor of $4/3$ which
was sufficient to overcome the apparent discrepancy with the data 
when the final two nucleons are in the $^3S_1$ state.   
The contributions to the pion-production amplitudes from subleading loops were 
derived  in Ref.~\cite{NNLOswave},  see  also Ref.~\cite{ksmk09} for an earlier study,  and are presented in 
Sec.~\ref{sec:swave}.
Early on it was realized that the Delta-isobar ($\Delta$) should be  explicitly  
included as a dynamical degree of freedom~\cite{cohen}. 
In the MCS expansion the Delta-nucleon ($\Delta$-N) mass splitting is
numerically of the order of $p$ and will contribute to s-wave pion 
production starting at NLO diagrams.
The general argument justifying the inclusion of the $\Delta$ resonance 
was confirmed numerically in phenomenological
calculations~\cite{jouni78,ourdelta,ourpols}, see also
Refs.~\cite{cohen,rocha,HanKai,NNpiMenu} where the effect of the $\Delta$ in
$NN\to NN\pi$ was studied within chiral effective field theory. 
We will elaborate on this topic in Sec.~\ref{sec:swaveD}. 
As discussed earlier, in Sec.~\ref{sec:pwave} we present one possible 
determination of the LEC $d$  by comparing the \NNLO{}  
calculation of Refs.~\cite{ch3body,newpwave} with the measured  observables. 
Finally, we mention that the reactions $NN\to NN\pi$ provide a direct experimental access to
charge symmetry breaking (CSB) operators, which are difficult to study otherwise,
--- a topic we will present in Sec.~\ref{sec:CSB}.  
In Table~\ref{table1} we indicate a CSB reaction to be discussed in  Sec.~\ref{sec:CSB}. 
In Secs. ~\ref{sec:twopiproduction}-\ref{heavyquarkonia} we will discuss several applications 
of the MCS scheme  in  $NN\to NN\pi$   to  two-pion production  in    nucleon-nucleon collisions (Sec.~\ref{sec:twopiproduction}),  $\pi d$ 
scattering at threshold  (Sec.~\ref{sec:pid-atom})  and decays of heavy  quarkonia  (Sec.~\ref{heavyquarkonia}).  The results 
of the review are summarized in Sec.~\ref{sec:outlook}.

\section{Formal aspects of the momentum counting scheme} 
\label{sec:formalism}

\subsection{ Lagrangian densities}
\label{sec:Lagrangian}

The heavy baryon ChPT Lagrangian is written as a series in increasing powers of derivatives. 
The underlying assumption in writing this series 
is that the higher order Lagrangian terms will only contribute 
smaller corrections to the 
dominant amplitudes which are generated by the lowest order Lagrangian terms. 
\begin{equation}
{\cal L}_{\rm{ch}} = {\cal L}^{(1)}_{\pi N}  
+ {\cal L}^{(2)}_{\pi N} + {\cal L}^{(2)}_{\pi \pi} 
+ {\cal L}^{(3)}_{\pi N} + \cdots\;\; , 
\label{eq:Lag0}
\end{equation}
where ${\cal L}^{(\bar{\nu})}$ ($\bar{\nu}=1,2, \ldots$)
denotes the number of derivatives and/or powers of $m_\pi$ in the Lagrangian.  
Note that within the MCS some terms from higher order Lagrangians might get promoted to
lower orders --- see e.g. the discussion below Eq.~(\ref{eq:la1}).
The effective chiral Lagrangian to
lowest-order (LO) in the $\pi N$ interaction terms read
in $\sigma$-gauge~\cite{OvK,ulfbible} when we expand 
Eq.(\ref{eq:lagU}) in powers of the pion field 
\begin{eqnarray}
 {\cal L}^{(1)}_{\pi\!N}  &=&
   N^{\dagger}\left[\frac{1}{4 f_{\pi}^{2}} \boldtau \cdot
         (\dot{\boldpi}\times{\boldpi})
         +\frac{\mathring{g}_A}{2 f_{\pi}}
         \boldtau\cdot\vec{\sigma}\left(\vec{\nabla}\boldpi
{+}\frac{1}{2f_\pi^2}\boldpi(\boldpi \cdot \vec \nabla \boldpi)
\right)
  \right]N
+\cdots \ .
\label{eq:la0}
\end{eqnarray}
The ellipses represent further terms which are not relevant for the presentation 
in this review. 
The next-higher order $\pi N$ interaction terms have the form 
\begin{eqnarray}\nonumber
 \hspace*{-0.2cm}{\cal L}^{(2)}_{\pi\!N}&=&
    \frac{1}{8\mN f_{\pi}^{2}}
    \bigg[ iN^{\dagger}\boldtau\cdot
        (\boldpi\times\vec{\nabla}\boldpi)\cdot\vec{\nabla}N + h.c.\bigg]\\\nonumber 
&{-}&\frac{\mathring{g}_A}{4 m_{N} f_{\pi}}\bigg[iN^{\dagger}\boldtau\cdot
\left(\dot{\boldpi}
{+}\frac{1}{2f_\pi^2}\boldpi(\boldpi \cdot  \dot{\boldpi})
\right)        \vec{\sigma}\cdot\vec{\nabla}N +  h.c.\bigg]\\\nonumber 
&-&\frac{\mathring{g}_A}{8 \mN \fpi^3} N^{\dagger} \boldpi \cdot 
         (\vec{\sigma} \cdot \vec{\nabla})
        (\dot{\boldpi} \times \boldpi) N
\nonumber \\\nonumber &+&\frac{1}{f_{\pi}^{2}}N^{\dagger}\bigg[ 
         \left(c_3+c_2-\frac{g_A^2}{8m_N}\right) \dot\boldpi^{2}
          - c_3 (\vec{\nabla}\boldpi)^{2}-2c_1\mpi^2 \boldpi^{2}\\
          &-&\frac{1}{2} \left(c_4 + \frac{1}{4m_{N}}\right)  
         \varepsilon_{ijk} \varepsilon_{abc}
          \sigma_{k} \tau_{c}
        \partial_{i}\pi_{a}\partial_{j}\pi_{b}\bigg] N 
   +\cdots  \, \ .
 \label{eq:la1}
\end{eqnarray}
\noindent 
To be  consistent  with chiral symmetry  the chiral Lagrangian is only  
allowed to have terms analytic in the  quark masses. 
Since the square of the pion mass depends linearly on the quark masses, 
only terms of even powers of $m_\pi$  appear in the hierarchy of the  
Lagrangian densities Eq.~\eqref{eq:Lag0}.

Note that in MCS some  nucleon recoil terms $\propto 1/m_N$ in the  
Lagrangian will appear at lower order than 
what is indicated by the Lagrangian order. 
For example, the WT term
leads to a $\pi N$ rescattering vertex proportional to $m_\pi$. 
Whereas in the standard chiral counting the corresponding WT recoil 
correction term, the 
first term in Eq.~(\ref{eq:la1}), would be of next order,
in MCS this  WT recoil correction term 
is proportional to $\vec{p}^{\, 2}/m_N \sim m_\pi$, i.e., 
it is   of the same order as the WT
vertex from the lower order Lagrangian  Eq.~(\ref{eq:la0}). 
A more complete discussion on this topic will be presented 
in a separate subsection, Sec.~\ref{sec:count}. 
We refrain from presenting the
complete  ${\cal L}^{(3)}_{\pi\!N}$ Lagrangian and refer to the 
literature~\cite{ulfbible,Fettes,Fettes2}.

The pion Lagrangian density ${\cal L}_{\pi\pi}^{(2)}$, which gives 
the leading 4$\pi$ vertex  needed for
the  calculation of  the reaction $NN\to NN\pi$  up to  \NNLO,  
reads in the sigma-gauge:
\begin{eqnarray}
 {\cal L}_{\pi\pi}^{(2)}  =
 \frac{1}{2\fpi^2} (\boldpi \cdot \partial^{\mu} \boldpi) 
    (\boldpi \cdot \partial_{\mu} \boldpi )
 - \frac{\mpi^2}{8\fpi^2} \boldpi^4 +\cdots \; , 
\label{eq:lapi}
\end{eqnarray}
and we will also refer to the following S-wave $NN$ Lagrangian, 
which specifies  the two LECs, $C_S$ and $C_T$:
\begin{eqnarray}
{\cal L}_{NN}&=&  - \, \frac{1}{2}C_S\, \left(N^\dagger N\right) 
       \left(N^\dagger N\right) 
-  \, \frac{1}{2}C_T\, \left(N^\dagger\vec{\sigma} N\right) 
       \cdot \left(N^\dagger\vec{\sigma} N\right).
\label{eq:NNlag}
\end{eqnarray}

The derivation of the analytical expression for the transition amplitude, 
${\cal A}_{\rm prod}$, for pion production 
in nucleon--nucleon collisions including \NNLO{} contributions 
will be presented in the next sections.   
At \NNLO{}  
new short range operators 
with the associated LECs describing the short distance processes 
not probed at energies associated with pion-production threshold  
have to be introduced.  
Parity conservation requires the presence of one derivative that
acts either on the pion field --- leading to a p-wave pion with
the final nucleon pairs in an S-wave --- or on a  nucleon field, leading
to a pion s-wave production, as reflected in Table~\ref{table1}. 
Then, however, there must be an additional pion
mass term
present for consistency with the Goldstone theorem --- this is why in the equations
below the pion field appears with a time derivative.
One may write
following Ref.~\cite{cohen} :  
\begin{eqnarray}
 {\cal L}_{NNNN\pi } & = & 
\frac{\tilde e_1}{2\mN\fpi} 
\left[ i \left(N^\dagger ( \boldtau\cdot \dot{\boldpi} )
\vec{\sigma}\cdot \vec{\nabla}N \right) \left(N^\dagger N\right) + h.c. \right] 
\nonumber \\ & &
+\frac{\tilde e_2}{2\mN\fpi} \left[
i(N^\dagger \left( \boldtau \cdot \dot{\boldpi} \right)
\vec{\sigma}\times \vec{\nabla}N)  \cdot
( N^\dagger \vec{\sigma} N) + h.c. \right] 
\nonumber \\ && 
-\, \frac{d}{f_\pi}\, \left(N^\dagger \boldtau  \vec{\sigma}\cdot(\vec{\nabla} \boldpi) N\right)
(N^\dagger N)
+\cdots
\label{eq:4Npi} 
\end{eqnarray}
where the LECs 
$ \tilde  e_i$, $i = 1,$ 2,  are of order ${\cal O}(1/\fpi^2\mN)$ 
and the LEC $d$ is the one in Eq.~(\ref{lag}). 
These LECs will be discussed in more detail in 
Secs.~\ref{sec:swave} and \ref{sec:pwave}, respectively. 
Note that  a pion in an   s-wave can be produced   in   channels  
either with  the  spin-triplet  or  spin-singlet 
$NN$ FSI,  as  shown  in Table \ref{table1}. As  a consequence,   
the  production operator for s-wave pions contains  at  N$^2$LO only
two independent  LECs called  $\tilde e_1$  and  $\tilde e_2$  in Eq.\eqref{eq:4Npi}.  
We also emphasize that the effective Lagrangian of Ref. \cite{cohen}
contains other $(NN)^2\pi$ contact operators both for  s- and p-wave pions that can be shown
to be redundant as a consequence of the Pauli principle \cite{NNLOswave,ch3body,pdchiral,park2}.

\subsection{The hybrid  approach  and the concept of diagram reducibility}
\label{sec:reduce}

\begin{figure}[t]
\includegraphics[scale=0.8]{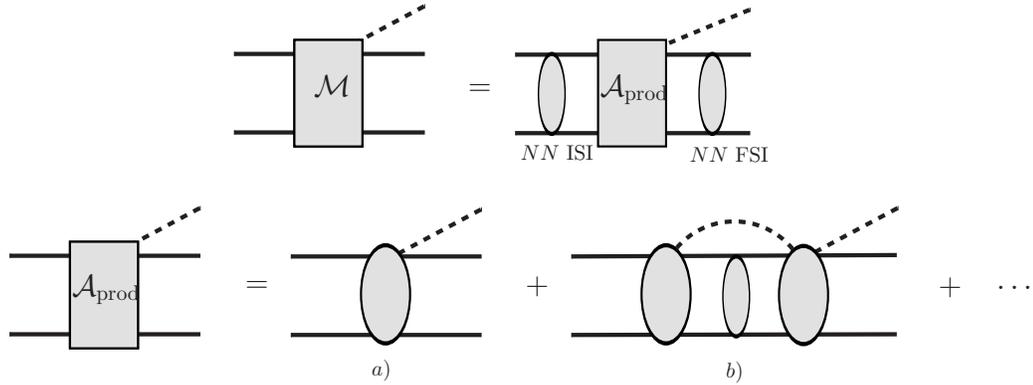} 
\caption{\label{DWBA} Graphical illustation of  the  approach  for  calculating  
the  amplitude $\mathcal{M}$ for $NN\to NN\pi$. 
The initial and final state $NN$ interactions sandwich the $NN\to NN\pi$ 
transition amplitude ${\cal A}_{\rm prod}$ evaluated using ChPT. 
}
\end{figure}

It is well known that perturbation theory is insufficient
to properly describe two or more nucleons at small relative momenta --- a finite sum
of diagrams can neither  produce the unnaturally large $NN$ scattering lengths 
nor bound nucleons in nuclei. 
In order to adopt the perturbative approach of EFT to  few nucleon systems,  Weinberg proposed to classify all possible diagrams 
as reducible or irreducible~\cite{wein90,weinberg1991,OvK,Epelbaum2005}. 
Those diagrams that have a two nucleon cut are classified as  reducible, 
whereas those  
which do not are considered irreducible. 
The latter diagrams make up the $NN$ potential which is to be constructed
according to the rules of ChPT. 
The former diagrams are generated by solving the
Schr\"odinger equation, using the ChPT evaluated  $NN$ potential as kernel. 
This scheme
acknowledges the special role played by the two--nucleon cuts.

Furthermore,  Weinberg suggested  how to calculate 
few-nucleon systems interacting with an  
external low-energy probe \cite{swein1}.   
First,  one should  note that the relevant transition
operators of such an external probe will act perturbatively    
and  as such the transition operators can be calculated using ChPT. 
 The  transition operators are then convoluted with the 
non-perturbative nuclear ($NN$) wave
functions. 
We will apply this scheme to pion  production  in  $NN$ collisions where    
the reaction amplitude  is  calculated  by  sandwiching the  
perturbative production operator, ${\cal A}_{\rm prod}$, 
with $NN$  wave functions  in the initial  and  final  states, 
as  illustrated  graphically   
in Fig.~\ref{DWBA}\footnote{The inclusion  of   diagram  a)   in Fig.~\ref{DWBA}  while neglecting diagram b) and others  is equivalent 
to  the so--called distorted wave Born approximation (DWBA) 
traditionally used in phenomenological calculations.  As will be discussed in the  next section,  for the  calculation 
up-to-and-including N$^2$LO in the MCS  it suffices to keep only  diagram  a). }.  
The transition  operator, ${\cal A}_{\rm prod}$, in  the second line in Fig.~\ref{DWBA},  
consists  of all  irreducible diagrams which   
means  that no  diagrams  with a two-nucleon cut  is part of the production operator.  
The $NN\to NN \pi$ reaction amplitude reads 
\begin{eqnarray}\label{suma}
\!\! \!\! \mathcal{M}(\bm p,\bm p',\bm q)\!\!  =
\int\frac{d^3l}{(2\pi)^3} 
\frac{d^3l'}{(2\pi)^3} 
{\Psi_{\rm ISI}}(\bm p,\bm l)
\maA_\mathrm{prod}(\bm l,\bm l',\bm q)\ 
{\Psi_{\rm FSI}}(\bm l',\bm p')%
\end{eqnarray}
where  the  $NN$  wave  functions in the initial  and  final 
states are 
\be
\Psi_{\rm ISI}(\bm p,\bm l)&=&(2\pi)^3 \delta(\bm p -  \bm l)+ 
\frac{\maM_{\rm ISI}(\bm p,\bm l,E)}{4 m_N^2 [l^2/m_N-E-i0]},\\
\Psi_{\rm FSI}(\bm l',\bm p')&=&(2\pi)^3 \delta(\bm p' -  \bm l')+ 
\frac{\maM_{\rm FSI}(\bm l',\bm p',E')}{4 m_N^2 [l'^2/m_N-E'-i0]}.
\ee
Here  $\maA_\mathrm{prod}$  denotes the production  operator  and  
$\maM_{\rm ISI} $ and  $\maM_{\rm FSI}$  stand for  the $NN$ amplitudes  
in  the  initial and  final states  evaluated at  the energies  
$E$ and $ E'$,  respectively.  
As  will  be discussed in Sec.\ref{sec:swaveD},
since  pion-production  operator  acquires important 
contributions from  diagrams  with the
$\Delta$-resonance  degrees of freedom,  
it is useful 
to generalize   the  
formalism described    to  include $N\Delta$  intermediate states.
This can  be implemented by utilizing    $NN$  models  
which include  explicitly  $NN$  and $N \Delta$ coupled channels 
as in,  e.g., Ref.~\cite{CCF}.  
The  generalisation of  the  above  expressions  is straightforward, e.g., 
see Appendices in Refs.\cite{newpwave,future} for the explicit  
results.

A potential  obstacle  for the application of  
this approach to the $NN\to NN \pi$ reaction is 
the large  momentum  transfer  between  initial and final  nucleons 
inherent in this reaction  already   at threshold
\beq
t \approx -(\bm p - \bm p')^2 \approx  - m_N \mpi.
\eeq 
As discussed in the introduction, this 
brings  a new scale into the problem  compared  to standard ChPT.  
However,   the  use  of the  MCS, which  will be discussed  
in detail in the next  paragraph,  
allows  one  to  built  a new  hierarchy  of   diagrams,  allowing 
a  systematic  perturbative calculation of the production operator.  
Ideally, the $NN$ wave functions should be evaluated within
the same framework, i.e ChPT.   
However, this pion-production process requires 
an $NN$ initial  state  interaction (ISI)  at  an energy which is beyond the 
applicability of todays $NN$ potentials constructed within chiral EFT. 
Therefore, for pragmatic reasons, in a hybrid calculation we make use of the modern 
realistic phenomenological $NN$ potentials~\cite{CCF,CDBonn,AV18} 
to generate the initial and final state 
$NN$ wave functions.   These modern  $NN$ potential models  are  successful in reproducing the  $NN$ phase shifts,  
not only  at  very low  energies  where  the  $NN$ amplitude  is   governed  by  
the scattering length and  effective  range  parameters,  
but  also  at   relatively  high  energies  around the  pion-production  threshold.  
Unlike chiral EFT, however, the  short range  interactions of the 
phenomenological  $NN$ potentials   are  parametrized in model-dependent ways.   
It is therefore  important  to  require  that the variations of the reaction amplitude 
${\cal M}$  for  $NN\to NN\pi$ due to  the  use of   different  $NN$ potentials  
are no larger than the variations expected from the next 
order MCS contributions to the reaction amplitude.

The intrinsic scheme-dependence inherent in the hybrid approach  needs to be quantified.
Specifically,  the hybrid approach, as   illustrated graphically in Fig.\ref{DWBA},  does not  
include the so-called stretched  diagrams,  
where  parts  of  the production operator   are  intimately connected to those from initial (or final) $NN$  interaction  via   
intermediate  states in  the   Time-Ordered Perturbation Theory (TOPT) version  of the  diagrams.
The study  of Ref.~\cite{toy}  revealed  that  the   stretched box contributions,  which do not  possess a two-nucleon cut,  
are   numerically suppressed.   In  addition,  corrections  due to  $NN$  wave  function orthonormalization~\cite{Krebs} 
are ignored in the  hybrid formalism. 
These  corrections normally  appear at relatively higher orders and are  
known to cancel   leading parts of    stretched diagrams in the static limit~\cite{Krebs,Eden}.   
Meanwhile,  a systematic  treatment of  these  effects  is  necessary.  
Such a treatment is , however,  only  possible  if  the  production operator  and  
$NN$ wave functions  are  constructed  within  the  same EFT
formalism.  For   some  attempts  to study  the mismatch 
between   the   construction of the $NN$ wave functions  and the  production operator 
we  refer to  Ref. \cite{BM}.
In  this review  we will not elaborate  further on  these topics.   

\subsection{Power  Counting}
\label{sec:count}
%

In ChPT the standard expansion parameter is $Q/\Lambda_{\chi}$, where
$Q$ is identified either with a typical momentum of the process or
$\mpi$. The key assumption for convergence of the theory is $Q \ll \Lambda_{\chi}$.
As mentioned earlier, the reaction $NN\to NN\pi$ at
threshold involves
momenta of  ``intermediate range'' $p\approx \sqrt{\mpi \mN}$ larger  than $\mpi$ but still
smaller than the $\Lambda_{\chi}\sim \mN$.
In the MCS we are thus faced with a two-scale expansion.
Furthermore, for near threshold pion production,
the outgoing two-nucleon pair has a very low relative three-momentum $p'$.
For counting purposes we therefore assign an order
 $\mpi$ to $p'$, and  the expansion parameter is written as: 
\begin{equation}
\chi_{\rm MCS}\simeq\frac{p'}{p}\simeq\frac{\mpi}{p}\simeq\frac{p}{\mN}.
\label{expansionpapar}
\end{equation}
To implement this scheme properly it is necessary to keep explicitly track of  
momenta and masses in the power counting of the diagrams.

\begin{figure}[t]
\includegraphics[width=12.5cm]{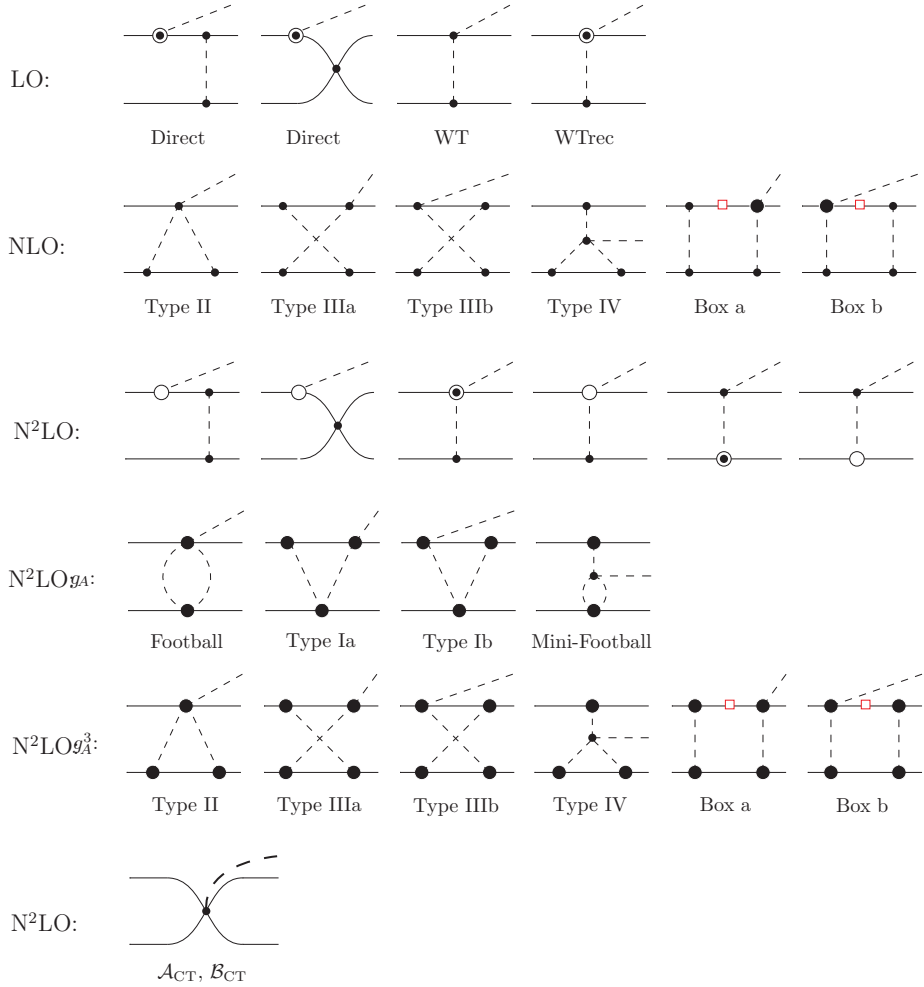}
\caption{\label{fig:allN2LO}
Complete set of diagrams up to N$^2$LO   (in the $\Delta$-less theory)  for  s-wave pions. 
Solid (dashed) lines denote nucleons (pions).
Solid dots correspond to the leading  vertices from
${\cal L}^{(1)}_{\pi\!N}$ and ${\cal L}_{\pi\pi}^{(2)} $,    
$\odot$ stands for the sub-leading vertices from  ${\cal L}^{(2)}_{\pi\!N}$
whereas the blob  indicates the possibility to have  
both leading and subleading vertices from  ${\cal L}^{(1)}_{\pi\!N}$ and  
${\cal L}^{(2)}_{\pi\!N}$,  
the opaque symbol $\circ$  stands for  the  
vertices $\sim1/m_N^2$ from  ${\cal L}^{(3)}_{\pi\!N}$.
The $NN$ contact interaction in the top row 
is represented by the leading S-wave LECs  $C_S$ and $C_T$ from ${\cal L}_{NN}$ whereas the 
contact five-point vertices in the bottom row 
are given by the LECs ${\cal A}_{CT}$ and ${\cal B}_{CT}$.
The red square on the nucleon propagator in the box diagrams indicates 
that the corresponding nucleon propagator cancels with
parts of the $\pi N$ rescattering vertex producing an  irreducible contribution from 
the two box diagrams, 
see Sec.~\ref{NLOloops} and discussion around 
Eq.~(\ref{eq:pipivert}) for further details.}
\end{figure}

In  Fig.~\ref{fig:allN2LO}  we  show  the  hierarchy  of  the diagrams   for  
s-wave pion production  in the MCS up to N$^2$LO.     
The first two LO diagrams in the first row of  Fig.~\ref{fig:allN2LO} 
are sometimes called the direct 
one-nucleon diagrams in the literature,
whereas the last   two are  called the rescattering (WT) diagrams. 
We will discuss both next.  
We stress  that  the one-pion exchange (OPE) and  the $NN$ contact  term  
in  the  direct single-nucleon  operator   are  considered 
parts of  the $NN$  wave  function  and   drawn  explicitly  
only  for the purposes  of power  counting.   
The   estimate  of  the  direct  operator   gives  (in the counting $g_A \sim 1$)
\be
 \left( g_A\, \frac{ \mpi}{\mN } \cdot\frac{p}{\fpi} \right) \cdot \, 
\frac{1}{\mpi}\, \cdot \frac{1}{f_\pi^2}   \sim \frac{1}{\fpi^3} 
 \frac{p}{\mN} \simeq  \frac{1}{\fpi^3}  \chi_{\rm MCS},
\ee
where   the   expression in  the first  bracket on the l.h.s   
corresponds  to  the  recoil  $\pi NN$ vertex given by 
${\cal L}^{(2)}_{\pi N}$ of Eq.~(\ref{eq:la1}),   
the  next term  reflects the energy $v\cdot p\sim m_\pi$ of  
the   nucleon propagator,   see  Sec.\ref{sec:Nprop} for  further details about  the treatment of  nucleon propagators in MCS.  
The last  term   corresponds  to the  estimate  of  the  OPE or the  contact term.  
Analogously,   the estimate of  the rescattering operator   reads
\be
 \left( \frac{ \mpi}{\fpi^2}\right) \cdot\frac{1}{p^2}  \cdot \, 
\left( g_A\, \frac{p}{\fpi}\right) \,   
\sim \frac{1}{\fpi^3}  \frac{\mpi}{p} 
\simeq  \frac{1}{\fpi^3}\, \chi_{\rm MCS}, 
\label{eq:treePC}
\ee
where  the first term on the l.h.s  stands  for  the  Weinberg-Tomozawa vertex 
given in ${\cal L}^{(1)}_{\pi N}$ of Eq.~(\ref{eq:la0}),  
the second term the pion  propagator and  
finally the $\pi NN$ vertex of ${\cal L}^{(1)}_{\pi N}$,  in order.  
Thus  the leading order  diagrams  are linear  
in  $\chi_{\rm MCS}$  for  s-wave pion production.
 
\begin{figure}[ht]
\begin{center}
\includegraphics[width=4.cm]{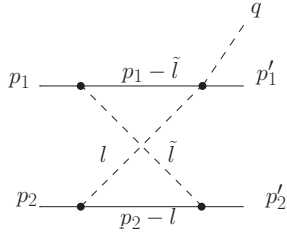}
\caption{\label{fig:ga3struct}
The kinematics  for the  example of the  loop  diagram Type IIIa.  }
\end{center}
\end{figure}

The second row of diagrams in Fig.\ref{fig:allN2LO}   represents  
the  contribution of  irreducible pion-nucleon  loops  at  NLO.   
In order to illustrate the power counting of the loop diagrams in MCS as 
an example we discuss in detail the power counting for diagram type IIIa 
in the second  row of  
Fig.~\ref{fig:allN2LO}. 
First we remind the reader that in heavy baryon ChPT the 
nucleon four-momentum $P^\mu$ is rewritten as 
$P^\mu = m_N v^\mu +p^\mu$ where the velocity is chosen to be 
$v^\mu = (1, \vec{0})$. 
Secondly, for the reaction $NN\to NN\pi$ close to threshold   
the initial nucleon, e.g., nucleon no. 1 
(see  Fig.~\ref{fig:ga3struct} for  the  notation)  
has four-momentum  
$p_1^\mu = (m_\pi /2, \vec{p})$,  meaning $v\cdot p_1 \sim m_\pi$.   
Analogously,   the  momentum  $p_2^\mu = (m_\pi /2, -\vec{p})$. 
It is also important for the following discussion to note that  the 
loop diagrams in the second row of Fig.~\ref{fig:allN2LO}  all have  
a large ``external" momentum  of order $p$ in the propagators     
and that all vertices are given by the lowest order 
Lagrangian in Eq.~(\ref{eq:la0}). 
Furthermore, a consistent power counting of  loop-diagrams requires 
the inclusion of the integral measure
$l^4/(4\pi)^2$ where all components of the loop four-momentum $l$ 
is of order $p$, i.e. $v\cdot l \sim |\vec{l}| \sim p$ for  
irreducible diagrams,   while  for
reducible diagrams that explicitly contain an $NN$ cut,  
the zeroth  component $v\cdot l $  is to be counted of order $m_\pi$,  
as  discussed  in  the  next subsection.   
In addition to this integral measure, in the loop-diagram IIIa   
one has to account for   
two pion propagators ($\sim 1/(p^2)^2$), two nucleon propagators ($\sim
1/(v\cdot l)^2\sim 1/p^2$), three $\pi NN$ vertices ($\sim (p/f_\pi)^3$, 
and one $\pi N$ rescattering vertex 
($\sim v\cdot l/f_\pi^2\sim p/f_\pi^2$). 
Combining 
all these factors and using $4\pi f_\pi\sim m_N$, 
one obtains the order estimate for this diagram as follows
\be
\frac{p^4}{(4\pi)^2} \, \cdot \frac{1}{(p^2)^2}\, \cdot\frac{1}{p^2}\, 
\cdot\left(\frac{p}{f_\pi}\right)^3\,\cdot \frac{p}{f_\pi^2} 
\sim \frac{1}{\fpi^3}\frac{p^2}{m_N^2}\simeq \frac1{f_\pi^3}\chi_{\rm MCS}^2.
\ee
This counting order estimate of diagram IIIa should be compared with 
the counting of the leading order diagrams for the $NN\to NN\pi$ 
reaction  vertex in Fig.~\ref{fig:allN2LO}  
estimated above.  
Thus we  find that diagram type IIIa in Fig.~\ref{fig:allN2LO} 
indeed contributes at NLO.

At this point the reader should note a peculiar feature of the MCS counting.
The order estimate of a loop diagram in MCS implies 
that this diagram also contributes at all higher orders in MCS. 
For example, diagram IIIa is estimated above to be of order NLO 
but, unlike standard chiral counting, this diagram also 
contributes at \NNLO{} and higher MCS orders.  
The reason for this behaviour lies in the two scale expansion employed there, since
the evaluation of the loop diagrams typically reveals expressions like
$\mpi \log(1+\mpi/p)$. In a theory where $p$ is of order $\mpi$ the given term
contributes to a single order only. However, in the MCS
$\mpi/p \sim \chi_{\rm MCS}$, and thus a Taylor expansion of the logarithm
becomes necessary.
We will return to this discussion in Sec.~\ref{sec:Nprop}. 

At  \NNLO{},  there are several  contributing tree-level diagrams  
which are  topologically   
similar  to those  at LO  but  with  sub-leading  vertices  from  
${\cal L}^{(2)}_{\pi\!N}$  and  even  ${\cal L}^{(3)}_{\pi\!N}$.
These diagrams are shown in the  third  row  in  Fig.~\ref{fig:allN2LO}.  
In  addition,   one needs   to  account  for  the  pion-nucleon  loops 
which, at   this  order,   can  be  combined  in  two  series of amplitudes, one  
proportional  to     $\gA^3$ with a topology like the NLO pion-nucleon loop diagrams   and    one proportional to $\gA$. 
These diagrams are given in rows four and five in    Fig.~\ref{fig:allN2LO}. 
 To the  order we  are  working,  it  suffices  to  include  the
  sub-leading vertex  from ${\cal L}^{(2)}_{\pi\!N}$  only once  
(but we have to consider all possible permutations of this sub-leading  vertex) 
in these loops while retaining  the  other 
   vertices  at   leading  order, see  Ref.\cite{future}  for further details.   
Also at N$^2$LO  for  s-wave pions  there  are  two    $(NN)^2\pi$ contact  operators  
${\cal A}_{CT}$ and ${\cal B}_{CT}$ which represent  contributions  
to the channels  with the spin-singlet  ($^3\!P_0 \to {^1 S_0} s $) 
and the spin-triplet  ($ ^3P_1 \to { ^3\!S_1} s$)  $NN$ FSI,   
in accordance with  the notation of  Table \ref{table1}.
The LECs  ${\cal A}_{CT}$ and ${\cal B}_{CT}$,  
shown in the last row of  Fig.\ref{fig:allN2LO}  
can be   expressed  in terms of  linear  combinations  of  
the  LECs  $ \tilde  e_1$  and $ \tilde  e_2$
in the Lagrangian \eqref{eq:4Npi} to be used in   Sec. \ref{sec:renorm1}.

Finally,   we  note  that  the  diagram  b)  in  Fig~\ref{DWBA},  
which  is  also a  part  of the  production  operator, 
starts  to  contribute at \NNNLO,   that  is  one  order  higher  
than  considered in  the  present  study.     
This  type  of 
operators  contains   the three-body  $\pi NN$  cut,   
the  contribution of  which  is   strongly  suppressed  near  pion 
production  threshold.

\subsection{The heavy nucleon propagator in ${\rm MCS}$} 
\label{sec:Nprop}
%
A major difference between standard chiral counting  
(see e.g. Ref.~\cite{ulfbible})
and MCS~\cite{hanhart04}
is associated with the different treatments of the
heavy nucleon propagator. In heavy baryon ChPT (HBChPT) the nucleon momentum $P^\mu$
is written as $P^\mu \!=\!m_N \, v^\mu\!+\!p^\mu$,  and 
it is assumed that $|p^\mu| \!\ll\! m_N$. 
In terms of $p_\mu$ the Feynman propagator 
for the heavy nucleon is expressed as 
\begin{eqnarray}
S_N(p) = i\; \frac{ \not\!\! P +m_N }{P^2-m_N^2+i0} =\frac{ i\; } 
{v\cdot p +\dfrac{p^2}{2m_N}+i0} 
 \!\!\left( 
\frac{1+\gamma_0 }{2} +\frac{\gamma_0\; p_0 }{2m_N} 
- \frac{{\vec\gamma}\cdot \vec{p} }{2m_N} 
\right) \; ,
\label{eq:Nprop}
\end{eqnarray}
where we introduced $v^\mu=(1,0,0,0)$.
In the standard HBChPT scheme,
the free heavy-nucleon Lagrangian is chosen 
in such a manner that the heavy-nucleon propagator 
is given by 
\begin{eqnarray}
S_N (p)= \frac{i}{v\cdot p+i0 } \; .
\label{eq:Nprop1}
\end{eqnarray} 
The difference between  the propagators in
Eqs.~(\ref{eq:Nprop}) and (\ref{eq:Nprop1}), which scales as $(p/m_N)^n$  with $n=1,2,...$,
is treated  as perturbative recoil corrections in standard HBChPT,
and is  included in  ${\cal L}_{\pi N}^{(2)}$  and higher-order  Lagrangians  
in Ref.~\cite{ulfbible},  cf.  e.g.   $1/m_N$ terms in the last two 
rows in Eq.~\eqref{eq:la1}. 
On the other hand, in some cases  the  use  of  Eq.~\eqref{eq:Nprop1}  
is not justified and one should utilize   Eq.~\eqref{eq:Nprop}. 
One example 
where   a careful  treatment  of the  nucleon propagator in 
Eq.~\eqref{eq:Nprop} is required   is  
the  two-pion-exchange  contribution   to  $NN$ scattering   
discussed  by  Weinberg  in Ref.\cite{weinberg1991}.
In this  case  the second iteration of  OPE  results in  
the box  diagram  with  an $NN$ intermediate state which is reducible.
A naive  evaluation of this  diagram  with  the standard HB  propagator  
Eq.~\eqref{eq:Nprop1}  would  yield   the  pinch  singularity.  
To properly  account  for  the two-body singularity   ($NN$ cut)
  one  should  therefore keep  the nucleon kinetic energy  
in  the denominator of Eq.~\eqref{eq:Nprop} unexpanded,  i.e.  
this HB propagator for both intermediate nucleons deviates from the strict HBChPT formulation and becomes 
\begin{eqnarray}
S_N(p) =   \dfrac{i}{
v\cdot p -\dfrac{{\vec p}^{\, 2}}{2m_N}+i0} \;,
\label{eq:NNprop}
\end{eqnarray}
while the  residual  terms  in Eq.~(\ref{eq:Nprop}) 
are treated as perturbations in the Lagrangian.  
Only then the $p_0$- integration leads to a proper two--nucleon
propagator that in particular shows a strong infrared enhancement
calling for a re--summation in, e.g.,  a Schr\"odinger equation ---
this is why they are called reducible. 
In contrast,  in $NN$ scattering  diagrams  which do not 
possess the two-nucleon  cut  (e.g.  in the  crossed boxes) the standard HBChPT
expansion of the  nucleon propagator  is justified,  i.e.  $v\cdot p \sim|\vec p\, |$.  

The treatment of the  nucleon propagator  in pion production is  
actually  quite similar  to  $NN$ interaction~\cite{subloops,lensky2,ksmk09}. 
The  calculation   of $NN$ reducible  diagrams  for energies near the pion
production threshold  requires   
a  non-perturbative treatment of  nucleon recoils  as given by Eq.~\eqref{eq:NNprop},  
which means   that  zeroth  components of   4-momenta    
are counted as    $v\cdot p \sim{{\vec p}^{\, 2}}/(2m_N) \sim \mpi$.
Meanwhile,  to evaluate  irreducible  loop  contributions  in  
MCS  it  suffices  to  use   Eq.\eqref{eq:Nprop1}  for the propagator  structure 
while  keeping  $1/m_N$  (and higher-order) corrections  perturbatively.  
Since there  is no $NN$ cut,   in the latter  case  
$v\cdot p \sim|{\vec p}\, | \sim\sqrt{{\mpi}{m_N}} \gg {{\vec p}^{\, 2}}/(2m_N) $
since the energy scale is introduced into the diagrams from the pion
progagators (cf. Appendix E of Ref.~\cite{hanhart04}).

It might be  also  instructive  to  discuss  the  special case  of   
the pion  emission from a   single nucleon 
characteristic for  the production process,  
see  the  first direct 
diagram in Fig.\ref{fig:allN2LO}  where  the OPE  is part of the  $NN$ FSI.  
Although  this diagram contains  an $NN$  intermediate  state,   it  can not  
go on shell    since a massive pion  can not  be produced  by  a free  nucleon.   
Therefore,  naively   one is  tempted  to expand  the  nucleon propagator 
using standard  HBChPT,  as   it was  done in Ref.\cite{novel}.  
Based on  this  expansion it was  argued 
  that the heavy  baryon scheme should not be used for pion production. 
We are now in a position to answer the critical comments in Ref.~\cite{novel} 
regarding  the non-convergence of the heavy nucleon propagator.
The expansion of the nucleon propagator in the first diagram of  Fig.~\ref{fig:allN2LO} 
would read, following Ref.~\cite{novel} 
$$ 
S_N(p_1-q) = 
\frac{1}{v\cdot (p_1- q)-\vec p_1^{\, 2}/(2m_N)}= - \frac{2}{\mpi}
\sum_{n=0}^{\infty} \left(-\frac{\vec p_1^{\, 2}}{m_N \mpi}\right)^n \ ,
$$ 
where  on the r.h.s. $v\cdot p_1=\mpi/2$ and $v\cdot q= \mpi$ 
was used at threshold ($\vec{q}=0$) and the $1/m_N$ terms 
were treated as correction. 
Clearly this series
does not converge, since $\vec p_1^{\, 2}\simeq m_N\mpi$. 
However,  had Ref.~\cite{novel} employed the on--shell condition for the
external nucleon,  $v\cdot p_1\simeq {\vec{p}_1^{\; 2} }/{2m_N}$, the
propagator would have taken the correct expression $-1/\mpi$ right from the start.
Thus, when applied properly, the heavy baryon prescription 
can be used for pion-production reactions~\cite{subloops}.

\section{s-wave pion production at threshold}
\label{sec:swave}

\subsection{ Operator amplitudes from tree-level  diagrams  up  to  \NNLO:   }
\label{sec:diagrams}

The LO tree-level diagrams were discussed in Sec.~\ref{sec:count}  
which includes the LO rescattering diagram containing the 
isovector Weinberg-Tomozawa vertex and its recoil contribution given by  
${\cal L}^{(1)}_{\pi N}$ and  ${\cal L}^{(2)}_{\pi N}$, respectively. 
The same two Lagrangians, Eqs.~(\ref{eq:la0}) and (\ref{eq:la1}),  
will contribute to \NNLO{} tree-level rescattering diagrams with the
same topology as the LO rescattering diagram in Fig.~\ref{fig:allN2LO}. 
In the MCS the LO diagrams are the direct diagrams and the rescattering diagrams, 
labelled "Direct'' and "WT'' in Fig.~\ref{fig:allN2LO}. 
 
\begin{figure}[h] 
\includegraphics[width=12.5cm]{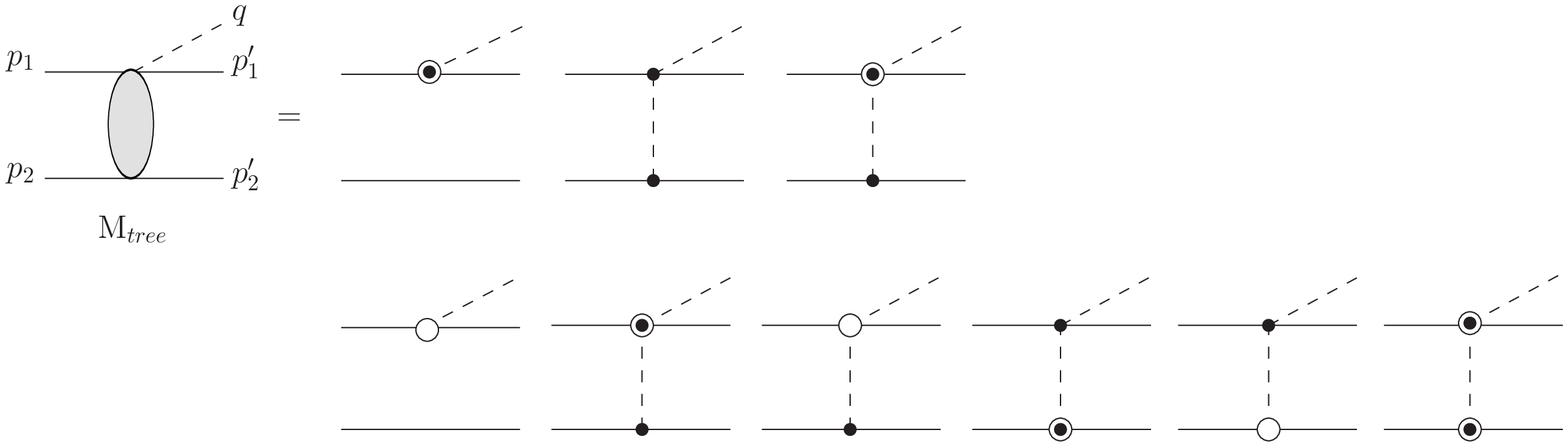}
\caption{ \label{fig:treeN2LO}
The single nucleon and the rescattering diagrams which contribute  to  
s-wave pion production up to \NNLO{}, are given.    
The meaning of the symbols is the same as in Fig.~\ref{fig:allN2LO}.    
The diagrams in the  first  row  on  the r.h.s  appear  at LO,   
the   diagrams in the second row  give the \NNLO{}  contributions.     
The last diagrams  in  the first and second  rows  involve the WT recoil correction,
whereas the second diagram in the second row involves 
 rescattering vertices proportional to the $c_i$.   
In the hybrid approach the first diagrams in each row, the single nucleon diagrams, 
will contribute to ${\cal A}_{\rm prod}$ in Eq.~(\ref{suma}) when convoluted with initial and final
$NN$ wave functions as discussed in the text. 
}
\end{figure}

The rescattering operator  at  LO  reads \footnote{ 
At this order it suffices to replace  $\mathring{g}_A$ by its 
physical value  $g_A$.  
The renormalization corrections to $\mathring{g}_A$ will 
be discussed in Sec.~\ref{sec:renorm2}.}: 
\begin{eqnarray}
	i M^\text{LO}_{\rm rescat} = i M_\text{WT}^\text{LO}+i M_\text{WT}^\text{recoil}
	=\frac{g_A}{2 f_\pi^3} 
	\frac{ v \cdot q}{k_2^2-m_\pi^2+i0}
		(S_2 \cdot k_2) \taux^a \, ,
\label{eq:treeLO}
\end{eqnarray}
where the superscript $a$ ($a$=1,2,3)  here  and in what follows    
refers to the isospin quantum number of  the outgoing pion field, 
and where the antisymmetric [symmetric] 
isospin operator is $\taux^a  = i (\boldtau_1 \times \boldtau_2)^a$ 
[$\taup^a = (\boldtau_1+\boldtau_2)^a$].   
The  momenta are defined  in Fig.\ref{fig:treeN2LO}, and  we define 
$k_2=p_2-p_2'$.   

The  rescattering operator  at  \NNLO{}  includes   the   $1/m_N$ corrections 
due to  the vertices given by ${\cal L}^{(2)}_{\pi\!N}  $   and 
also the corrections  $\propto 1/m_N^2$  from ${\cal L}^{(3)}_{\pi\!N}$. 
We  call these amplitude operators 
 $M_{\rm rescat1}^\text{N$^2$LO}$ and  $M_{\rm rescat2}^\text{N$^2$LO}$, respectively.  
The  explicit expressions are:
\begin{eqnarray}
	i M_{\rm rescat1}^\text{N$^2$LO} &=& \frac{g_A}{f_\pi^3} \frac{ (S_2 \cdot k_2) 
                 \tau_2^a }{k_2^2-m_\pi^2+i0}
		\left[  
			4c_1 m_\pi^2 - v \cdot q \, v \cdot k_2 
                         \left( 2c_2+2c_3-\frac{g_A^2}{4 m_N} \right) 	
		\right]  
\nonumber 
\\ 
            &-& \frac{g_A}{f_\pi^3}
	\frac{(v\cdot q )\,  \taux^a }{k_2^2-m_\pi^2+i0}
           \frac{S_2 \cdot (p_2+p_2^\prime)}{4 m_N}  (v \cdot k_2 )
            +   (1 \leftrightarrow 2),  
\label{treeNNLO1} 
\\
	i M_{\rm rescat2}^\text{N$^2$LO} &=& 
	\frac{g_A}{f_\pi^3}
	\frac{v\cdot q }{k_2^2-m_\pi^2+i0} 
		\Bigg\{ 
	- \tau_2^a \, (S_2 \cdot k_2) 
	  \frac{k_2 \cdot (p_1+p_1^\prime)}{m_N^2} \left( m_N c_2 -\frac{g_A^2}{16} \right) 
\nonumber
\\ 
	+ && \hspace*{-0.5cm}\taux^a \, (S_2 \cdot k_2) 
		\left[
			\frac{\vec{p}_1^{\,2}+\vec{p}_1^{\,\prime 2}}{16 m_N^2}
			+ \frac{1+g_A^2+8m_N c_4}{8 m_N^2} 
                 \left( [S_1\cdot k_2, S_1 \cdot (p_1+p_1^\prime)] + \frac{k_2^2}{2}\right) 
		\right]
\nonumber
\\ 
	&-& \frac{\taux^a }{8 m_N^2} 
		\left[  
			(S_2 \cdot p_2^\prime) p_2^2 -  (S_2 \cdot p_2) p_2^{\prime 2}
		\right] 
		\Bigg\} +  (1 \leftrightarrow 2),
\label{treeNNLO2}
\end{eqnarray}
%
The  first  two  terms  in the curly bracket  in Eq.~\eqref{treeNNLO2}   
are  due to recoil corrections  to  the $\pi \pi N  N$  vertices from ${\cal L}^{(3)}_{\pi\!N}$ 
while  the  last   one stands  for  the  analogous correction  to the  
$\pi NN$ vertex  at the  same order.  
Both  amplitudes $M_{\rm rescat1}^\text{N$^2$LO}$  
and $M_{\rm rescat2}^\text{N$^2$LO}$ contribute  to  the   isoscalar   and   isovector part of the production amplitude 
${\cal A}_\mathrm{prod}$ in Eq.~(\ref{suma}).

In  addition to the rescattering operators,  
one  has  to   account  for  the  contributions 
to the transition operator amplitude ${\cal A}_{\rm prod}$ from the  pion  
emission from  a single nucleon,  the so-called  direct  diagrams  
which are  shown  in  
the first and third rows of Fig.\ref{fig:allN2LO}.  
Note     that  in the hybrid approach  
the  OPE  or the $NN$ contact term in these direct diagrams, 
which  appear together  with the $\pi NN$ vertex  for the 
outgoing pion  in Fig.\ref{fig:allN2LO},   will   be  considered  as   part 
of the final (or initial)  $NN$  wave  function as 
discussed in  Sec.~\ref{sec:reduce}.  
In Fig.~\ref{fig:treeN2LO} the OPE and the $NN$ contact term are 
therefore not shown in the direct diagrams.  
The  explicit expressions for these  diagrams are derived in Ref.~\cite{future}.
In Sec.~\ref{sec:NNLO-3leveldia} we will discuss the influence of these direct 
diagrams in more details.

The  LO and \NNLO{} operator amplitudes generated by the 
diagrams in Fig.\ref{fig:treeN2LO} contribute  to the  \NNLO{} s-wave pion-production operator 
 ${\cal A}_{\rm prod}$ from pion-nucleon diagrams. 
We postpone the discussion of the  s-wave pion-production operator  
until the end of Sec.~\ref{sec:swaveD}, where the $\Delta$
degrees of freedom will be added to this operator.

\subsection{A  discussion of the tree-level diagrams up to ${\rm N^2LO}$ }
\label{sec:NNLO-3leveldia}

The LO diagrams, which contain only pion and nucleon degrees of freedom
that contribute to the reaction $NN\to NN\pi$ for s-wave pions, 
are shown in the first line of Fig.~\ref{fig:allN2LO}. 
Traditionally, the 
direct diagrams have been evaluated numerically by including the
pion propagator in the  distorted $NN$ wave functions, i.e. only the
single nucleon pion-production vertex gives the transition operator. 
 Numerically, in the traditional distorted wave Born approximation approach,
the direct diagram appears to be significantly smaller than the estimate based
on our naive MCS's dimensional analysis.  
The suppression of the direct diagram in Fig.~\ref{fig:allN2LO} 
comes from two sources:
first,  there is the momentum mismatch between the initial and final
distorted nucleon wave functions~\cite{hanhart04} --- see also 
Refs.~\cite{BM,nogi2002}
for  more detailed discussions. 
Secondly, there are accidental 
cancellations from the final state interaction present in both channels,
$pp\to pp\pi^0$ and $pp\to d\pi^+$, that are not accounted for in the power
counting. 
Specifically,
the $NN$ phase shift in the final $NN$ 
$^1S_0$ partial wave relevant for $pp\to pp\pi^0$ crosses zero  at an energy
close to the pion-production threshold \cite{SAID,ARN2007}.  
All realistic $NN$
scattering potentials that reproduce this feature show in the half-off-shell
amplitude at low energies a zero at off-shell momenta of a similar magnitude.
The exact position of the zero varies between different models, such that the direct
production amplitude turns out to be quite model dependent.
The suppression mechanism of the direct term for the reaction $pp\to d\pi^+$
comes from  a strong
cancellation between the deuteron S-wave and D-wave components.
Thus, it is not
surprising that numerically the direct diagrams in both channels
are about an order of magnitude smaller than the  LO 
rescattering   amplitudes  
which are consistent with the dimensional analysis in the MCS.
Since this LO rescattering contribution is forbidden by selection rules for $pp\to
pp\pi^0$ while allowed for $pp\to d\pi^+$, 
one gains an immediate theoretical understanding why
it is a lot more difficult to reproduce the measured 
cross section for the former reaction.

\subsection{Two-pion  exchange diagrams  at  ${\rm NLO}$ }
\label{NLOloops}

The observation that the LO contributions to  the pion s-wave reaction 
$pp\to pp\pi^0$ 
are highly suppressed~\cite{cohen,park,sato}   
makes this particular reaction 
an ideal probe of the higher order contributions, 
which give smaller corrections to most $NN\to NN\pi$ reaction amplitudes. 
Among the next order terms illustrated in Fig.~\ref{fig:allN2LO},  
two-pion exchange loop diagrams were evaluated~\cite{DKMS,Ando} 
within  standard ChPT and  some 
two-pion exchange diagram contributions were found to be very 
large compared to the tree-level contributions.
However, for the channel $pp\to pp\pi^0$ the sum of 
NLO diagrams type II\footnote{
The amplitude for diagram type II in Fig.~\ref{fig:allN2LO} as given in 
Ref.~\cite{DKMS} has the wrong sign.}, III and IV in
Fig.~\ref{fig:allN2LO} is zero due to a cancellation between 
individual diagrams \cite{HanKai}, 
where diagrams  Box a and b  in Fig.~\ref{fig:allN2LO}   
are absent since the WT vertex does not  contribute to this channel. 
Some aspects of these diagrams and this cancellation was further 
discussed in, e.g., 
Refs~\cite{ksmk09,ksmk07}. 
The cancellation among the diagrams for $pp\to pp\pi^0$ was confirmed 
in Ref~\cite{NNLOswave}.  Meanwhile,  
the sum of diagrams II -- IV gives a finite answer for the channel
$pp\to d\pi^+$, as was noted in Ref.~\cite{HanKai}.   
Since the net contribution of these operators depends linearly on the $NN$
relative momentum,   a large sensitivity of the observables 
to the short-distance $NN$ wave functions was found in Ref.~\cite{Gardestig}.  
This linear $NN$ momentum sensitivity to short-distance 
$NN$ forces at NLO, ${\cal O}(m_\pi / m_N)$, is 
troubling since  
chiral symmetry requires interactions  to be proportional to 
an even power of the pion mass, $m_\pi$. 
Specifically,  at NLO  there is no five-point contact interactions that could absorb this linear 
momentum behaviour of the loop operators.  
This puzzle was resolved in Ref.~\cite{lensky2}, 
where it was demonstrated
that for the deuteron channel there is an additional 
contribution at NLO, namely the
box diagrams in Fig.~\ref{fig:allN2LO}, 
due to the time-dependence of
the Weinberg--Tomozawa pion-nucleon  vertex.    
The cancellation of the NLO loop diagrams found in Ref.~\cite{lensky2}  reinforces the 
importance of applying ChPT to the $NN\to NN\pi$ reactions.

\begin{figure}[t]
\includegraphics{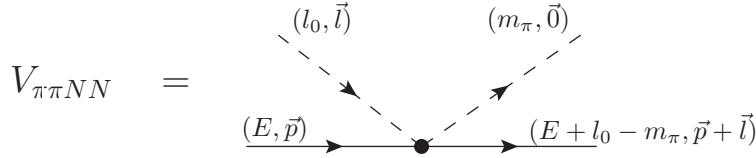}
\caption{\label{fig:vpipinn}
The $\pi N$ rescattering vertex:
definition of kinematic variables as used in Eq.~(\ref{eq:pipivert})
where, e.g., $v\cdot l = l_0$.}
\end{figure}
To demonstrate that the irreducible parts of the box diagrams 
in Fig.~\ref{fig:allN2LO} contribute to
the NLO amplitude,  
we write the expression for the WT $\pi N$ rescattering vertex, $V_{\pi\pi NN}$, 
in the notation of Fig.~\ref{fig:vpipinn}, where one pion is the outgoing pion 
of four-momentum $(m_\pi , \vec{0})$ and the 
other pion is part of the loop with four-momentum $l$: 
\begin{eqnarray}
V_{\pi\pi NN}&=& 
l_0{+}\mpi {-}\frac{\vec l\cdot(2\vec p+\vec l)}{2\mN}
\nonumber \\
&=& {2\mpi}+ {\left(l_0{-}\mpi {+}E{-}\frac{(\vec l+\vec p)^2}{2\mN}+i\nolik \right)}-
{\left(E{-}\frac{\vec p\, ^2}{2\mN}+i\nolik\right)}
\ . 
\label{eq:pipivert}
\end{eqnarray} 
We keep the leading WT vertex and its nucleon recoil correction,
which are of the same order in the MCS, as explained below Eq.~(\ref{eq:la1}). 
For simplicity, we omit the isospin dependence of the vertex.
The first term in the last line is the WT-vertex for  kinematics with the 
on-shell incoming and outgoing nucleons,
the second term the
inverse of the outgoing nucleon propagator while the 
third one is the inverse of the
incoming nucleon propagator; compare 
the nucleon propagator expression in Eq.~(\ref{eq:NNprop}). 
Note that for on-shell 
incoming and outgoing nucleons 
(as in diagram WT in Fig.~\ref{fig:allN2LO}), 
the expressions in brackets 
in Eq.~\eqref{eq:pipivert} vanish, and
the $\pi N\to \pi N$ transition vertex
takes its on-shell value $2\mpi$ 
(even if the incoming pion is off-shell).
This is in contrast to standard phenomenological 
treatments, e.g., Ref.~\cite{koltun},
where $l_0$ in the first line of  \eqref{eq:pipivert} is
identified with $\mpi/2$, the energy transfer in the on-shell 
kinematics for $NN\to NN\pi$,
but the recoil terms in Eq.~(\ref{eq:pipivert}) are not considered.
However, as argued earlier 
$\vec{p}^{\ 2}/\mN\approx \mpi$ so that
the recoil terms are to be kept in the vertices and also in
the   nucleon propagator as discussed in Sec.~\ref{sec:Nprop}. 
The MCS is designed to properly keep track 
of these recoil terms.
A second consequence of Eq.~(\ref{eq:pipivert}) is that  only the
first term gives a reducible diagram. 
The second and third terms in Eq.~(\ref{eq:pipivert}), however,
lead to irreducible contributions, since one of the nucleon propagators is
cancelled by parts of the adjacent WT $\pi N$ rescattering vertex expression. 
These irreducible contributions are illustrated by red squares
on the nucleon propagators in the two box diagrams of 
Fig.~\ref{fig:allN2LO}. 
It was shown explicitly in Ref.~\cite{lensky2} that these
induced irreducible contributions
cancel exactly the finite remainder of the NLO loop diagrams (II -- IV)
in the $pp\to d\pi^+$ channel.
As a consequence, there are no contributions at NLO  for both
$\pi^0$ and $\pi^+$ s-wave  production from two initial nucleons.

\subsection{Total  cross section in $NN\to d\pi$ at threshold  up to ${\rm NLO}$}
\label{sec:obs}
\begin{figure}[t!]
\includegraphics[width=10.5cm]{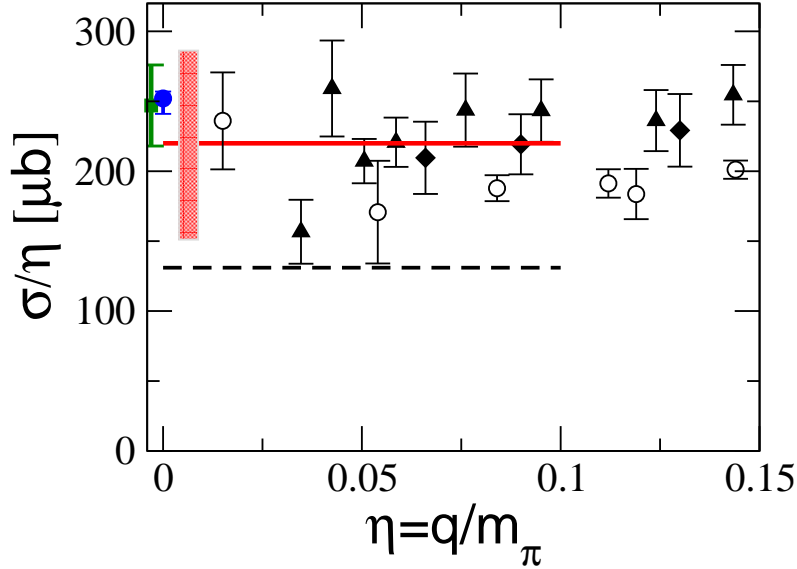} 
\caption{\label{fig:obs} 
Comparison of the results of Ref.\cite{lensky2} 
with experimental data for total cross section in $NN\to d\pi$. 
The data sets are from Refs.~\cite{Hutcheon,Heimberg,Drochner,dgotta,pidexp}.
The dashed line corresponds to the model of Koltun and Reitan~\cite{koltun}. 
The solid red curve represents the results  for   s-wave pion-production 
at  threshold, i.e.  the coefficient  $\alpha$ in Eq.\eqref{totXS}, 
as given in Ref.\cite{lensky2}.  
The filled red rectangular box indicates the theoretical uncertainty of the NLO calculation. 
}
\end{figure}
In this subsection we briefly discuss  the status of  the  description to  NLO of  
threshold pion-production data  
when the final two nucleons are in the spin-triplet ($^3S_1 - ^3 {\! D_1}$)  channel.  
The  total cross section in the  reaction $NN\to d\pi$    
close to threshold  is historically written in  the form
\be
\sigma=  \alpha \eta + \beta \eta^3 + \dots \, ,
\label{totXS}
\ee
where the dimensionless variable $\eta$ is defined in the 
center-of-mass system as the pion 
momentum in units of the pion mass, $\eta = |\vec{q}|/m_\pi$  
and  ellipses stand  for the  terms of higher order in $\eta$. 
In Fig.~\ref{fig:obs}  we compare  the chiral 
 EFT  evaluation of  $\alpha$ by Ref.~\cite{lensky2}   with
 experimental data.    
Note that the green square and 
the blue circle correspond to 
the most recent measurements~\cite{dgotta,pidexp} of  the 
coefficient $\alpha$ in Eq.~\eqref{totXS}, 
see also Sec.~\ref{sec:pid-atom}.  
The value for  $\alpha$  in Refs.~\cite{dgotta,pidexp} was
extracted  from the high-precision  lifetime measurement  of the $\pi^-d$ atom at  PSI.  
 The dashed  curve  in Fig.~\ref{fig:obs} 
 corresponds to the  calculation  of Ref.~\cite{koltun},  
in which  the nucleon recoil  corrections to the Weinberg-Tomozawa 
 operator were  neglected.   
As     pointed  out in the Introduction  and  discussed earlier in this section, the important 
 observation of  Ref.~\cite{lensky2} was  that    
the  nucleon recoil corrections  $\propto 1/m_N$ contribute  
in MCS    at  lower  order 
  than  what  is  indicated naively by the order of the  Lagrangian.  
Including the WT recoil correction, labelled ``WTrec" in  Fig.~\ref{fig:allN2LO},  in Eq.\eqref{eq:treeLO}  
enhances the WT operator by a factor of 4/3. 
This enhancement resulted in an increase  of  the cross section  by  
about the missing  factor 2 already  at  LO.   
Furthermore,   as  was shown in Ref.~\cite{lensky2}  and  
discussed in detail in the previous subsection,  all  loops   at  NLO  cancel. 
 Given  the expansion  parameter  in the MCS,   
the  theoretical  uncertainty  
for  $\alpha$ is  about   2$\chi_{\rm MCS}$ at NLO,  
as  shown  by  the  filled rectangular box  in Fig.~\ref{fig:obs}.


\subsection{Two-pion exchange diagrams to ${\rm N^2LO}$ }
\label{sec:2pions}
At \NNLO{} there are new loop diagrams contributing to the 
$NN\to NN\pi$ s-wave pion reaction amplitude. 
The \NNLO{} two-pion exchange diagrams 
shown in Fig.~\ref{fig:allN2LO} have been evaluated 
in Ref.~\cite{NNLOswave}. 
One finds that also at this order a similar cancellation
pattern appears as  discussed in the 
previous subsection for the NLO contributions.
However,  at \NNLO{} the cancellation is not complete and leaves us with
a non-zero 
result, which originates from diagrams proportional to 
$g_A$ of type I and the one called 
``mini-football", see the fourth row of diagrams in Fig.~\ref{fig:allN2LO}, 
and from diagrams proportional to 
$g_A^3$ of type III and IV in the fifth row of diagrams 
in Fig.~\ref{fig:allN2LO}. 
The other loop diagrams give no net contribution to the 
amplitude for  s-wave pion production.

There are some fundamental reasons why at least some of  these 
cancellations are to be expected. 
The Goldstone boson nature of the pion requires the 
pion to decouple from the nucleon field as 
its four-momentum components become zero in the chiral limit.
This  chiral symmetry  requirement is  exemplified by the 
Lagrangian densities in the so-called sigma gauge   
given in Eqs.~(\ref{eq:la0}), (\ref{eq:la1}) and (\ref{eq:lapi}). 
Another aspect of ChPT,  which forces us to consider classes of diagrams, 
is that it is the  $U$ matrix which enters in the interactions, 
and the $U$ matrix    
depends non-linearly on the pion field. 
 The chiral matrix-field $U(x)$ is an SU(2) matrix that 
has standard chiral transformation properties, 
see e.g. Ref.~\cite{ulfbible}. 
There are several  equivalent ways one can define the pion field 
content of the matrix-field $U$. 
The $U$ matrix  can be written in terms of the pion field a la 
Weinberg~\cite{weinberg1991}, on an exponential form used in, 
e.g.~\cite{ParkMinRho93}, 
or as in Ref.~\cite{ulfbible}'s  sigma-gauge form   
which is chosen in this review.    
This $U(x)$ field is expanded  
in terms of $[{\vec \pi}(x)/f_\pi]$ in order to 
generate the Lagrangian expression in, e.g., Eq.~(\ref{eq:la0}). 
The different ways to categorize $U$ in terms of the pion field 
have consequences for interaction terms involving three or more pions,
e.g., the last term in Eq.~(\ref{eq:la0}).
Even the nucleon fields have a certain arbitrariness 
since we can change $N(x)$ to $N^\prime (x)$ 
as long as this 
transformation is consistent with chiral symmetry, 
see e.g., Ref.~\cite{BLee1972}. 
This arbitrariness of defining the fields in the Lagrangian forces one to
consider not individual diagrams, but classes of diagrams, and 
at a given order only the sum 
of the diagrams is  
independent of the definitions of the fields in the Lagrangian.  

It should be remarked that 
similar to what occurs in HBChPT,  
the cancellation of the sum of diagrams for $NN\to NN\pi$ is also found 
using Gasser {\it et al.}~\cite{gss88}'s covariant formulation of ChPT  
as shown by Filin~\cite{Arseniy2012}.   
The pattern of cancellations in this case is even more elegant than in HBChPT, where  to see 
the same cancellations one needs  to keep track of all $1/m_N$  terms in the expansion.   
Specifically, several not obvious cancellations among the 
\NNLO{} two-pion exchange diagrams in Fig.~\ref{fig:allN2LO}
were observed within the MCS formulation of HBChPT in Ref.~\cite{NNLOswave}. 
First, all recoil terms in the vertices ($\propto 1/(2m_N)$) cancel at \NNLO{},
and since the NLO amplitude is zero, as discussed in Sec.~\ref{NLOloops}, 
the nucleon recoil terms 
in the heavy nucleon propagator, Eq.~(\ref{eq:NNprop}), 
also do not contribute at \NNLO{}. 
Secondly, none of the low-energy-constants (LECs), 
$c_i$, $i = 1, \cdots , 4$ of Eq.~(\ref{eq:la1}) contribute 
in the loop diagrams 
at \NNLO{}  
to the s-wave pion-production amplitude. 
This means none of  
the interactions in ${\cal L}^{(2)}_{\pi N}$ of Eq.~(\ref{eq:la1})  contribute to the  amplitude from the pion loop diagrams 
in Fig.~\ref{fig:allN2LO} at this order. 
Finally, as a surprise, the result in Ref.~\cite{NNLOswave} shows that 
the isoscalar s-wave amplitude, which gives the s-wave $pp \to pp \pi^0$ 
cross section,  receives contributions at \NNLO{} only 
from the two cross-box diagrams, diagrams Type III, in 
the fifth row of Fig.~\ref{fig:allN2LO}. 
In particular this implies that earlier phenomenological 
studies~\cite{Lee,HGM,Hpipl,jounicomment,eulogio,unsers}, 
neither of which include the
crossed box diagrams, did not include the appropriate long-range physics 
as we will discuss in the final section. 

The evaluations of the two-pion exchange \NNLO{} loop diagrams 
in Fig.~\ref{fig:allN2LO} 
result in the following s-wave pion-production operator~\cite{NNLOswave}:    
\begin{eqnarray}
i M^{\text{\NNLO{}}} &=& \frac{\gA \ (v \cdot q)}{\fpi^5}
\taux^a  (S_1 + S_2)\cdot k_1
\left[ \frac16 I_{\pi\pi}(k_1^2) - \frac{1}{18} \frac{1}{(4 \pi)^2} \right] + 
\nonumber \\ &+&       \frac{\gA^3 \ (v \cdot q)}{\fpi^5} \bigg\{
    \taup^a i \varepsilon^{\alpha\mu\nu\beta} v_\alpha k_{1\mu} S_{1\nu} S_{2\beta}
    \left[-2 I_{\pi\pi}(k_1^2)\right] \nonumber \\
    && + \taux^a  (S_1 + S_2) \cdot k_1
    \left[ -\frac{19}{24}  I_{\pi\pi}(k_1^2) + \frac{5}{9} \frac{1}{(4 \pi)^2} \right]
    \bigg\} \, ,
\label{eq:Mga3final}
\end{eqnarray} 
where in the
center-of-mass system $\vec{p}_1 = -\vec{p}_2 = \vec{p}$, 
$k_1=p_1-p_1^\prime = -k_2+q$,  
and the approximate relation $k_1^2 \simeq k_2^2$ is used 
(higher-order terms being ignored). 
The symmetric combination of the two nucleon spins $S_1+S_2$ are 
combined with the isovector operator 
$\taux^a$,  whereas the antisymmetric 
two nucleon spin operator goes with 
$\taup^a$. 
Note that  $M^{\text{\NNLO{}}}$ 
is proportional to the outgoing pion energy 
$v\cdot q \simeq \mpi$ as expected. 
The integral expression for $I_{\pi\pi}(k_1)$, which contains two pion propagators
and is  UV divergent,  is given in the appendix. 
As will be discussed in the next subsection,  
the UV divergence has to be regularized, however, 
the finite long-range \NNLO{} loop amplitude behavior of 
Eq.~(\ref{eq:Mga3final}), 
is renormalization scheme independent.

\subsection{Regularization  and renormalization of loop diagrams}
\label{sec:renorm1}

The form of the threshold operator  with s-wave pions and 
two nucleons in a final S-state is 
\be 
M &=& \taup (\vec{\sigma}_1\times \vec{\sigma}_2)\cdot \vec{p}\,  
{\cal A}
-i\taux (\vec{\sigma}_1+ \vec{\sigma}_2)\cdot \vec{p}\, 
{\cal B}
\label{eq:Mthr} 
\ee
The contributions of the loops to the amplitudes
${\cal A}$ and ${\cal B}$ in Eq.~\eqref{eq:Mthr} 
are separated into  singular and  finite parts,
where the singular parts are given by the  
 UV divergences appearing in the integral, $\ipipi(k_1^2)$,  
in Eq.~(\ref{eq:Mga3final}).  
\begin{eqnarray}
{\cal A}&=& \frac{\mpi}{(4\pi \fpi)^2\fpi^3}(\tilde {\cal A}_\text{singular}
+\tilde{\cal A}_\text{finite} ), \nonumber \\
{\cal B}&=&\frac{\mpi}{(4\pi \fpi)^2\fpi^3} (\tilde {\cal B}_\text{singular}
+\tilde {\cal B}_\text{finite} ).  
\label{eq:loops1}
\end{eqnarray}
The UV divergences are 
absorbed into  LECs accompanying the $NNNN\pi$ amplitudes 
${\cal A}_\text{CT}$ and ${\cal B}_\text{CT}$,  
given in the last row  in  Fig.\ref{fig:allN2LO}. 
By renormalization,  the singular parts of the loop amplitudes 
are eliminated and we 
are left with the renormalized finite LECs, 
${\cal A}^r_\text{CT}$ and ${\cal B}^r_\text{CT}$, which will be added 
to the finite parts of the loop amplitudes. 
Based on the renormalization scheme of Ref.~\cite{NNLOswave}, 
the finite parts of the pion-nucleon loops  are 
\begin{eqnarray}
\label{ABfin}
\tilde {\cal A}_\text{finite}(\mu)&=& -\frac{g_A^3}{2}
\left[ 1 - \log \left( \frac{\mpi^2}{\mu^2} \right)
-  2 F_1 \left(\frac{-{\vec p\,}^2}{\mpi^2}\right) \right],
 \\
 \tilde {\cal B}_\text{finite}(\mu)&=& -\frac{g_A}{6}
\left[ -\frac{1}{2}\left(\frac{ 19}{4} g_A^2 -1\right)
\left( 1 - \log \left( \frac{\mpi^2}{\mu^2} \right)
-  2 F_1 \left(\frac{-{\vec p\,}^2}{\mpi^2}\right) \right)+ 
    \frac{5}{3}g_A^2-\frac16
 \right] \ , 
\nonumber 
\end{eqnarray}
where  $\mu$ is the scale and 
the  function $F_1$  is  defined in the appendix, Eq.~(\ref{eq:F1}). 
In general, as discussed in Ref.~\cite{NNLOswave}, 
the finite parts of the loops 
$ \tilde {\cal A}_\text{finite}$ and  $ \tilde {\cal B}_\text{finite}$
can be further decomposed into the {\it short}- and {\it long}-range parts.
The former one is just a (renormalization scheme dependent) constant 
to which  all terms in Eq.~\eqref{ABfin} except $F_1$ contribute.
On the other hand, the long-range part of the loops is scheme-independent. 
By expanding the function $F_1(-{\vec p\,}^2/m_\pi^2)$, Eq.~(\ref{eq:F1}), 
which includes the only long-range piece in 
Eq.~\eqref{ABfin}, in the kinematical
regime relevant for pion production, i.e.~$({\vec p\,}^2/m_\pi^2) \gg 1$, 
one obtains 
up--to--and--including  terms at \NNLO{} 
\begin{eqnarray}
\tilde {\cal A}^\text{long}_\text{finite}&=&-\frac{g_A^3}{2} 
\log\left(\frac{\mpi^2}{{\vec p\,}^2}\right) + 
{\cal O}\left(\frac{\mpi^2}{{\vec p\,}^2}\right),
\nonumber \\
 \tilde {\cal B}^\text{long}_\text{finite}&=& \frac{g_A}{12}
\left(\frac{ 19}{4} g_A^2 -1\right)
\log\left(\frac{\mpi^2}{{\vec p\,}^2}\right)+ 
{\cal O}\left(\frac{\mpi^2}{{\vec p\,}^2}\right).
\end{eqnarray}

A numerical evaluation of these terms   using    $g_A=1.27$  gives 
$ \tilde {\cal A}^\text{long}_\text{finite}=2.2$ and 
$ \tilde {\cal B}^\text{long}_\text{finite}=-1.5$.  
In Ref.~\cite{NNLOswave} these  numbers were  compared    
with  those from  the most important  
phenomenological contributions which were proposed in   
Refs.~\cite{Lee,HGM,Hpipl,jounicomment}  in order to resolve the discrepancy 
between phenomenological 
calculations and experimental data.  
Using the resonance saturation mechanism, shown to be applicable for the two--nucleon 
system in Ref.~\cite{EMGE} and  proposed for pion--production in $NN$ collisions    
in Refs.~\cite{Lee,HGM,Hpipl,jounicomment} to 
estimate the LECs $\tilde{e}_1$ and $\tilde{e}_2$ in Eq.~(\ref{eq:4Npi}), 
the values for the two finite renormalized $NNNN\pi$ amplitudes 
${\cal A}^r_\text{CT}$ and ${\cal B}^r_\text{CT}$  are obtained:  
$\tilde{\cal A}^r_\text{CT}\simeq 2$ and $\tilde{\cal B}^r_\text{CT}\simeq 1$ 
in the  same units.   
Based  on this,  it   was  concluded in \cite{NNLOswave} 
 that  the  scheme-independent  long-range  contributions 
  of  pion-nucleon loops  
    are  comparable in size  with the  contribution  
needed to  bring    theory in agreement  with    experiment. 
Hence, the importance of the long-range \NNLO{} pion loop effects, 
not included in the previous studies, 
raises serious doubts on the physics interpretation 
behind the phenomenologically  successful models  of
Refs.~\cite{Lee,HGM,Hpipl,jounicomment}.

In the standard ChPT counting scheme 
the tree-level diagrams will usually be 
renormalized by including nucleon self-energy pion-loop diagrams as well 
as vertex loops. 
However, when we only consider the pion and nucleons 
fields in the MCS, 
the loop diagrams which contribute to the renormalization of,
e.g., the nucleon mass $\mN$ and the
axial coupling constant $g_A$, do not involve a loop  momentum
of order $p \sim \sqrt{m_\pi m_N}$.
Consequently, these diagrams,  driven by a loop momentum  $\propto \mpi$, 
contribute in the MCS at order \NNNNLO{}
which is beyond the level considered.                    
To illustrate the MCS counting of such loop  diagrams, 
we concentrate on the  LO rescattering diagram in Fig.~\ref{fig:allN2LO},   
which in our naive MCS 
counting is of order $\sqrt{\mpi / \mN}$. 
Consider now  pion loops in this diagram at a vertex or on a nucleon line.    
In standard ChPT counting these pion loops 
would require a renormalization  of the
vertex and the nucleon mass. 
However, these pion loops contain  only ``small" loop momenta $l\sim m_\pi$, 
i.e.  
the loops do not contain any 
external momenta of order $p \sim \sqrt{m_\pi m_N}$. 
Therefore, these type of   
loop diagrams will increase the MCS order
by a  factor $(\mpi / \mN)^2$ as shown in, e.g.~Ref.~\cite{hanhart04}, Table 11. 
In other words,  at 
\NNLO{}, we only have to consider the loop diagrams which have already been discussed. 
We will return to these considerations in Sec.~\ref{sec:renorm2} where we include the 
$\Delta$-resonance in the loops.

\section{ Delta-resonance  induced contributions to s-wave pion production }
\label{sec:swaveD}

In the energy region of  the
pion-production threshold the $\Delta$-resonance 
is not heavy enough to be parametrized just by $\pi N$ LECs. 
In fact, 
the $\Delta$ should   
be explicitly included 
in the loops as virtual nucleon excitations in order for the effective theory 
to properly describe the physics in this energy region. 
Whereas the mass difference $\delta = m_\Delta - m_N$ 
is non-zero even in the chiral limit of the theory 
 (when $m_\pi \to 0$),    the physical value  of
$\delta$,  $\delta\approx  $ 300~MeV,   
is numerically  very close   to  the
``small''  scale   in the MCS,  i.e.  
the  momentum  scale $p \sim \sqrt{m_\pi  m_N}$.   
This observation prompted   Hanhart and Kaiser~\cite{HanKai} to argue that, 
as a practical consistency in MCS,     
$\delta$ should  be counted as   $p$.  
In this section  we will outline the operator structure   
due to the inclusion of 
explicit $\Delta$ degrees of freedom for the $NN \to NN \pi$ reaction.

The LECs,  $c_2$, $c_3$ and $c_4$,   
which are determined from pion-nucleon data,  
have to be re-evaluated when the $\Delta$ is explicitly included.  
Consequently,  
one obtains  the  LECs  in which  the  
$\Delta$ contribution is  subtracted. 
These  residual     LECs  enter  the calculation of  the  
pion-production  operator  derived in Sec.~\ref{sec:diagrams}, see 
Eqs.\eqref{treeNNLO1} and \eqref{treeNNLO2}. 
When the  $\Delta$-field is explicitly included 
in the Lagrangian,  the 
parameter  $\delta\sim p$ will appear in loops containing the $\Delta$ propagator 
and the  resulting  loop four-momentum will naturally be
of the order  of  $p$,  i.e.,  
these loop diagrams will then contribute at NLO and \NNLO{} in the MCS.

In the MCS with a $\Delta$ explicitly included,  
loop diagrams with topologies different from those
discussed in previous sections have to be considered. 
Some of these additional loop diagrams containing a $\Delta$ 
propagator will renormalize 
LO diagram vertices as well as 
the nucleon wave function. 
This is in contrast to  the loop diagrams with only nucleon and pion propagators,  which
contribute  to  renormalization of the vertices at \NNNNLO{}  only,  
as  argued in Ref.~\cite{NNLOswave}.  
This is a higher order contribution and is not considered any further. 
These considerations will be explained in detail in the next subsections,   
where we also will outline the regularization of the UV-divergence of 
this effective field theory containing explicit 
$\Delta$ degrees of freedom.

\subsection{The Lagrangian with $\Delta$ interactions}
\label{sec:DeltaLag}

The  effective  $\Delta$  Lagrangian~\cite{Hemmert,Hemmert:1997ye} written in the 
sigma-gauge is the basis for the   evaluation of   the operator contributions 
\begin{eqnarray}
{\cal L}_{\pi N \Delta} &=& 
-\Psi_\Delta^\dagger (i v\cdot\partial - \delta)
\Psi_\Delta
 + \frac{g_1}{\fpi}  \Psi_\Delta^\dagger \, {\Sdd}^\mu S^\beta \Sd_\mu \, T_i \boldtau 
\cdot \partial_\beta \boldpi  T_i 
\Psi_\Delta
\nonumber
\\ 
&&- \frac{1}{4 f_{\pi}^{2}} 
\Psi_\Delta^\dagger \Big[ 
         (\dot{\boldpi}\times{\boldpi})  \cdot  {T}_i^\dagger \boldtau  { T}_i + 2i
         \left(   ({\bf T}^\dagger  \cdot\boldpi)  ({\bf T} \cdot
           \dot{\boldpi}) -
( {\bf T}^\dagger \cdot \dot{\boldpi})  ({\bf T} \cdot  \boldpi)
\right)     \Big]   \Psi_\Delta
\nonumber
\\ 
&&- \frac{h_A}{2f_\pi} \Big[ 
N^\dagger {\bf T}\cdot \left(\partial^\mu \boldpi
{+}\frac{1}{2f_\pi^2}\boldpi (\boldpi \cdot \partial^\mu \boldpi) \right)
      \Sd_\mu\, \Psi_\Delta
+ h.c. \Big] 
\nonumber
\\
&&+
\frac{h_A}{2m_N f_\pi} \Big[ i
N^\dagger {\bf T}\cdot \dot{\boldpi}\
    \Sd \cdot \partial \Psi_\Delta
+ h.c. \Big]+\cdots \, ,
\end{eqnarray} 
Here   $g_1$  is the  $\pi \Delta\Delta$  
coupling  constant, 
$h_A$ is the leading $\pi N \Delta$ coupling constant 
and $\Sd$ and ${\bf T}$ 
are the spin and isospin transition matrices, 
normalized such that 
\begin{eqnarray}
\Sd_\mu \Sd_\nu^\dagger &=& g_{\mu \nu} - v_\mu v_\nu - \frac{4}{1-d} S_\mu S_\nu,	
\quad
T_i T_j^\dagger = \frac{1}{3}\left( 2\delta_{ij}-i\epsilon_{ijk}\, \tau_k\right) 
\, ,  \quad i,j=1,2,3 .
\nonumber 
\end{eqnarray} 
The value for the  $\pi \Delta\Delta$ coupling constant 
$g_1= 9/5 g_A$ is obtained in the  large $N_c$ limit~\cite{FettesMesissner2001}.
An estimate of the $\pi N \Delta$ coupling
constant  $h_A= 2g_{\pi N\Delta} = 3g_A/\sqrt{2} =2.7$ is
derived from large $N_c$ arguments~\cite{Kaiser1998},
whereas a dispersion-theory analysis gives $h_A=2.1$~\cite{Hohler1983}.

\subsection{Tree-level  diagrams with $\Delta$-resonance}
\label{sec:Deltatree}

\begin{figure}[t]
\includegraphics[scale=0.6]{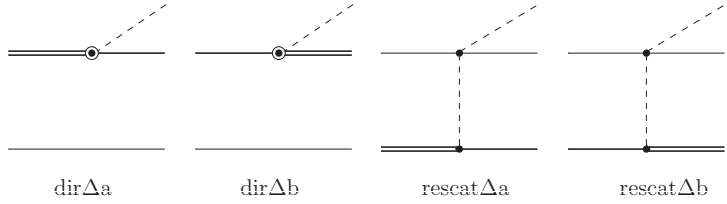}
\caption{\label{fig:treeDelta}   
Single baryon and rescattering diagrams with 
$\Delta$ contributions which appear  as  building blocks  in the  construction of 
the pion-production amplitude.  In the last  two rescattering 
diagrams only  the   on-shell  part  of   the  $\pi N$  vertex  
(2\mpi)  from  Eq.~\eqref{eq:pipivert}  
should  be  included.}
\end{figure}

Since  the $NN$  and $N\Delta$  states 
are coupled  in the  $NN$ models \cite{CCF,CDBonn} 
which will be  used  in the hybrid approach to  
evaluate  the  initial  and  final state $NN$  
wave  functions in $NN\to NN\pi$,  
the  full pion-production amplitude  also obtains  contributions   from  
the  building  blocks  containing  $N\Delta$  states as    shown in 
Fig.~\ref{fig:treeDelta}.   
In  full  analogy  to the  single nucleon direct diagrams in Fig.~\ref{fig:allN2LO} 
and as   discussed  in Sec.~\ref{sec:diagrams},    
the single baryon diagrams  
shown  in  Fig.~\ref{fig:treeDelta}    do  not  contribute  to  
the on-shell  pion-production operator  but  are relevant  only  when convolved 
with an  $NN-N\Delta$  amplitude  either  in the  initial  
or in the  final state.  
%
The explicit  expressions for the  
$\Delta$ contributions to the pion-production operator from  
the  diagrams in Fig.\ref{fig:treeDelta} are
\begin{eqnarray}
i M_{\rm Dir \Delta a}  &=& 
    -\frac{\gpind}{\mN \fpi} T_1^a \vdotq (\Sd_1 \cdot p_1) 
    \delta(\vec p_2- {{\vec p}_2}^{\,\prime}) ,   
    \nonumber 
\\
i M_{\rm Dir\Delta b}  &=&	
    -\frac{\gpind}{\mN \fpi} T_1^{\dagger a} \vdotq (\Sdd_1 \cdot p_1^\prime) 
        \delta(\vec p_2- {{\vec p}_2}^{\,\prime}) ,       \nonumber
\\
i M_{\rm rescat\Delta a}&=& 
    \frac{\gpind}{2 \fpi^3} \vdotq \, i \epsilon^{bac} \tau_1^c T_2^b 	
    \frac{1}{k_2^2-\mpi^2 + i0} (\Sd_2 \cdot k_2) ,  
    \nonumber
\\
i M_{\rm rescat\Delta b}&=&	
    \frac{\gpind}{2 \fpi^3} \vdotq \, i \epsilon^{bac} \tau_1^c T_2^{\dagger b} 	
    \frac{1}{k_2^2-\mpi^2 + i0} (\Sdd_2 \cdot k_2) ,      \label{Mdel}
\end{eqnarray}
These  tree-level  pion-production amplitudes with a $N\Delta$ initial or final
state can  not  by definition  contribute  to the  $NN\to NN \pi$ 
irreducible production operator.  
However,  the amplitudes in Eq.~\eqref{Mdel} give  nonzero contributions 
to the full pion-production amplitude
  when they are inserted as  building blocks into
those of FSI and ISI  diagrams that have $N\Delta$ as an intermediate state 
as mentioned in Sec.~\ref{sec:reduce}.   
The DWBA  convolution  results in  the NLO  and 
N$^2$LO  contributions to $NN\to NN\pi$  from 
the direct and rescattering diagrams in Fig.~\ref{fig:treeDelta}, respectively. 
This DWBA-approach was used by  Ref.~\cite{NNpiMenu} 
in the evaluation of these  operators for the reaction 
$pp\to d\pi^+$.

\subsection{Role of the tree-level diagrams with $\Delta$ resonance}

The effect of the tree-level $\Delta$ operators has been   studied in the literature 
using both  the
phenomenological framework and  EFT with  quite  contradictory  conclusions.
In  a phenomenological study\cite{jounicomment} it was shown that the inclusion of
the $\Delta$ isobar leads to an enhancement of the total cross section in $pp\to
d\pi^+$  by almost a factor of 3 mainly due to the influence of 
diagrams  "rescat  $\Delta$a  and b" in Fig.\ref{fig:treeDelta}.  
This enhancement found in Ref.~\cite{jounicomment} was not confirmed 
in a model calculation by Ref.~\cite{ourdelta}. 
Furthermore, an EFT evaluation~\cite{rocha} of the direct  tree-level diagrams found 
results which differ from the findings of Refs.~\cite{jounicomment,ourdelta}. 
We notice, however, that  the static $\Delta$ propagator was used in Ref.~\cite{rocha}
which led to the large model dependence of the results. 
A similar problem with the use of the static $\Delta$ propagator in
 $\pi d$ scattering was investigated in Ref.~\cite{piddelta} with the result 
that  the recoil  in the $\Delta$ propagator  is needed  in  the $N\Delta$
 intermediate  states to obtain scheme independent results.    
To illustrate the difficulties with the evaluation of the diagrams 
in Fig.~\ref{fig:treeDelta} 
we  focus on  the $\pi N$ rescattering diagrams  
with the $\Delta$, diagrams "rescat~$\Delta$a  and b" in the figure. 
These rescattering diagrams contain the energy
dependent WT vertex, and thus the method developed in the previous
section, see Eq.~(\ref{eq:pipivert}), can be applied. 
In particular, the  $\pi N$ rescattering vertices in 
diagrams  "rescat~$\Delta$a  and b"  convoluted  with   $NN$ wave  functions 
can be  divided into two parts:  the first one is an
on-shell $\pi N$ rescattering  vertex ($\propto 2\mpi$)
 and the  second one  cancels   the nucleon propagators adjacent to 
this vertex  yielding  irreducible loop diagrams (cf.  diagrams 
 $\Delta$Box a  and b in Fig.~\ref{fig:allDLoops}).
 Refs. \cite{HanKai,future} showed that these  irreducible  contributions 
take part in a cancellation with other loop diagrams,  as  will  be  
discussed in the next  subsection. 
Thus, the only  contribution of the tree-level diagrams  up to  
N$^2$LO originates from the direct pion production  
and the rescattering process with the on-shell $\pi N$ rescattering vertex, 
shown in Fig.~\ref{fig:treeDelta}.
The hybrid EFT calculation of  Ref.~\cite{NNpiMenu}, where the $\Delta$ propagator  
was treated  similar  to the 
nucleon one  (see Sec.\ref{sec:Nprop}), 
revealed that each of these diagrams yields about  10-15\% corrections 
to the transition amplitude discussed in Sec.\ref{sec:obs} 
but they enter with opposite signs.   
While  
the direct diagrams increase the cross section in line with 
the finding of Ref.~\cite{ourdelta}
the rescattering diagrams lead to a reduction by almost the same amount. 
This significant cancellation between different diagrams
  resulted in a very small net contribution from 
the tree-level diagrams in  Fig.~\ref{fig:treeDelta} 
to $pp\to d\pi^+$  (see Fig.~\ref{fig:obs}).
The calculation in Ref.~\cite{ourdelta} 
was done with the CCF~\cite{CCF} and the Hannover~\cite{hannover} 
coupled-channel $NN$ models, and the 
pattern of cancellation was the same for both models although the individual contributions 
were slightly different.

\subsection{Loop diagrams with $\Delta$ propagators}
\label{sec:Deltaloops}

\begin{figure}[t]
\includegraphics[scale=0.47]{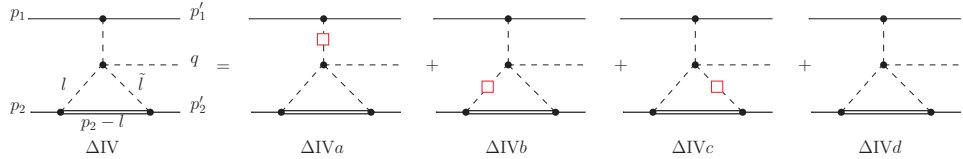}
\caption{\label{fig:DIV}
An example  of the  loop  diagrams  with  the explicit $\Delta$. 
Double lines denote the $\Delta$ propagator,  
remaining notation is as in Fig.~\ref{fig:allN2LO}.
The red squares on the pion propagators  
indicate that for each diagram, the pion propagator
cancels  parts of the four-pion vertex expression,
as explained in the text.} 
\end{figure}

In order to illustrate the power counting of the loop diagrams with 
$\Delta$ in MCS we, 
as an example, discuss in detail the power counting for diagram type $\Delta$IV  in
Fig.~\ref{fig:DIV}.   
First,  note that the  four-pion  vertex of the leftmost diagram 
in Fig.~\ref{fig:DIV} 
can  be  rewritten  as  a  linear combination of the   three pion 
propagators  adjacent  to this  vertex  plus  a residual  vertex term\footnote{ 
While  the  first  three terms  depend explicitly on 
the  parameterisation (or ``gauge'') of  the pion  field,  
the  residual  term  is  
pion-gauge  independent \cite{subloops}.}  
(see, e.g., appendix A.4 in  Ref.~\cite{NNLOswave}). 
This results in a separation of diagram  $\Delta$IV  
in   four  parts:  for  the  diagrams   
$\Delta$IV  a-c the pion propagator cancels   corresponding parts of the  
four-pion vertex,  as  indicated by the red  square 
in   Fig.~\ref{fig:DIV},   while  diagram $\Delta$IV d  appears 
as  the residual part in this  separation. 
%
%
To estimate  the magnitudes of the amplitudes of these diagrams,  
we first remind the reader that  for the reaction $NN\to NN\pi$ close to threshold   
the initial nucleons have four-momentum  
$p_1^\mu = (m_\pi /2, \vec p \,)$ and  $p_2^\mu = (m_\pi /2, -\vec p \,)$  
with  $p=|\vec p \,|\approx \sqrt{\mpi m_N}$  (see  Fig. \ref{fig:DIV} for  the  notation). 
Secondly,  we  note that  the 
loop diagrams with  the  explicit $\Delta$ all involve  
the  $\Delta$-$N$ mass  difference  $\delta\sim p$  in the $\Delta$ propagator. 
The power counting for  the loop  diagrams also requires 
the inclusion of the integral measure
$l^4/(4\pi)^2$ where all components of the loop four-momentum $l$ 
are of order $\delta\sim p$, i.e. $v \cdot l \sim |\vec{l}| \sim p$.   
In addition to this integral measure, in the diagrams $\Delta$IVa--c  
one has to account for  one  $\Delta$ propagator ($\sim
1/(v\cdot l)\sim 1/p$),   three pion propagators ($\sim 1/(p^2)^3$),    
the $4\pi$ vertex ($\sim p^2/f_\pi^2$) and   
two $\pi N\Delta$ and one  $\pi NN$ vertices ($\sim (p/f_\pi)^3$).
Combining 
all these factors and using $4\pi f_\pi\sim m_N$, 
one obtains the order estimate for this diagram as follows 
\begin{eqnarray}
\frac{p^4}{(4\pi)^2} \, \frac{1}{p} \, \frac{1}{(p^2)^3} \, \frac{p^2}{f_\pi^2}\,  
\left(\frac{p}{f_\pi}\right)^3
\sim \frac{1}{\fpi^3}\frac{p^2}{m_N^2} \simeq \frac{1}{\fpi^3}\chi_{\rm MCS}^2.
\end{eqnarray}
These dimensional arguments give the above
order estimate of diagrams $\Delta$IVa--c and 
should be compared with the  estimate 
of a LO rescattering diagram for the $NN\to NN\pi$ 
reaction   
given  by  Eq.~\eqref{eq:treePC}.  
We  find that the diagrams  $\Delta$IVa--c   
in Fig.~\ref{fig:DIV}    start  to  contribute at  NLO.   
Meanwhile,  the  pion-gauge independent diagram $\Delta$IV~d  starts  to  contribute  at  
\NNLO{} only. 
%
The reason is    that the  residual  pion-gauge independent  four-$\pi$ vertex  is suppressed 
compared  to the leading four-$\pi$ vertex contributions.  
Notice  that   the  \NNLO{}  expression for diagram $\Delta$IV  contains more terms
 than the corresponding pure pion-nucleon  diagram IV.   
 As  explained in
Appendix A.4 in Ref.~\cite{NNLOswave} 
the contributions  similar to  type 
 $\Delta$IVb  and  $\Delta$IVc  are  strongly  suppressed 
in the  pion-nucleon  case.

\begin{figure}[t]
\includegraphics[scale=0.75]{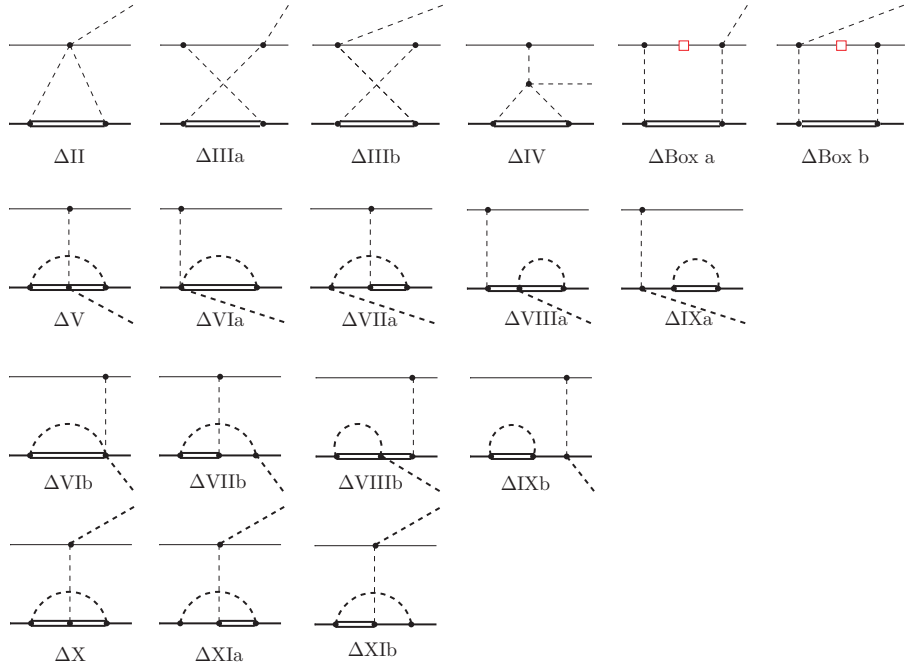} 
\caption{\label{fig:allDLoops}
Loop diagrams with  the $\Delta$ degree  of  freedom contributing to 
s-wave pion production up to \NNLO{}.
Double lines denote the $\Delta$ propagator,  
remaining notation is as in Fig.~\ref{fig:allN2LO}. Again,
the red square on the nucleon propagator in the two box diagrams indicates 
 that the corresponding nucleon propagator cancels  parts of 
the Weinberg--Tomozawa $\pi N$ rescattering vertex leading to 
an irreducible contribution of the box diagrams 
as discussed in Sec.~\ref{sec:diagrams}. 
} 
\end{figure}

Fig. \ref{fig:allDLoops} shows the 
loop  diagrams involving  $\Delta$  which  contribute  
to  s-wave pion  production up
to  \NNLO{}. 
In the  first row of  Fig.~\ref{fig:allDLoops}, we have
the two-pion exchange diagrams with topologies 
completely  analogous to the pion-nucleon $g_A^3$-graphs  in 
the second row in 
Fig.~\ref{fig:allN2LO}. 
The two-pion exchange diagrams in the first row 
of Fig.~\ref{fig:allDLoops}   start individually to contribute  at NLO. 
However,  these NLO  diagrams cancel completely  in the sum  for 
the same reason as do the NLO pion-nucleon  ones in Fig.~\ref{fig:allN2LO}.  
In fact, it is relatively straightforward to show that,  
on the operator level, 
this cancellation of the NLO level diagrams is  independent of whether we
have a nucleon or a $\Delta$ propagator on the lower baryon line in
Figs.~\ref{fig:allN2LO} and \ref{fig:allDLoops}. 
Consequently, in MCS there are no 
contributions from these two-pion exchange diagrams at NLO. 
Moreover,  the \NNLO{}  contributions  of the diagrams in the first row 
in Fig.~\ref{fig:allDLoops} also  show  cancellations   analogous to the purely pion-nucleon  case as shown in Ref.~\cite{future}.  
It  should be mentioned  that   the  diagrams 
 in  the  first  row  of     Fig.~\ref{fig:allDLoops}  
also  obtain corrections  from 
higher-order  vertices  $\propto  1/m_N$ and $c_i$ (Delta-subtracted) from   
${\cal L}^{(2)}_{\pi\!N}$.  
Those  corrections,  however,  again  cancel  completely at \NNLO{} in a full 
analogy  to the cancellations among the  corresponding pion-nucleon loop  contributions,  
see  Ref.~\cite{NNLOswave}   and  the discussion  in  Sec.\ref{sec:2pions}. 
As shown in Ref.~\cite{future}, the 
sum of the \NNLO{} diagrams in the first row of 
Fig.~\ref{fig:allDLoops} receives
contributions only from   diagrams  
 $\Delta$IIIa and   $\Delta$IIIb,     
where  the  Weinberg-Tomozawa $\pi N$ vertices are on-shell, 
and from  the pion-gauge independent 
$\Delta$IVd shown in Fig.\ref{fig:DIV}.

In addition,  there are several new loop diagrams containing $\Delta$ propagators 
where one effectively has  a pion being exchanged between the two
nucleons,  see diagrams   $\Delta$V--$\Delta$XI in  Fig.~\ref{fig:allDLoops}.  
Surprisingly,                        
diagrams $\Delta$V--$\Delta$IX  in
rows two and three   undergo  significant cancellations 
as detailed in Ref.~\cite{future}, and   only  the part  of  the diagram $\Delta$V,  that is proportional to the on-shell   
$\pi \Delta-\pi\Delta$  vertex (equal to $2\mpi$),  remains.  The three one-pion-exchange $\Delta$ loop diagrams 
in the  last  row  of   Fig.~\ref{fig:allDLoops}, 
which  have to be taken into account at  \NNLO{}, 
contribute only  to the  renormalization of  $g_A$  at \NNLO{}, 
see the next subsection  where  we will discuss the renormalization of these 
$\Delta$ loops and give the \NNLO\ transition amplitude operator 
as obtained in Ref.~\cite{future}.

\subsection{Regularization of UV singularities and renormalization}
\label{sec:renorm2}

The loop diagrams  with explicit $\Delta$  are  
UV-divergent at \NNLO{}. 
Contrary to the pion-nucleon case discussed in Sec.~\ref{sec:renorm1}, 
these $\Delta$ loop diagrams require that   
the couplings and masses appearing in the Lagrangian  are renormalized.  
In particular,  up  to \NNLO{}  in  the MCS there are two  relevant
renormalization  corrections:
one involves the  $\pi N $  coupling constant  $g_A$  and 
the other  the  nucleon  wave  function renormalization factor $Z_N$.  These renormalization corrections  of order  
$\delta^2/\Lambda_\chi^2 \sim \chi_{\rm MCS}^2$
were   evaluated in Ref.~\cite{Bernard:1998gv,Fettes,Fettes2}  
  for $\pi N$ scattering with explicit
$\Delta$ in the loop using dimensional  regularization. 
These results were confirmed in Ref.~\cite{future}.  
In order to  calculate the  production operator  up to  NLO  it  suffices to   use  
$Z_N =1 $  and  $\mathring{g}_A =\ga $.  
However,  in a theory  with  explicit  $\Delta$  degrees   of   
freedom, renormalization  corrections  to the tree-level  diagrams at LO 
in Fig.~\ref{fig:allN2LO}  generate  \NNLO{}  contributions. 
At  \NNLO{}, the nucleon  fields ($N$) in  
the Lagrangian   
must  be  renormalized, i.e.,   $N \to  N \sqrt{Z}$,  
and, similarly, for the axial constant $\mathring{g}_A \to g_A$   
(i.e., $\mathring{g}_A$  deviates  from the  physical  value),   
due  to the loop corrections with  explicit  $\Delta$.           
The explicit evaluations  of the diagrams in Fig.~\ref{fig:allDLoops} 
reveal that the  contributions  of  diagrams  
$\Delta$X and  $\Delta$XI   
reproduce the \NNLO{}  correction to the  tree-level rescattering  
diagram in Fig.~\ref{fig:allN2LO}  
due to renormalization of  $\mathring{g}_A$  and $Z_N$,    meaning  that  at \NNLO{} there is no genuine contributions  
of diagrams  $\Delta$X and  $\Delta$XI   in MCS.

As outlined in Ref.~\cite{future} the 
\NNLO{} contribution from  $Z_N$ to the  WT  vertex  
is included  in  the
rescattering  operator  together  with  the  residual  contributions
of the  diagrams  $\Delta$III, $\Delta$IV and  $\Delta$V and  
gives non-vanishing  correction     at   \NNLO{}.   
The individual non-vanishing contributions  of the    
$\Delta$ loop diagrams  in Fig.~\ref{fig:allDLoops} are   
 expressed  in terms of four  scalar integrals,   
$\jpid$, $\ipipi$, $\jpipid$, and $\jpipind$, which are defined in  the appendix. 
The two integrals $\jpid$ and $\ipipi$ contain UV singularities from the 
$\Delta$ loop diagrams which as shown in Ref.~\cite{future} are absorbed by 
the $(NN)^2\pi$ five-point contact terms.

Before we present the final  contribution for s-wave pion production 
from $\Delta$ loop diagrams, there is one issue which deserves attention. 
In a  theory  containing a ``heavy"  resonance $\Delta$,   it  is not  sufficient to  require  just  the  
regularization of the UV  divergent  
terms with the corresponding LECs. 
The integrals  $\jpid$ and $\ipipi$,                 
which are multiplied by the factor  $ \delta^2/k_1^2$, 
pose an additional problem as discussed in Ref.~\cite{future}.  
Such polynomial behavior  would give       
divergences  if the $\Delta$ resonance was infinitely heavy, i.e., if  $\delta\to \infty$.
Therefore,    to find the most natural finite values of the renormalized LECs,   
the explicit decoupling 
renormalization scheme  was  introduced~\cite{Appelquist,Bernard:1998gv}.
In such a scheme, the finite parts of  LECs are 
chosen  such that the renormalized contribution 
from diagrams with $\Delta$ loops vanish in the limit $\delta\to \infty$.
Ref.~\cite{future}  showed  that  the following combinations 
of loop integrals (up-to-and-including \NNLO{} in MCS) do vanish when $\delta\to \infty$:    
\begin{eqnarray}
\label{intcomb1}
		&&\intL, 
		\\
\label{intcomb2}
		&&\left( \intI  + \frac12 \intJ + \intT + \intC{2} \right),
		\\
\label{intcomb3}
		&&\frac{\delta^2}{k_1^2} \left( \intI  + \frac12 \intJ + \intT + \intC{2}\right) 
	    	- \frac{1}{12} \left(  \intI + \frac12 \intJ + \frac13 \intC{2} \right).
\end{eqnarray}
Further, 
Ref.~\cite{future} found that the combination of the two integrals 
$\jpid$ and $\ipipi$ in Eqs. (\ref{intcomb1})-(\ref{intcomb3}),  
$\ipipi + \frac{1}{2\delta}\jpid $,  
cancels the  UV divergences of  the  individual integrals, 
as shown  in Eq.\eqref{intfinite} in the Appendix, which means that
the expressions in  
Eqs.~\eqref{intcomb1}-\eqref{intcomb3} are all  UV  finite  
and  vanish  when $\delta\to \infty$.

The fully renormalized, finite  \NNLO{} $\Delta$ loops contribution to the 
s-wave pion  production amplitude  
 derived in Ref.~\cite{future}  is  
\begin{eqnarray}
\label{delsymmampstart}
\nonumber
		&& \hspace*{-0.5cm} i M_{\Delta\text{-loops}}^{\text{\NNLO{}}} 
       = \frac{\ga \gpind^2}{\fpi^5} \vdotq \,
	    \tau_{+}^a \left( i \varepsilon^{\alpha \mu \nu \beta} 
                v_\alpha k_{1\mu} S_{1 \nu} S_{2 \beta} \right) 
	    \\
          \nonumber
	 \times&&  \hspace*{-0.cm}%
	    \Bigg\{ 
	    	    \frac29 \left( \intI + \frac12 \intJ + \intT + \intC{2} \right) 
	    	  + \frac{1}{18} \intL   
	    \Bigg\} 
	    \\ \nonumber\label{delsymmampend}
           \nonumber
   +&& \hspace*{-0.cm}      \frac{\ga \gpind^2 }{\fpi^5} \vdotq \, \tau_{\times} 
          (S_1 + S_2) \cdot k_1 
   \\     \times&&\hspace*{-0.cm}%
        \Bigg\{
 	    	\frac{5}{9} \left(  \intI + \frac12 \intJ + \intT + \intC{2} \right) 
	    	+ \frac{1}{18} \intL   
	    	\\  \nonumber 
	+&&\hspace*{-0. cm}      \frac{8}{9} \frac{\delta^2}{k_1^2} \left( \intI  + \frac12 \intJ 
                            + \intT + \intC{2}\right) 	- \frac{2}{27} \left(  \intI + 
          \frac12 \intJ + \frac13 \intC{2} \right) 
	        \Bigg\}. 
\end{eqnarray} 
This expression should be added to the finite s-wave transition  operators 
presented in Sec.~\ref{sec:renorm1}.

\subsection{Comparison of  the  $\Delta$  and  pion-nucleon loop  contributions}
\label{sec:Delta_N} 

In Sec.~\ref{sec:renorm1} 
(see also discussion  in  Ref.~\cite{NNLOswave})
we argued  that the  long-range scheme-independent  part
of   the  pion-nucleon  loops  at  \NNLO{} is sizable  and   could  
resolve  the problem  with the description of  pion  production data
in the  neutral  channel, $pp \to pp \pi^0$.  
The question to be answered in this subsection is 
what happens when we 
add the long-ranged $\Delta$ contribution. 
First,  note  that 
the spin-isospin structure of the    $\Delta$-loops  
in Eq.~(\ref{delsymmampstart}) 
is exactly
the  same as for the  pion-nucleon  case  in  Eq.~\eqref{eq:Mga3final},  
as required by the threshold decomposition 
of the pion-production amplitude, Eq.\eqref{eq:Mthr}.  
Meanwhile,  the
dimensionless  integrals are different  and   the 
coefficients  in front  of the spin-isospin operators  also differ.  
Ref.~\cite{future} compared the resultant amplitudes  from the 
nucleon and $\Delta$ loop diagrams for $NN$   
relative distances  relevant  for
pion production,  i.e.  for $r\sim 1/p \simeq 1/\sqrt{m_\pi m_N}$.
In Ref.~\cite{future}   
the long-range scheme-independent  contributions 
of  the  $\Delta$-loop expressions are separated from  the short-range ones 
by making  
a  Fourier  transformation of the expressions in 
Eqs.~\eqref{eq:Mga3final} and \eqref{delsymmampend}.   Specifically,  Ref.~\cite{future}  Fourier  transformed   the   integral
combinations  in the  curly   brackets  in  Eq.~\eqref{delsymmampend}
(multiplied by $\gA \gpind^2$)
corresponding to   $\boldtau_+$ (final $NN$ spin-singlet) 
and  $\boldtau_{\times}$ (final $NN$ spin-triplet)  channels. 
Ref.~\cite{future}  compared the resulting Fourier transformed $\Delta$ loop amplitudes  with  
the Fourier  transformed amplitudes of  the  corresponding 
pion-nucleon  contribution given in Eq.~\eqref{eq:Mga3final},  namely 
$-2 \gA^3 I_{\pi\pi}$ and  $(-19/24 g_A^3+1/6 g_A)I_{\pi\pi} $.  
It was found that in the  spin-singlet $NN$  channel,  the
long-range part of the $\Delta$ contribution   constitutes  
less  than  20\% compared 
to  the  pion-nucleon loop amplitude, whereas in the spin-triplet $NN$ 
channel  the  $\Delta$- loop contribution  to the 
s-wave pion-production amplitude is  almost of
the same  magnitude (roughly 60\%).  
The  conclusion  of  Ref.~\cite{future}    is as follows:    
 the  importance
 of  the pion-nucleon  loops in explaining $pp\to pp\pi^0$ 
appears to be  only slightly modified by the  $\Delta$ loop contributions 
and the \NNLO{} loop contributions are significant, 
whereas   for the final spin-triplet $NN$ channel the 
net \NNLO{} amplitude from the nucleon and $\Delta$ loop diagrams is  
considerably 
reduced in importance  compared to the spin-singlet channel.   
The pattern that emerged from the \NNLO{} loop diagrams 
is therefore exactly what is necessary 
to quantitatively describe the data on both $pp\to pp\pi^0$ and $pp\to d\pi^+$
 near threshold. 
For the former reaction there persists a huge discrepancy
between data and the chiral perturbation theory calculation to NLO, 
while the latter already at NLO describes the experimental data 
 quite well~\cite{lensky2}.  
As discussed, for the $pp\to pp\pi^0$ reaction the LO diagrams are
suppressed both kinematically as well as dynamically~\cite{hanhart04}, 
the net \NNLO{} loop diagram results presented provide 
a dynamical understanding of why it was so much harder to understand phenomenologically
the $pp\to pp\pi^0$ reaction compared to the other channels.

\section{p-wave pion production}
\label{sec:pwave}

In this section we will first present  evidence that for 
pion p-wave production  
the perturbative series, which is ordered according  to the MCS,   
converges as expected. 
Then we will 
focus on a specific feature of 
chiral symmetry which connects the $NN\to NN\pi$ reactions to different
reactions at low energies as discussed in the introduction, Sec.~\ref{sec:intro}. 
In other words, we will
  discuss 
how the reaction $NN\to NN\pi$ 
can be used to pin down the $(N N)^2 \pi$ 
LEC, $d$, which  plays an 
important role  connecting many very different few-nucleon reactions 
as illustrated in Fig.~\ref{ct}.   As follows from Eqs.~(\ref{lag})-(\ref{u}) the  
contact operator associated with the LEC $d$ 
 is proportional to the pion derivative. 
Therefore, it should contribute to  the production of 
p-wave pions in $NN\to NN\pi$ while connecting S-wave nucleons. 
This implies that for all partial waves where there are $NN$ P-waves in
the final state, the chiral perturbation theory calculation is parameter
free up to \NNLO{}.

\begin{figure*}[!htb]
\begin{center}
\includegraphics[scale=.58]{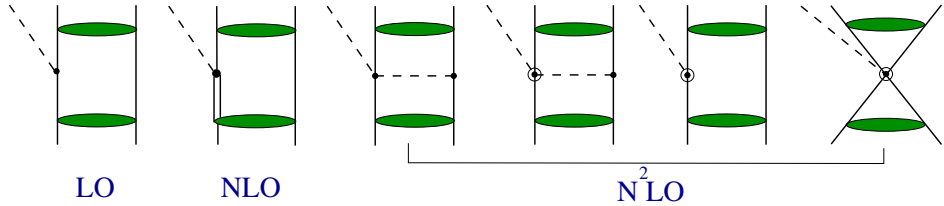}
\caption{
\label{diag}  Diagrams that contribute to p-wave pion production up to N$^2$LO
in the MCS where the shaded ovals indicate initial and final $NN$ state interactions.} 
\end{center}
\end{figure*}

In contrast to pion s-wave production,  the p-wave pion-production operator  up
to  N$^2$LO in  MCS consists  of tree level graphs
only, shown in Fig.~\ref{diag}.    
In particular,  the  only  operator at LO  corresponds to the direct emission
of  a pion  from  one of the  nucleons.   
The  estimate for  this operator  based on
the naive dimensional analysis  outlined in Sec.~\ref{sec:count} gives 
$|\vec{q} |/{(\fpi^3\mpi)}$, where $\vec{q}$ is the pion three-momentum.   
The pion can also be produced  through  
the de-excitation of a 
$\Delta$ resonance.  
However, since the $\Delta$ propagator 
contains the mass-difference $\delta\sim p$, it is suppressed 
by  one order  of   $\chi_{\rm MCS}$  as compared to the nucleon one, and 
the corresponding  diagram starts  to contribute at  NLO.  
To this order  the production operator  does not involve  any  free
parameters.  
All \NNLO{} diagrams in the MCS have 
tree-level  topologies and include  pion rescattering  operators  with
$\pi\pi NN$  vertices  from  
${\cal L}^{\rm(1)}_{\pi N}$ of Eq.~(\ref{eq:la0}) and  
${\mathcal L}^{\rm(2)}_{\pi N}$ of Eq.~(\ref{eq:la1}),  and 
recoil  correction to  the   $\pi NN$ vertex at LO,  
and finally 
the    contact operator   with the associated  LEC $d$, all shown 
in  Fig.~\ref{diag}.  

\begin{figure}[!t]
\begin{center}
\includegraphics[scale=.5]{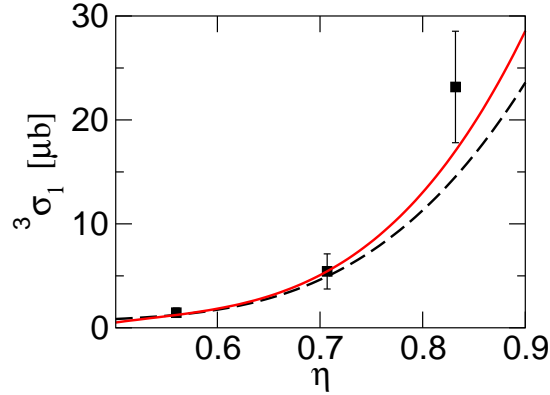}
\caption{
\label{3sig1} Results for the double polarization observable
$^3\sigma_1$ versus $\eta$ ($\eta=|\vec{q}|/m_\pi$ in the center-of-mass system) 
as presented in Ref.~\cite{ch3body}. 
The dashed line denotes
the result at leading and next--to--leading order, while the
solid line shows the result at \NNLO{}. For this calculation a value
of $c_3=-2.95$ GeV$^{-1}$ was used. 
The data are from Refs.~\cite{complete
}.}
\end{center}
\end{figure}

In order to demonstrate the rate of convergence of the
series defined within the MCS, we will concentrate on 
the evaluation of the  
spin cross section $^3\sigma_1$ for $pp\to pp\pi^0$.
This observable is a linear combination 
of the reactions total cross section and the double polarization 
observables $\Delta\sigma_T$ 
and $\Delta\sigma_L$ tailored such that only the $pp$ spin 1 with 
spin projection $+1$ in the initial state contributes. 
The measurement and detailed discussion of this observable can be found in Ref.~\cite{complete}.
The spin-1 cross section $^3\sigma_1(pp\to pp\pi^0)$  
appears to provide a  test of the convergence of the theory   
up to \NNLO{}.   The reason is that the unknown contact term $d$
does not contribute to  $^3\sigma_1(pp\to pp\pi^0)$ 
as pointed out in Ref.~\cite{ch3body,hanhart04}   
since, due to selection rules,    a p-wave pion can  only be produced in
combination with the final $NN$-pair in at least a P-wave.  
The lowest final state partial waves contributing in the final state are 
$pp$ P-waves together 
with pion p-waves and the $pp$ S-wave with a pion d-wave. 
A comparison of the results of Ref.~\cite{ch3body}
with the data is given in Fig.~\ref{3sig1} as a function of the
variable $\eta$, 
which here equals  the maximum pion momentum allowed in units of the pion mass. 
Note that for this reaction a value of $\eta=0.9$  equals  
an excess energy  $Q=50.5$~MeV, 
which is  quite large.
The dashed line in Fig.~\ref{3sig1} is the result of leading and next-to-leading order, 
namely the nucleon and the $\Delta$ direct production terms, 
 the first two diagrams of Fig.~\ref{diag}  
(in Ref.~\cite{ch3body} both counted as leading order), 
while the solid line contains the result at \NNLO{}
providing an excellent agreement with data. 
In the calculation of Ref.~\cite{ch3body}, 
pion d-waves were neglected. 
This is not necessarily in
disagreement with the discussion in Sec.~\ref{COSY_data}, 
where it is
argued that pion d-waves 
are  important for such large excess energy: 
the observables discussed in that section are
defined with the final $NN$ energy limited to very small values 
at the same time maximizing the
pion momentum and therefore the pion d--waves. 
In contrast to the restricted $NN$ final energy selection presented in Sec.~\ref{COSY_data}, 
the observable $^3\sigma_1$ is
fully integrated with respect to the final $NN$ energies and  
it is therefore expected that the final $NN$ P--waves
together with the pion p--wave (Pp final state) become dominant.  
In order to  provide  further  evidence,
we refer to  the experimental  analysis of  $pp\to pp \pi^0$ in  
Ref.~\cite{bilger}   where  in Table 6 it is shown    
that the  Pp final state contribution to the total cross section completely
dominates  over  Sd final state  at the energy of interest.  
We  conclude from Fig.~\ref{3sig1} that, at least for this spin observable,
the perturbative MCS expansion converges nicely and is in agreement with experiment.

In the second part of this section we will demonstrate how  the strength of
the contact operator $d$ in Fig.~\ref{diag} could be extracted from experimental data. 
There are   only  two  reaction channels  with $NN$ S-waves
in both the final and the initial state  where this \NNLO{} contact
interaction 
can contribute.
One  corresponds to  the  case   with
the spin-triplet  S-wave $NN$  final state  interaction  (FSI),  
which as seen in Table~\ref{table1} is
realized  in the $pp\to pn\pi^+$ and  $pp\to
d\pi^+$  reactions, and the other   goes  
with the spin-singlet  $NN$ FSI in an S-wave,  
which appears in  $pn\to pp\pi^-$.  
In general, when we restrict the final two nucleons to be in S-wave, 
the relevant  partial waves  for production of  p-wave pions are: 
$^1S_0\to  {^3 S_1}p$  and  $^1D_2\to  {^3 S_1}p$ for  $\pi^+$ production
and  $^3S_1\to  {^1 S_0} p$ and  $^3D_1\to  {^1 S_0} p$  for producing a 
$\pi^-$,  as  can be
read off from   Table \ref{table1}.  
This imposed limitation reflects recent $NN\to NN\pi$ experiments,
which will be presented in the next section, where 
the final two nucleons relative energy was  below 3~MeV.

\begin{figure*}[h]
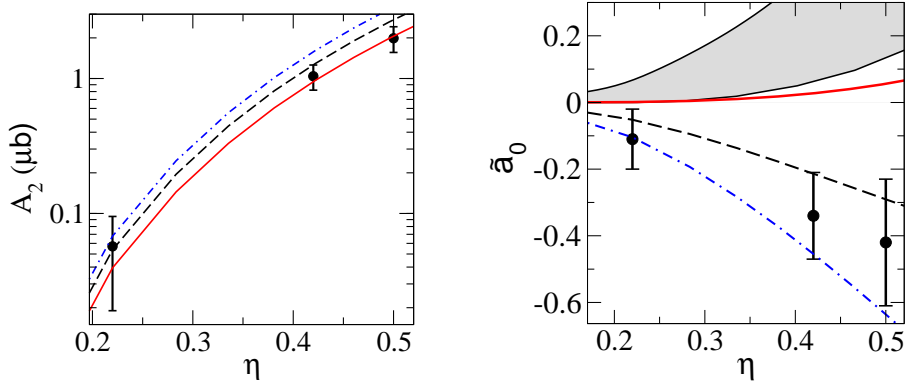
\hspace*{-0.3cm}
\begin{center}
{\includegraphics[scale=0.44,clip=]{./Figures/Fig7a.eps}}
{\hspace*{0.8cm}\includegraphics[scale=0.42,clip=]
{./Figures/a0ct_with_pf_paperv3.eps}} 
\caption{ \label{A2}
Results for the magnitude $A_2$ (left panel)
and the  partial wave amplitude  $\tilde a_0 (\sqrt{\mu b})$ 
representing the relevant transition 
$^1S_0\to {}{^3 S_1}p$  (right panel) for 
$pp\to pn\pi^+$ for different values of the LEC $d$  
(in units  $1/(f^2_{\pi}M_N)$). Shown are $d=3$ (red solid line), 
$d=0$ (black dashed line), and $d=-3$ (blue dot-dashed line). 
The grey band corresponds to the results of Ref.~\cite{nakamura} 
which incorrectly 
were  interpreted as  a  failure to  connect  
the solar fusion  reaction  with pion production, 
see discussion in the text. 
 The data are from Ref.~\cite{Flammang}.
}
\end{center}
\end{figure*}

The first extraction of the value of LEC $d$ from $NN\to NN\pi$ 
was perfomed in Ref.~\cite{ch3body}, 
where the EFT  calculation  was confronted with the results of  the
partial  wave analysis (PWA) of  the   experimental data
in the channel $pp\to pn\pi^+$ \cite{Flammang}.    An attempt to connect  the primary solar fusion reaction, 
$pp\to d e^+\nu$, and $pp\to pn\pi^+$ through
the LEC $d$ was done in Ref.~\cite{nakamura}  with the result
that no simultaneous  solution is possible.   However, the results of Ref.~\cite{nakamura}  were  compared to the
 results  of the PWA of Ref.\cite{Flammang}  which
 turned  out  to  be incorrect,  as  pointed out in Ref.\cite{newpwave}. 
Specifically,  the PWA needed the  contribution of the spin-singlet $pn$-state
and in   Ref.\cite{Flammang} this contribution was extracted   from  
$pp\to pp\pi^0$ data. 
Unfortunately, in   Ref.~\cite{Flammang} this contribution was not  corrected  for  the  
significant difference between  the $pp$
and $pn$ final  state  interactions in S-wave. 
In  Ref.\cite{newpwave}  the  LEC  d
was therefore adjusted to reproduce the differential cross section and analyzing
power in  $pp\to pn\pi^+$ directly.  
Flammang {\it et al.}~\cite{Flammang} parametrized their measured  differential 
cross section for $pp\to pn\pi^+$ as 
\be
\frac{d\sigma}{d\Omega} = A_0 + A_1 \, {\rm cos}\theta + A_2 P_2( {\rm cos}\theta )
\label{diffXS}
\ee 
where  $P_2$ is the second Legendre polynomial. 
Assuming only s- and p-wave pions,  Ref.~\cite{Flammang}  determined that $A_1\simeq 0$. 
The results of the calculation in  Ref.\cite{newpwave} for the magnitude $A_2$
 are compared with the experimental values  
  in the left  panel of Fig.~\ref{A2}.    
The problem of the PWA of  Ref.~\cite{Flammang} 
is  illustrated in the right panel of 
Fig.~\ref{A2} where we also show the results of the calculation for the
relevant partial wave $^1S_0\to {}{^3 S_1}p$, 
denoted below by  $\tilde a_0$.  Although  the red  curve representing the best  fit  is in
agreement with  the  data presented in Ref.~\cite{Flammang} 
(see left panel in Fig.~6 in Ref.\cite{newpwave} for details), 
the partial wave amplitude $\tilde a_0$ extracted from the very same data is not described at all. 
Note that Ref.~\cite{nakamura} stressed
that a positive value for $\tilde a_0$ is necessary in order to achieve a result for 
pion production that is consistent with the ones for the solar fusion rate. 
This is confirmed by the result of  Ref.\cite{newpwave}   based on
the data for $pp\to pn\pi^+$ (cf. red  curves in Fig.\ref{A2}). 

The purpose  of the study  in   Ref.~\cite{newpwave}  was , however, more 
ambitious  than  just  to reanalyse the  data  in $pp\to pn\pi^+$.
The idea was  to  perform a combined analysis of all 
p-wave pion-production amplitudes contributing 
to the three different channels, namely  $pn\to pp\pi^-$, 
 $pp\to d\pi^+$ and  $pp\to pn\pi^+$  
 [To properly analyze these  reactions  near threshold  
the  pion s-wave amplitudes are needed.
In Ref.~\cite{newpwave} the  pion s-wave amplitudes 
were taken  
from  $\alpha$, Eq.~\eqref{totXS}, measured in Ref.\cite{pidexp}  
 and from the measured  $pp\to pp\pi^0$ total cross section 
close to threshold~\cite{cosytof}].
This should provide a   non-trivial test of  the  approach  since even
in the different channels 
of the production process, the contact
term connects $NN$ wave functions in very different 
kinematic regimes. For
the first channel the p-wave pion is produced  along with the
slowly moving protons in the ${^1\!}S_0$ final state whereas for the other
two channels the initial ${^1\!}S_0$ $pp$ state is to be evaluated at the
relatively large initial momentum.  
%
\begin{figure*}[!htb]
\centering
\includegraphics[width=0.8\columnwidth,angle=0]{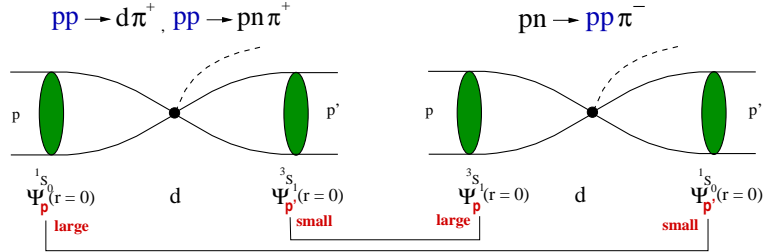}
\caption{Illustration of the contribution of the LEC $d$, 
which only appears when we have initial and final  $NN$ 
S-wave states, for   different   
$NN\to NN\pi$ channels. 
The labels "small" and "large"  indicate that the $NN$ wave 
functions in the same partial wave contribute
to different production channels in different kinematic regimes.}
\label{ct_NN}      
\end{figure*}
%
 Fig.~\ref{ct_NN} illustrates  that the contribution of the contact operator 
gets multiplied with the product of the $NN$ wave functions 
at the origin. 
 Each of these wave functions, in turn, may be represented 
by the inverse of the corresponding Jost function~\cite{goldbergerwatson}, 
which has an
 energy  dependence  
that is fixed by the on-shell $NN$ data --- up to a polynomial with coefficients 
known to the required order within MCS.  

Since the $^3S_1$ and $^3D_1$ partial waves are coupled,  a rigorous treatment calls
for a coupled channel Jost function.  In the reasoning below, however, we will neglect this channel coupling. 
Note that a recent analysis of data
found strong evidence that the channel coupling indeed is weak 
as will be discussed 
in Sec.~\ref{COSY_data}. 
In addition, even if the channel coupling were significant
it would not change, 
the general line of reasoning to be presented.  
The $NN$ wave functions at the origin can  be
represented as an integral 
over the relevant partial wave phase shifts $\delta_{NN}$ 
by means of the so-called Omn\`es
function~\cite{Omnes} (see also the discussion in Ref.~\cite{Ashot}) 
\be
\nonumber
\hspace*{-0.7cm}\Psi_p(r=0)&=&1+ m_N\int\limits_0^{\infty} d^3p'  
                \frac{T(p',p)}{p^2-{p'}^2+i0}\\
&\approx & { C} \exp{\left \{\frac{s}{\pi}\int\limits_{4m_N^2}^{\infty}
               \frac{d s'}{s'}\frac{\delta_{NN}(s')}{s'-s(p)+i0}\right\}},
\ee
where $T$ is the $NN$ T-matrix, $s(p)\simeq (2m_N+p^2/m_N)^2$ for 
the nonrelativistic nucleons and $C$ is 
a model-dependent constant (up to higher order terms in $p/m_N$). 
Thus the whole momentum dependence of the contact term contribution 
is model-independent,
since it is fully determined by the product of the two Omn\`es 
functions for the $^1S_0$ and $^3S_1$ partial waves.  The product 
of the constants $C_{1S0}$ and $C_{3S1}$ is absorbed into 
the strength (the LEC, $d$) of the contact term  for all   $NN\to NN\pi$ channels. 
\begin{figure*}[h]
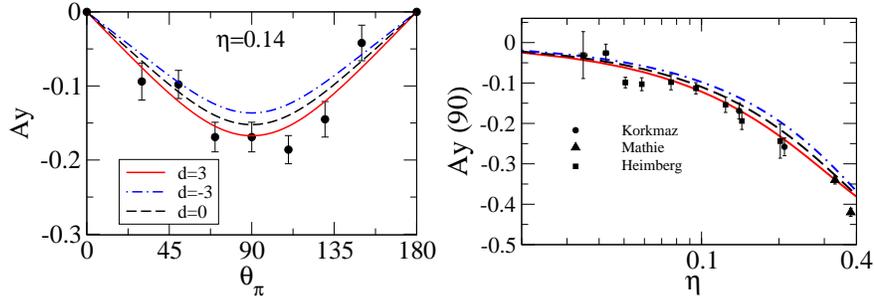

\begin{center}
\vspace*{.2cm}
\includegraphics[width=0.45\columnwidth,clip=]{./Figures/Ay_0_14.eps}
\includegraphics[width=0.45\columnwidth,clip=]{./Figures/Ay90dpi_with_pf_fully_eta04v0.eps}
\caption{Results for the analyzing power at $\eta$=0.14 (left panel)
 and  the analyzing power at 90 degrees
(right panel) for the reaction $pp\to d\pi^+$ for different
values of the LEC $d$  (in units  $1/(f^2_{\pi}M_N)$)
of the $(N N)^2 \pi$  contact operator. Shown are $d=3$ (red solid line), 
$d=0$ (black dashed line), and $d=-3$ (blue dot-dashed line). 
The data are from Refs.~\cite{Ritchie,Heimberg,Drochner,Korkmaz,Mathie}.
}
\label{Ay_dpi}
\end{center}
\end{figure*}
In Ref.~\cite{newpwave} 
the value of the LEC $d$ was varied   
to achieve qualitatively the best  
overall description of the differential cross sections and 
analyzing powers in the different  $NN\to NN\pi$ channels.  
In particular,  Fig.\ref{Ay_dpi} shows   the results    for the
analyzing power for the reaction $pp\to d\pi^+$\footnote{Note also that the 
first  data  on  the  transverse correlation coefficient $A_{xx}$ in $pn \to d\pi^0$ have been measured   recently\cite{Shmakova}.}  
  for various
values of $d$.
It is found that the data prefer a positive value of 
$d \sim 3/(f^2_{\pi}M_N)$  when the CCF model~\cite{CCF} of $NN$ interaction is used. 
 A similar pattern can be observed in the reaction $pn\to pp\pi^-$, 
as illustrated in Fig.~\ref{res_pppimin}.
Again the data show a clear preference of the positive value for the LEC $d$. 
Thus,  the    conclusion  of Ref.~\cite{newpwave}  was 
that it is possible to qualitatively describe the data available
at that  time with the same value of the LEC $d$. 
\begin{figure*}[t]
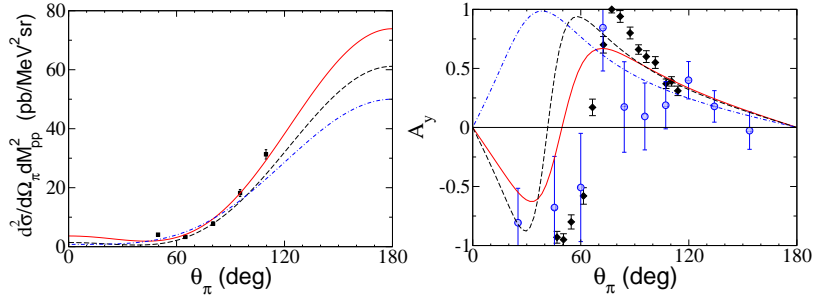

\begin{center}
\includegraphics[width=0.42\textwidth,clip=]{./Figures/Fig6a.eps}\hspace*{-0.12cm}
\includegraphics[width=0.425\textwidth,clip=]{./Figures/Fig6b.eps}
\caption{ 
Results for $d^2\sigma/d\Omega_{\pi}dM^2_{pp}$ (left panel) and $A_y$
(right panel) for $pn\to pp(^1S_0)\pi^-$. 
Shown are the results for $d=3$ (red solid line), $d=0$ (black dashed line) and $d=-3$
(blue dot-dashed line). 
The data are from  TRIUMF~\cite{hahn,duncan} (black squares) and from
PSI~\cite{Daum} (blue circles) .} 
\label{res_pppimin}  
\end{center}
\end{figure*}
However, as discussed in   Ref.~\cite{newpwave},  
in their analysis there are uncertainties which  prevent a more definite conclusion. 
First of all, the relevant $^1S_0\to \!\!\phantom{1}^3S_1p$  
partial wave for p-wave pions is of minor importance to describe the observables 
in $pp\to pn\pi^+$ and $pp\to d\pi^+$. 
Secondly, 
at the relevant measured energies the final $NN$ P-waves may 
already contribute significantly~\cite{bilger,complete,pwpi0}. 
These higher partial waves were ignored  
in Ref.~\cite{newpwave}. 
Finally, the channel $pn\to (pp)_S\pi^-$  has been measured at TRIUMF~\cite{hahn,duncan} 
and  only  those  events were selected that  correspond to  the energy of the 
proton  pair smaller than $1.5$~MeV.  
This  garantees  that  the $pp$-system  is  purely in an
S-wave and that  no higher $NN$ partial waves in the final state 
can spoil  the  analysis. 
However, the measurement   was done with $T_{\rm lab}=353$~MeV
($\eta$=0.66), which means that pion  d-waves  are 
important.  
This is confirmed by a 
very recent  measurement  at COSY \cite{ANKEpi0} to be presented in Sec.~\ref{COSY_data}.   
Note that the pion d-wave production  was discarded in the analysis in  Ref.~\cite{newpwave}.

\section{A combined  amplitude  analysis of recent measurements of   $pn\to pp\pi^-$
  and   $pp\to pp\pi^0$ at COSY}
\label{COSY_data}

In the  framework of  the experimental program   devoted to the   study  of
pion production  at the   COSY  accelerator  in  J\"ulich,  the ANKE
collaboration  has carried out a  combined  measurement 
of  two  reaction  channels   \pnpppi\    and     \pppi\
\cite{ANKEpi0,ANKEpi-}.  
Here  the  ${\{pp\}}_S$  denotes  a  proton-proton  pair  with very low
excitation energy,  $E_{pp} < 3$~MeV,  which  is  
considered to be primarily in the
$^1S_0$ state. 
The  differential cross section and   analysing power  for  both
reactions were measured   at   $T_{\rm lab}$=353~MeV, 
which  allows a  direct comparison  with the
earlier TRIUMF  data~\cite{hahn,duncan}. 
Whereas  the  TRIUMF  experimental data cover only the 
pion's central  angular region,   
the ANKE data  extend over the whole  angular domain and have much higher statistics.   
In  addition,   the   results of  Refs.\cite{ANKEpi0,ANKEpi-} 
were  extended  in  Ref.\cite{ANKEAxx}  where 
the  transverse spin-correlation parameters $A_{xx}$ and $A_{yy}$  
for the reaction  \vnvpppppi  were  extracted  from the measurement  of  
$\pol{d}\pol{p}\to p_{\text{spec}}\{pp\}_{\!s}\pi^-$  
at  COSY-ANKE at  the  same energy.

In what   follows   we summarize the amplitude   analysis of  these data,
further  details  of these measurements  can be found in
Refs.\cite{ANKEpi0,ANKEpi-,ANKEAxx}.  
In particular,   a combined measurement  of  the 
differential cross section and
  analyzing power in \pppi\     allowed  the determination of the 
 pion s- and d-wave amplitudes
 with only minimal theoretical assumptions.   
Meanwhile,   the global partial wave analysis 
 of  all   data sets  yielded  several   
almost  equivalent  solutions  for the p-wave amplitudes  
  with  approximately  the  same  value  of $\chi^2$.     
The only  possibility  of  resolving  this  ambiguity, 
as  argued in\cite{ANKEAxx},   would be  
through an  additional  measurement    of  the  
mixed spin-correlation parameter    $A_{xz}$.  
In what  follows  we  discuss  these issues in some detail.

As long as the final $pp$-system is purely in an S-wave, 
the most general structure of the reaction 
amplitude can be written as
\be 
\label{Mgen}
M= A\ {  \mathcal {\bf S}}\cdot \hat{\bf p}  + B\   {   {\mathcal
    {\bf S}} \cdot\hat{\bf q}}\,,
\ee
where 
$ \hat{\bf p}$ is the unit vector of the initial nucleon  momentum in
the overal center of mass system (cms), 
$ \hat{\bf q}$ is the unit vector of the  pion momentum and 
$ \hat{\bf p}\cdot \hat{\bf q}= \cos \theta_{\pi}$. 
Here 
${\mathcal{\bf S}}=\chi^T_2\frac{\sigma_2}{\sqrt{2}} {\bf \sigma}\chi_1$
denotes the normalized spin structure corresponding 
to the initial spin-triplet  state
and $\chi_{1,2}$ stand for the spinors of the initial nucleons.

It was pointed out in\cite{ANKEpi0,ANKEpi-,ANKEAxx} that  the  data 
do not support  polynomial terms  proportional to  
$\cos^4 \theta_{\pi}$  and higher.  
This  fact   suggests  that  the  partial  waves
higher  than  d-waves for  the pion  can safely be ignored. 
Thus,   up--to--and--including pion d-waves 
the reaction amplitude for \pnpppi\ can be written as 
\begin{eqnarray}\nonumber
M&=& {M}^{P}_{\rm s} \,  {{ {\mathcal {\bf S}}}\cdot \hat{\bf p}} + 
 {M}^{S}_{\rm p} \, { { {\mathcal {\bf S}}}\cdot \hat{\bf k}}
+{M}^{D}_{\rm p} \, \left( {{ ({\mathcal {\bf S}}}\cdot \hat{\bf p}})\,({\hat{\bf k}\cdot \hat{\bf p})-
\frac13 { { {\mathcal {\bf S}}}\cdot \hat{\bf k}}} \right)\\\nonumber &+ &
{M}^{P}_{\rm d} \, \left( ({{ {\mathcal {\bf S}}}\cdot \hat{\bf k}})\,(\hat{\bf k}\cdot \hat{\bf p})-
\frac13 { { {\mathcal {\bf S}}}\cdot \hat{\bf p}} \right)\\
&+&{M}^{F}_{\rm d}\,  \left( {{ ({\mathcal {\bf S}}}\cdot \hat{\bf p}})\,({\hat{\bf k}\cdot \hat{\bf p})^2-
\frac25({{ {\mathcal {\bf S}}}\cdot \hat{\bf k}})\,(\hat{\bf k}\cdot \hat{\bf p})-
\frac15 { { {\mathcal {\bf S}}}\cdot \hat{\bf p}}} \right),
\label{Mspin}
\end{eqnarray}
where the superscript 
in the amplitudes $M^{L}_{\rm l}$ refers to the partial wave of the initial 
nucleons and the subscript
corresponds to the pion partial wave  in the overall cms. 
Altogether  we include in the analysis   
one  s-wave (${M}^{P}_{\rm s}$), 
two p-wave  (${M}^{S}_{\rm p} $  and ${M}^{D}_{\rm p} $)  
and  two  d-wave (${M}^{P}_{\rm d}$  and ${M}^{F}_{\rm d}$)  amplitudes.
These amplitudes  correspond to
the transitions      $^3P_0 \to  {^1 S_0} s$,  $^3S_1 \to  {^1 S_0} p$,  $^3D_1
\to  {^1 S_0} p$, $^3P_2 \to  {^1 S_0} d$,  and  $^3F_2 \to  {^1 S_0} d$,
respectively.

Comparing (\ref{Mspin}) with (\ref{Mgen}) one gets the expressions for $A$  and  $B$ 
in terms of the partial wave amplitudes for \pnpppi\
\be\nonumber
 A&=& {M}^{P}_{\rm s} +{M}^{D}_{\rm p}
\cos \theta_{\pi} - \frac13{M}^{P}_{\rm d}
+ {M}^{F}_{\rm d}\left (\cos^2\theta_{\pi}-\frac15\right)\,,\\  
B&=& {M}^{S}_{\rm p} -\frac13 {M}^{D}_{\rm p} +
\left({M}^{P}_{\rm d} 
-\frac25 {M}^{F}_{\rm d}\right) \cos \theta_{\pi}\,.  
\label{ABpi-}
\ee
For  \pppi\  one has to omit the pion p-wave amplitudes in the expression above to arrive at
\be\nonumber
A&=& {M}^{P}_{\rm s} - \frac13{M}^{P}_{\rm d}
+ {M}^{F}_{\rm d}\left (\cos^2\theta_{\pi}-\frac15\right)\,,\\ 
B&=& \left({M}^{P}_{\rm d} 
-\frac25 {M}^{F}_{\rm d}\right) \cos\theta_{\pi}\,.  
\label{ABpi0}
\ee

The observables in terms of these amplitude can be written as
\be\nonumber
\frac{d \sigma}{ d\Omega }&=&  |A|^2+|B|^2+ 2\mathrm{Re}\, [A B^*]
\cos \theta_{\pi}  \,,\\
A_y\frac{d \sigma}{ d\Omega }&=& 2 \,\mathrm{Im}\, [A B^*] \sin
\theta_{\pi} 
\,.
\label{obs}
\ee 
Omitting the squares of the
d-wave amplitudes, one finds  the following relation between 
the  observables 
in the  \pppi\  channel
\begin{eqnarray}
\nonumber
a_0&=&|{M}^{P}_{\rm s}|^2-
\fmn{2}{3} \mathrm{Re}\left[{M}^{P*}_{\rm s}({M}^{P}_{\rm d} +
\fmn{3}{5}{M}^{F}_{\rm d})\right]\\\nonumber
a_2&=& 2 \mathrm{Re}\left[{M}^{P*}_{\rm s}({M}^{P}_{\rm d} +\fmn{3}{5}{M}^{F}_{\rm d}) \right]\\
b_2&=& -2 \mathrm{Im}\left[
{M}^{P*}_{\rm s}({M}^{P}_{\rm d} -\fmn{2}{5}{M}^{F}_{\rm d}) \right].
\label{relationpi0}
\end{eqnarray}
where   $a_i$  and $b_i$  are the coefficients  in the  expansion 
 in powers  of  cos $\theta_{\pi}$ for  
the  differential  cross section  and the analyzing power,  i.e.   
\be
\frac{d\sigma}{d\Omega} &=& \frac{q}{4p} \sum_{n=0} a_n  
 {\rm cos}^n \theta_{\pi}     
 \nonumber \\
A_y\frac{d\sigma}{d\Omega} &=& \frac{q}{4p} \sin\,
\theta_{\pi}\sum_{n=0} b_{n+1} {\rm cos}^{n+1} \theta_{\pi}
\label{fits}
\ee
where  again $p$ is  the  incident proton momentum and $ q $  is  the momentum of the
outgoing pion.  
The  analogous  expressions connecting  
the partial wave  amplitudes  with the  observables  
 in the  \pnpppi\  channel  are given in\cite{ANKEpi-}. 
The coefficients  $a_i$  and $b_i$  
were extracted  from  the  best  fits  to   the data in both reactions  
\pppi\  and  \pnpppi.

\begin{table}[h]
\centering
\begin{tabular}{|c|r|r|r|} \hline
Amplitude  &  Real\phantom{xxll} & Imaginary\phantom{l}& Im/Re \phantom{1} \\ \hline
\multicolumn{4}{|c|}{Solution 1: $\chi^2/\textit{d.o.f.}=101/82$ } \\ \hline
$M^P_s$ & $  53.4 \pm    1.0$ & $ -14.1 \pm    0.3$ & \\ \hline
$M^P_d$ & $ -25.9 \pm    1.4$ & $  -8.4 \pm    0.4$ & \\ \hline
$M^F_d$ & $  -1.5 \pm    2.3$ & $   0.0 \pm    0.0$ &\\ \hline
$M^S_p$ & $ -37.5 \pm    1.7$ & $  16.5 \pm    1.9$ &$-0.44\pm0.06$\\ \hline
$M^D_p$ & $ -93.1 \pm    6.5$ & $ 122.7 \pm    4.4$ &$-1.32\pm0.11$ \\ \hline\hline
         \multicolumn{4}{|c|}{Solution 2: $\chi^2/\textit{d.o.f.}=103/82$} \\ \hline
$M^P_s$ & $  52.7 \pm    1.0$ & $ -13.9 \pm    0.3$ &  \\ \hline
$M^P_d$ & $ -28.9 \pm    1.6$ & $  -9.4 \pm    0.5$ &  \\ \hline
$M^F_d$ & $   3.4 \pm    2.6$ & $   0.0 \pm    0.0$ &  \\ \hline
$M^S_p$ & $ -63.7 \pm    2.5$ & $  -1.3 \pm    1.6$ &$0.02\pm0.03$  \\ \hline
$M^D_p$ & $-109.9 \pm    4.2$ & $  52.9 \pm    3.2$ &$-0.48\pm0.03$  \\ \hline \hline
         \multicolumn{4}{|c|}{Solution 3: $\chi^2/\textit{d.o.f.}=106/82$} \\ \hline
$M^P_s$ & $  50.9 \pm    1.1$ & $ -13.4 \pm    0.3$ &  \\ \hline
$M^P_d$ & $ -26.3 \pm    1.5$ & $  -8.5 \pm    0.5$ &  \\ \hline
$M^F_d$ & $   2.0 \pm    2.5$ & $   0.0 \pm    0.0$ &  \\ \hline
$M^S_p$ & $ -25.4 \pm    1.9$ & $  -7.3 \pm    1.5$ &$0.20\pm0.07$  \\ \hline
$M^D_p$ & $-172.2 \pm    5.6$ & $  92.0 \pm    6.2$ &$-0.53\pm0.04$  \\ \hline
\end{tabular}
\caption{\label{tab:fit}
Values of the real and imaginary parts of the amplitudes for five
lowest partial waves deduced from fits to the ANKE \pppi\ and \nppppi\
measurements at 353~MeV. Also shown are the ratios of the imaginary to real
parts of the p-wave amplitudes that have been freely fitted. 
The other three ratios for s- and d-wave amplitudes 
are fixed by the Watson theorem.} 
\end{table}

As  seen from Eq.\eqref{Mspin},  there  are  five complex  
partial  wave  amplitudes, $M^{L}_{\rm l}$, 
 that  need  to  be  determined  from a  global  fit  to the ten 
experimentally determined coefficients $a_i$  and $b_i$ in  Eq.(\ref{fits}).   
Fortunately the  number  of parameters  can be  reduced
if one imposes  the Watson  theorem  to  fix the  phases  of the
amplitudes. 
This  theorem states  that  the  phase  of the production
amplitude in  the  absence of coupled channels  is  fully determined
by  phases of  the   $NN$  interaction  in the  initial  and final
states.  
Hence, it  allows  one    to   fix    the  phase  of   the
$^3P_0\to  {^1 S_0} s$  transition which  removes  one  parameter.   
In   other partial  waves the initial $NN$  interaction appears in 
 coupled channels,    
and  it is  known  from  the phase shift analysis  of  $pp$
data   that  the  coupling  between  $^3S_1$  and    $^3D_1$  partial  
waves is sizeable for energies  around
pion-production  threshold\cite{SAID,ARN2007}. 
Meanwhile, for   $^3P_2-^3F_2$  transitions the  mixing parameter,  as
well  as  the inelasticites,   are still  negligibly small  at
$T_{\rm lab}=$353~MeV\cite{SAID,ARN2007}.   
This means  that  one  may  
 neglect  the coupled channel effect in   $^3P_2-^3F_2$  partial waves
 and  use  the Watson theorem to remove  another two
 parameters.  
Under  this 
 assumption, the s-wave  and   two  d-wave
pion-production amplitudes were  extracted  in Ref.\cite{ANKEpi0} 
based on  the  partial  wave  analysis of  the  \pppi\   data.
Simultaneously,  the fit to the combined \pppi\ and \pnpppi\ data
sets  has been  performed in Refs.\cite{ANKEpi-},   with
$\chi^2/\textrm{d.o.f.}=89/82$ for the  best  fit.  
The results for  s-  and  d-waves  from  the global  fit is
in good  agreement  with the  values  extracted  from the  analysis
of  the $\pi^0$ data only.  
\begin{center}
\begin{figure}[t]
\begin{center}
\includegraphics[width=0.8\textwidth, angle=0]{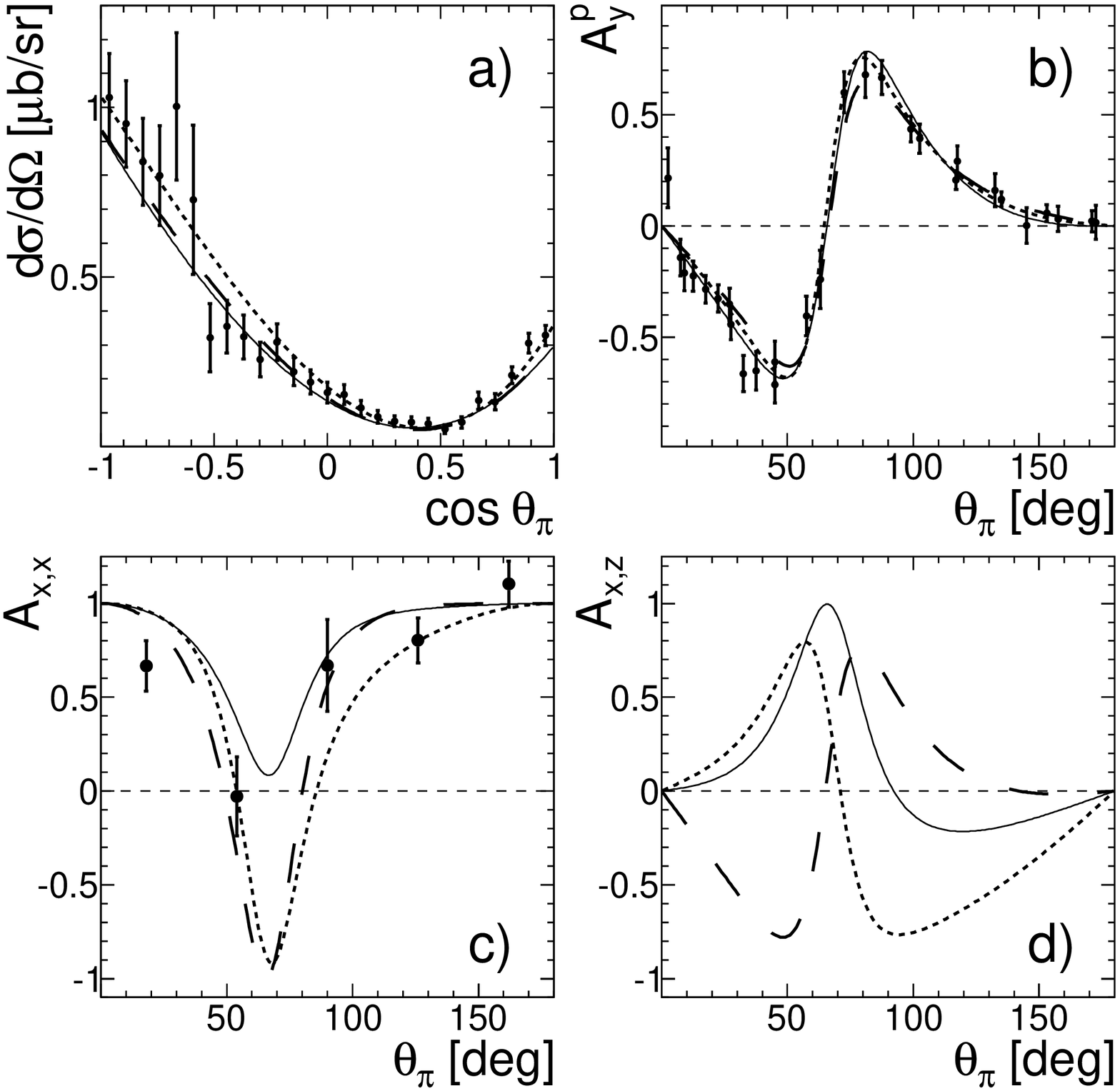}
\end{center}
\caption{ \label{fig:pwaAll4} Predictions of the partial 
wave analysis for the \pnpppi\ reaction
at 353~MeV with  the ${E}_{pp}<3$~MeV cut.
 Also shown are the ANKE experimental data with statistical
errors. The full, long-dashed, and short-dashed lines correspond to solutions
1, 2, and 3, as noted in Table~\ref{tab:fit}. a) Differential cross-section
taken from Ref.~\cite{ANKEpi0,ANKEpi-}, b) $A_y^p$ data from 
Ref.~\cite{ANKEAxx}, c) $A_{xx}$
data from Ref.~\cite{ANKEAxx}, d) $A_{xz}$, for 
which there are yet no experimental
data.  The  figure  is  taken  from Ref.{\cite{ANKEAxx}}. } 
\end{figure}
\end{center}
 
 In the  recent  study  \cite{ANKEAxx}  the results  of the
partial  wave  analysis  for p-waves \cite{ANKEpi-}  were  questioned.
In particular,   the  important  question  if  the global  solution,  
as  shown in Table  \ref{tab:fit}   solution 1,    
is  indeed  the true  one   was raised.  
In addition  to  the  global  minimum,  several local  minima
with only  slightly  larger  $\chi^2$ were  found     
yielding  completely
different  amplitudes  for the p-waves, shown as  
solutions 2  and 3  in  Table  \ref{tab:fit}.  
 The data on the differential cross section and  the analysing power   only
do not allow one to discriminate between the different  solutions,  
see  the  upper panels  in Fig.\ref{fig:pwaAll4}.  
The   transverse spin-correlation parameter  $A_{xx}$ 
could  resolve this p-wave issue,  since  the  curves for  
$A_{xx}$  corresponding  to  the different solutions   are,   in  principle,  distinguishable, 
 as   shown in    the lower  left panel of  Fig.\ref{fig:pwaAll4}, 
provided  the  high  resolution   
  data  existed in the whole angular  domain.  
However,  the  statistical uncertainty  of  the   recently  measured  data   \cite{ANKEAxx} 
  did not  allow  to distinguish between the minima.
As  argued  in\cite{ANKEAxx}, see also Fig.\ref{fig:pwaAll4} lower right panel,  
the double polarization  measurement  of  $A_{xz}$ in \pnpppi\  could  be  
useful in order  to  lift  the  ambiguity with the different $\chi^2$ solutions. 
The different p-waves scenarios  could  be  disentangled  either just by
determining the  sign  of  $ A_{xz}$       
or  by its  magnitude.

It is interesting to note  that  the phases  of  the p-wave amplitudes  
corresponding to  different  solutions in Table \ref{tab:fit}   
turned out to  
differ  significantly   from  each  other,  
as  can be seen  from  the last  column of this Table.  
On the  other  hand,  in spite  of   the   sizeable  coupling  between  
the $^{3\!}S_1$ and $^{3\!}D_1$ partial waves,  
one  would naively  assume that  the  phases   of  
the   p-wave  pion-production amplitudes 
should not  differ  drastically  from the  elastic  $NN$  phases.     
The   values  in    the last column of   Table \ref{tab:fit} should  be  compared 
with  nucleon-nucleon phase-shift analysis values of  
$(\tan\delta_{^{3\!}S_1},\tan\delta_{^{3\!}D_1})=(0.03,-0.46)$~\cite{ARN2007,SAID},
and to the values from the theoretical analysis of  the 
p-wave  pion  production amplitudes  of  $(0.04,-0.61)$~\cite{newpwave}.  
This  comparison  reveals  a  clear  preference  
against  the  solution 1    and   possibly  in favour  
of solution 2,   as  pointed out in \cite{ANKEAxx}.

 The experimental  measurements~\cite{ANKEpi0,ANKEpi-}  are  extremely
useful  to  improve  our knowledge about pion production: 
the new  high-accuracy  measurements  of   the  differential  cross
section and  analyzing power  in  \pnpppi\  in  the  whole  angular
domain  at  present constitute a  challenge  for  the theory.    
The
data  on \pppi\  clearly  demonstrated  that  the d-wave amplitudes
are  sizable  already  at  $T_{\rm lab}$=353~MeV and that  neglecting 
these amplitudes, as  was  done in  Ref.\cite{newpwave}  is not  justified.  
The partial  wave analysis  of the  \pppi\  data led  to  a  successful  extraction
of  the  s-  and  d-wave pion  production  amplitudes in the  isospin
1  channel  and  should  be
confronted soon with  the theory  calculations  within  EFT where the
complete  production operator up to  N$^2$LO for  s-wave pions  has been  already
derived,  cf.  Secs.~\ref{sec:swave} and \ref{sec:swaveD} for details.

\section{Charge symmetry breaking  in pion-production reactions}
\label{sec:CSB}

Isospin symmetry is normally realized to a few percent accuracy in
hadronic reactions. 
On the QCD level the symmetry is broken only by the different 
up and down quarks masses (both very small compared to typical hadronic
masses) and by their different charges. 
The largest isospin symmetry breaking effect that emerges
is the pion mass difference, 
$$2(m_{\pi^+}-m_{\pi^0})/(m_{\pi^+}+m_{\pi^0})\sim 4\% \ ,$$
which originate almost completely from electromagnetism.  Typical isospin-violating effects
are usually dominated by this pion mass difference as seen, e.g., in the 
spectacular energy dependence in neutral pion photo-production off protons near
threshold~\cite{e0p_data,Bernard:2007zu}. 
Due to this pion mass difference it is therefore difficult to quantify  
the isospin violations due to the up and down quark mass difference 
in hadronic observables.

The exceptions are observables that violate  charge symmetry (CS).
While general isospin symmetry or  charge independence  implies  invariance  of     interactions  
under  any rotation in isospin space,  CS  
requires invariance only   under an isospin rotation by 180$^\circ$ 
around the  $"2"$-axis. 
  As a  consequence,  when 
charge symmetry  is broken,   isospin  invariance is  also violated   
but   the  converse is not  correct.  
The pion  mass  difference term   is  charge symmetric, since the charge symmetry operation transforms 
a $\pi^+$ into a $\pi^-$ (and they have identical masses as a consequence of CPT invariance). 
CS is  violated by the up and down quark mass  difference as well as 
residual  electromagnetic interactions  (virtual photons)  after  effects of 
the pion mass  difference  are removed. 
The impact of virtual photons          
has been 
systematically studied in EFT~\cite{Meissner:1997ii,KnechtUrech,GMueller99,Fettes:2000vm
,Gasser:2002am,HKM,HKMlong}
and is a well  understood and calculable isospin violation. 

CSB effects manifest themselves in many different physical phenomena. 
Some  examples of CSB consist in 
  the mass splitting of hadronic isospin multiplets
(e.g.~$ m_n\ne m_p$~\cite{} and $M_{D^0}\ne M_{D^+}$~\cite{ourD}),
$\eta$--decays (for a recent two-loop calculation, see~\cite{Hans} and
references therein), the
different scattering lengths of $nn$ and $pp$ systems after removing
electromagnetic effects in $pp$ scattering (see, e.g.~the review
article \cite{Miller2006}), 
neutron-proton elastic scattering at
intermediate energies \cite{NNelast}, hadronic mixing 
(e.g.~$\rho^0-\omega$ \cite{Barkov} or $\pi^0-\eta$ \cite{coon} mixing) and
the binding-energy difference of mirror nuclei known as Nolen-Schiffer
anomaly \cite{Nolen}.  

In this section we will discuss  the manifestation of  CSB in  pion-production reactions.    
Recently, experimental evidence for CSB was found in reactions
involving the production of neutral pions. At IUCF a non-zero value for the
$dd\to \alpha \pi^0$ cross section was established \cite{Stephenson:2003dv}, and 
at TRIUMF a forward-backward asymmetry of the differential cross section for 
$pn\to d\pi^0$ 
was reported to be $A_{fb}=[17.2
\pm 8 {\rm (stat.)} \pm 5.5 {\rm (sys.)}] \times 10^{-4}$~\cite{Opper:2003sb}.

In what follows  we  will concentrate our discussion on the    
reaction $pn\to d\pi^0$ before we comment  
on the reaction $dd\to \alpha \pi^0$    at  the  end  of this section.  
Following the arguments presented in Ref.~\cite{filin}, 
we will show that at leading CSB order within the MCS scheme, 
only the strong (quark-mass induced) contribution to  the  proton-neutron
mass difference enters   the CSB  pion-production operator in $pn\to d\pi^0$.  
This quantity  was extracted in  Ref.~\cite{filin} based on the    
$A_{fb}(pn\to d\pi^0)$ data.          
        It is noteworthy  that 
the effect of  $\pi -\eta$ mixing, which 
was believed to completely dominate this CSB observable, 
is a sub-leading CSB effect as shown in Ref.~\cite{Niskanen:1998yi}.  

In order to see that  $A_{fb}(pn\to d\pi^0)$ is indeed a CSB observable, 
note first of all that the final $d \pi$-state is an isospin-1 state. 
This means that, in an isospin symmetric world, 
also the initial states is in isospin-1.  
Since an isospin-1 $pn$ pair behaves
like two identical nucleons $pp$ or $nn$ --- the forward and backward asymmetry is absent.  
However, since CSB  distinguishes  a proton from  a neutron, 
due to CSB interactions the initial $NN$ state acquires a small isospin-0 admixture, 
and the interference of this small isospin-0 component combined 
with the charge symmetry conserving contributions lead 
to a forward--backward asymmetry.

The forward-backward asymmetry is defined as
\begin{equation}
\label{Afb_definition_CS}
A_{fb} = \frac{ \int\limits_0^{\pi/2} \left[ 
\frac{d \sigma}{d \Omega} (\theta) - \frac{d \sigma}{d \Omega} (\pi - 
\theta)\right] \sin\!\theta d\theta}
              { \int\limits_0^{\pi/2} \left[ 
\frac{d \sigma}{d \Omega} (\theta) + \frac{d \sigma}{d \Omega} (\pi - 
\theta)\right] \sin\!\theta d\theta}
 = \frac{A_1}{2 A_0} \ ,
\end{equation}
where   Eq.~\eqref{diffXS} was used  in the last equality.  
The experiment at TRIUMF was done  at
$T_{{\rm lab}}=279.5$~MeV, which is very close to threshold and is 
equivalent to an excess energy of about
2~MeV. 
At this energy the total cross section 
$\sigma=4\pi A_0 = \alpha \eta + \beta \eta^3 $,  Eq.~(\ref{totXS}),  
is largely dominated by the isospin-conserving s-wave pion-production
amplitude in the final $NN$ spin-triplet channel,  cf. Table \ref{table1}.  
 In their analysis 
Ref.\cite{filin} used the experimental value for  $\alpha$ 
extracted by Ref.~\cite{pidexp} from the extremely precise PSI measurement 
of  the lifetime of the pionic deuterium atom.  
 Note that  unlike   scattering experiments the  measurement at PSI does not 
suffer  from normalization uncertainty.  
In addition,  in the $pn\to d\pi^0$ experiment close to threshold 
the  contribution from  pion p-waves,  calculated to N$^2$LO in
 Ref.~\cite{newpwave}   and discussed in Sec.~\ref{sec:pwave}, 
was taken into account in  Ref.~\cite{filin}.  
This p-wave contribution to the relevant final $NN$ 
spin-triplet channel\footnote{ Note  that  pion p-waves in the spin-triplet channel relevant here  
are  well under control,  cf.  Fig.~\ref{Ay_dpi}  in Sec.~\ref{sec:pwave}. 
On the other hand, the analysis in Sec.~\ref{COSY_data} discussed the uncertainty regarding the pion 
p-wave for a final $NN$ spin-singlet channel. } 
determines $\beta$ and amounts  to 10\% of the total cross section 
at the energy of the TRIUMF experiment. 
Thus the  total amplitude is determined to be 
$A_0=10.0^{+0.2}_{-0.4} \cdot \eta  \ + (47.8\pm 5.7)\cdot \eta^3 \  [\mu$b].

The coefficient $A_1$  in   Eq.~\eqref{Afb_definition_CS}  
proportional  to the interference of the CS  and CSB  pion-production amplitudes, 
 reads 
\begin{equation}
\label{eq_A_one}
A_1 = \frac{1}{128\pi^2} \frac{\eta \, m_\pi}{p(\mpi + m_d)^2} 
\, {\rm Re}\left[\left(M^{\rm CS,p}_1+ \frac23 M^{\rm CS,p}_2\right) 
M^{\rm CSB,s^\dagger}\right]+ \dots, 
\end{equation}
where  $m_d$ is the deuteron mass.  The amplitudes  $M^{\rm CS,p}_1$  and   
$M^{\rm CS,p}_2$  stand for the charge symmetric  amplitudes 
corresponding  to    production of  p-wave pions in  the  
$^1S_0\to {}{^3 S_1}p$ and $^1D_2\to {}{^3 S_1}p$ partial
waves, while $M^{\rm CSB,s}$ is the corresponding amplitude for the
charge symmetry breaking s-wave production in the $^1P_1\to {}^3S_1s$ partial wave. 
Furthermore, the ellipses stand  for  
the  other  term  in which the  CSB  amplitudes  for   a p-wave pion  
interfere  with the CS  amplitudes  for an s-wave pion.  
Since  the other term  is  suppressed 
by  $\chi_{\rm MCS}^2$,  it was ignored  in  the 
leading-order calculation by Ref.\cite{filin}.

As the charge symmetry conserving contributions were
already  discussed in great detail in the previous sections,  
we will here exclusively focus on the CSB contributions.

\begin{center}
\begin{figure}[t]
\begin{center}
\includegraphics[width=0.5\textwidth, angle=0]{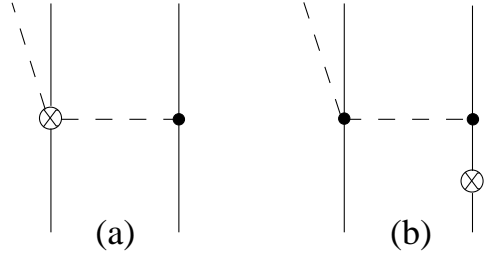}
\end{center}
\caption{\label{fig:diagLO_CSB}
The leading order CSB diagrams for $pn\to d\pi^0$. The crosses  correspond to  isospin-violating  vertices.}
\end{figure}
\end{center}

 At  leading-order  
 the  isospin-violating (IV) strong and electromagnetic 
operators  have the  form  
\beq
\mathcal{ L}_{\rm iv} =  \frac{\delta m_{N}}{2}
\; N^{\dagger}\tau_3 N 
-\frac{\delta m_{N}^{\rm str}}{4 f_{\pi}^2} 
\; N^{\dagger}\boldtau\cdot\boldpi \pi_3 N 
-\frac{\delta m_{N}^{\rm em}}{4 f_{\pi}^2} 
\; N^{\dagger}(\tau_3\boldpi^2-\boldtau\cdot\boldpi\pi_3 ) N 
+\ldots \, ,
\label{liv}
\eeq 
where the neutron proton mass difference is  
$\delta m_N=\delta m_N^{\rm str}+\delta m_N^{\rm em}$ and
the ellipses stand for further terms which are not relevant here. 
The $\delta m_N^{\rm str}$ is at this order given by the LEC  
$c_5 \propto m_d-m_u$, which  
 is part of ${\cal L}_{\pi N}^{(2)}$ in Eq.~(\ref{eq:la1}), and 
$\delta m_N^{\rm em}$ is given by the LEC $f_2$.
The precise definitions of the LECs  $c_5$ and $f_2$ can be found in 
Ref.~\cite{Meissner:1997ii}. 
At leading order in CSB 
the diagrams that contribute to the  transition amplitude from an initial $NN$
isospin-0 state to the final isospin-0 $NN$ state, are
shown in Fig.~\ref{fig:diagLO_CSB}. 
Diagram (a) corresponds to a pion
rescattering diagram where the $\pi N$ 
scattering vertex breaks charge symmetry. 
This CSB vertex is due to the last two terms in Eq.~(\ref{liv}). 
Diagram (b) on the other hand includes the $\pi N$ 
rescattering vertex given by 
the Weinberg-Tomozawa operator [the first term in 
${\cal L}_{\pi N}^{(1)}$, Eq.~(\ref{eq:la0})].
The isospin-violating term in diagram~(b) is, as indicated in 
Fig.~\ref{fig:diagLO_CSB}, given by 
the   proton-neutron mass difference,  which 
plays a pivotal role in the arguments below.

In the following we will show that due to the interplay of diagram~(a) and 
diagram~(b) in Fig.~\ref{fig:diagLO_CSB}, the electromagnetic term  
cancels, and  hence  the forward-backward asymmetry 
of the reaction $pn\to d\pi^0$  depends only on $\delta m_N^{\rm str }$. 
 In order to obtain this result, 
it is sufficient
to focus on the two $\pi N$ rescattering vertices on nucleon~1 
while we only keep the isospin structure from the  
pion-production vertex on nucleon~2. 
The other components  are identical for both diagrams. 
The relevant parts of diagram~(a), 
where a charged pion is exchanged between the two nucleons, 
read  
\beq
\hat I_{\rm (a)}= -i\frac{ \delta m_N^{\rm str}}{4\fpi^2} \left( \boldtau^{(1)} \cdot
  \boldtau^{(2)} + \tau^{(1)}_3 \tau^{(2)}_3\right) +i\frac{
  \delta m_N^{\rm em}}{4\fpi^2} \left( \boldtau^{(1)} \cdot \boldtau^{(2)}
  -\tau^{(1)}_3 \tau^{(2)}_3\right)\, ,  
\label{Iaop}
\eeq 
where $\boldtau^{(i)}$ is the isospin operator of the i'th nucleon. 
This expression 
represents the complete rescattering contribution included in
Ref.~\cite{jouni}. 
Diagram~(b) of Fig.~\ref{fig:diagLO_CSB} requires more scrutiny. 
First, observe    
that due to the isospin structure of the Weinberg-Tomozawa $\pi N$ scattering vertex, 
only charged pions can be exchanged between the two nucleons to produce a neutral pion. 
Thus as argued above for CSB reactions in general, the pion mass difference 
does not enter.
However, the proton-neutron mass difference generates an energy difference  
of the exchanged 
charge pion  in diagram~(b), 
depending on whether a $\pi^+$ is produced at the vertex on nucleon~2 
 via a $p\to n$ transition, 
or a $\pi^-$, produced via a $n\to p$ transition.
The energy dependence of the  Weinberg-Tomozawa vertex in diagram~(b) is  
sensitive to this energy difference and gives rise to the CSB amplitude from diagram~(b). 
On the operator level we obtain analogously to 
Eq.~(\ref{Iaop}), 
\begin{eqnarray}\nonumber
\hat I_{\rm (b)}&=&\frac{\delta m_N}{2}\frac1{4\fpi^2} 
\epsilon^{3bc} \tau^{(1)}_c\left(\tau^{(2)}_b \tau^{(2)}_3
-\tau^{(2)}_3\tau^{(2)}_b\right)   \\
&=& -i\frac{
  \delta m_N}{4\fpi^2} \left( \boldtau^{(1)} \cdot \boldtau^{(2)}
  -\tau^{(1)}_3 \tau^{(2)}_3\right) \ ,
\label{Iaop2}
\end{eqnarray} 
where the two terms in the first row include both an insertion of the CSB nucleon mass
term before and after the pion emission vertex.
It follows that in the sum of the two diagrams, the term
from the electromagnetic contributions cancels 
and the leading isospin-violating transition amplitude is proportional to $\delta m_N^{\rm str}$. 
The relevant transition matrix element for the $NN$ states is~\cite{filin}  
 \beq \langle I_f=0|\hat I_{(\rm a)}+\hat I_{(\rm b)}
|I_i=0\rangle = \frac{i}{4\fpi^2}\ 6\, \delta m_N^{\rm str}.
\label{sum}
\eeq 
Compared to the expression in Eq.~(\ref{Iaop}), 
which was used in  Ref.~\cite{jouni},  
the rescattering operator in Eq.~(\ref{sum}) 
is enhanced by about 30\% when the 
standard values  $\delta m_N^{\rm str}\approx 2$~MeV and 
$\delta m_N^{\rm em}\approx -0.76$~MeV~\cite{} are used.

An alternative method to derive the same result is 
obtained by redefining the pion and nucleon fields in the  Lagrangian as discussed in
Refs~\cite{friar,friar2,Epelbaum:2007sq} --- see also 
Ref.~\cite{Epelbaum:2004xf} which uses the  unitary
transformations. 
The pion and nucleon
fields are redefined in order to eliminate the first term in the
effective Lagrangian in Eq.~(\ref{liv}) 
and this allows one to work with
nucleons as indistinguishable particles. 
The key point is that the 
terms in a Lagrangian
are invariant under this transformation {\it except} 
the ones involving a time derivative. 
The Weinberg-Tomozawa operator 
in the $\pi N$ rescattering vertex in Eq.~(\ref{eq:la0}) 
includes a time derivative of the field which generates
an additional isospin-violating $\pi N\to \pi N$ vertex $\propto
\delta m_N$ that cancels exactly the electromagnetic contribution to
the  $\pi N$ vertex $\propto \delta m_N^{\rm em}$, see the last  term in Eq.~\eqref{liv}.

\begin{center}
\begin{figure}[t]
\begin{center}
\psfrag{xxx}{$pn{\to}d\pi^0$} 
\includegraphics[width=0.6\textwidth, angle=0]{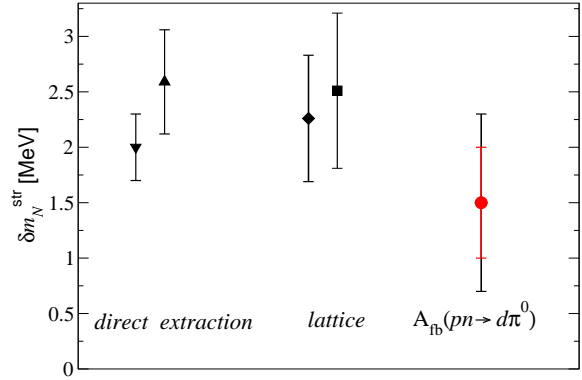}
\end{center}
\caption{\label{fig:delm}
The proton-neutron mass difference extracted with different methods:
a {\it direct extraction} via the Cottingham sum rule (triangle down\cite{Gasser:1982ap},     
triangle up\cite{Walker-Loud}),
a determination using $lattice$ QCD (diamond~\cite{lattice},  square\cite{lat2}) and an
extraction from $pn\to d\pi^0$~\cite{filin}.  
For the last value the inner (red) error bar indicates the size of the theoretical uncertainty.  
Note  also  that  
the uncertainties  of  lattice data  are dominated by  systematic errors.}
\end{figure}
\end{center}

To summarize, within the leading order CSB calculation the result  for  the
forward-backward asymmetry is  calculated to be~\cite{filin}  
 \beq 
A_{\rm fb}^{\rm LO} = (11.5 \pm 3.5)\times 10^{-4} \ 
\frac{\delta m_N^{\rm str}}{\rm~MeV} \, ,
\label{AfbLO}
\eeq 
where the theoretical uncertainty  was estimated in  \cite{filin}
to be $\chi_{\rm MCS}^2\simeq$15\% which 
is doubled in Eq.~(\ref{AfbLO}) in order to give a conservative error.  
This  result  was confirmed in Ref.~\cite{bolton} within error bars.   
At next order in CSB  photon-loop diagrams contribute,   
and in  analogy with the pion s-wave NLO loop amplitudes presented 
in Sec.~\ref{sec:swave}, the sum of these CSB loop amplitudes is 
zero~\cite{Gardestig:2004hs}, i.e. 
the contribution to the CSB amplitude from the next order diagrams vanishes 
(see also Ref.\cite{bolton} where some CSB 
tree-level operators at  N$^2$LO were also  taken into account). 
Comparing the experimental result for $A_{\rm fb}$ \cite{Opper:2003sb} 
with Eq.~(\ref{AfbLO}) allows an extraction of 
$\delta m_N^{\rm str}$:
\begin{equation}
\delta m_N^{\rm str} = 
\left(1.5 \pm 0.8 \ {\rm (exp.)} \pm 0.5 \ {\rm (th.)}\right) \ {\rm~MeV} \ .
\end{equation} 
This result is consistent with other evaluations of
this contribution to the neutron proton mass
 difference  in Refs~\cite{Gasser:1982ap,Walker-Loud}  
based on the Cottingham sum rule \cite{Cottingham}  and  
 in Refs. \cite{lattice,lat2,lat3} based on lattice 
simulations,   cf. Fig.~\ref{fig:delm}. 
This consistency is a  non-trivial and encouraging test of the
 applicability of ChPT to  $NN\to NN\pi$ reactions.

\begin{center}
\begin{figure}[t]
\begin{center} 
\includegraphics[width=1.0\textwidth, angle=0]{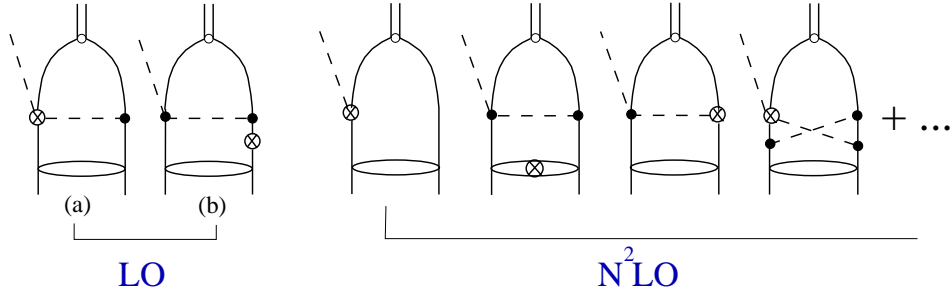}
\end{center}
\caption{\label{fig:diagN2LO_CSB}
CSB operators  for $pn\to d\pi^0$ to N$^2$LO.  The double  line  stands for  the  deuteron,  
the  crosses  correspond to the isospin-violating  vertices. } 
\end{figure}
\end{center}
 
As usual in EFT, the  next  step  in the study  of  $A_{fb}(pn\to d\pi^0)$  
should  consist  in the  extension 
of  the CSB  pion-production operator to  higher orders.   
As  already  said,  all   CSB operators  at NLO vanish.   
Some examples  of  CSB operators  at   N$^2$LO  are shown in 
Fig.~\ref{fig:diagN2LO_CSB},  see  Ref.~\cite{Gardestig:2004hs}  for further details.
In particular,     the  CSB  operator  at   N$^2$LO involves  
the  one-nucleon term  with  the  CSB  $\pi NN$ 
vertex   accompanied  by two unknown LECs which were modelled   
in Ref.\cite{kolckIV}  by  the $\pi-\eta$ mixing mechanism. 
In addition,   at this order,   CSB  in  the $pn$  potential generates  
a  contribution   to   CSB pion-production.
Furthermore,  there  are many different  loop contributions  analogous   
to those  discussed in Sec.~\ref{sec:swave} for  CS  pion production 
in an  s-wave.  
It remains  to be seen if  cancellations  which were operative  in the CS  case,  
will  also  take place  here.   
The proper evaluation of    $A_{fb}(pn\to d\pi^0)$ at this  order  
would also  require  the  calculation  of   CSB  pion production in a p-wave at LO, 
since  the  interference of  the CSB  operator for a  p-wave pion  
with the  CS  s-wave  pion-production amplitude  starts  to  
contribute at  N$^2$LO,  as  discussed.  
A  valuable  help  in understanding  of higher order  CSB mechanisms in 
$pn\to d\pi^0$    could  be  obtained  if  this reaction were  studied 
  in combination  with  the $dd\to \alpha\pi^0$ reaction.

The IUCF measurement~\cite{Stephenson:2003dv} of  the $dd\to \alpha\pi^0$ reaction 
inspired the 
first theoretical evaluation of this process~\cite{Gardestig:2004hs,nogga}.  
These studies  showed that the relative
 importance of the various charge symmetry breaking effects are very different
 for the $pn\to d\pi^0$ and $dd\to \alpha\pi^0$ reactions.   
For example, photon exchange in
 the $dd$ initial state could significantly enhance the cross sections for 
$dd\to \alpha\pi^0$~\cite{timo}.      In addition,  the rescattering mechanism  with  the CSB  $\pi N$ vertices at LO  
seems  to  play  only a subleading role in  $dd\to \alpha\pi^0$ due to 
symmetries in the  wave  functions, as  advocated in  Refs. ~\cite{Gardestig:2004hs,nogga}. 
Furthermore,  the  calculations  of Refs.\cite{nogga,jerrynew}  revealed  significant  
sensitivity to  the  nuclear interaction model. 
However, it was demonstrated in Ref.~\cite{jerrynew} that a simultaneous analysis of 
CSB in the two nucleon sector and in
$dd\to\alpha\pi^0$ strongly constraints the calculations of the latter. 
The main uncertainty in the evaluation of  $dd\to \alpha\pi^0$ is 
the treatment of the initial state interactions. 
At pion threshold energies the $dd\to \alpha\pi^0$ reaction 
requires reliable wave functions for $dd\to 4N$ in low partial waves 
at relatively high energies around pion-production threshold. 
A successful theoretical description of the data for the 
isospin-conserving reaction $dd\to ^3$He$\, n \, \pi^0$~\cite{Adlarson}, 
which partially shares the same initial state as $dd\to \alpha\pi^0$,  
could serve as a valuable test of the theoretical approach to the 
$dd\to \alpha\pi^0$ reaction.  
  
There is another important test    that could  be used to check   consistency of  the planned combined 
analysis of the recent CSB experiments. 
 Namely, once the LECs that enter CSB  s-wave pion production  at N$^2$LO  are fixed, 
 the p-waves in $dd\to \alpha \pi^0$ can be
predicted parameter free to leading and next-to-leading order. 
The IUCF experiment~\cite{Stephenson:2003dv},  being carried out 
very close to threshold,  was consistent with the contribution of pion
s-waves only.  Thus, the same experiment at somewhat higher
energy (but still well below the $\Delta$ region, e.g. at $Q\approx 30-60$~MeV) is  desirable.

\section{Remarks on the reaction $NN\to NN\pi\pi$}
\label{sec:twopiproduction}
The  reaction $NN\to NN\pi\pi$ provides a contribution to  the nucleon-nucleon  inelasticity  
next in importance after   one-pion production.  
The proximity  of  the  $\Delta$ resonance   mass  to    the   $\pi\pi N$ threshold   suggests  
the  dominance of the   $\Delta$ resonance
in  the two-pion production mechanism.    
However,  as will be argued below,  in the  channel  where  the  final  
isoscalar $\pi\pi$-pair  is  in an s-wave,  for example  in  $p n\to d\pi^0\pi^0$ 
near threshold,    the  contribution of the  $\Delta$    is suppressed   due to isospin    
compared to  the other two-pion production 
channels.   
The reaction channel   $p n\to d\pi\pi$ is  of particular  interest 
due to  the so-called  ABC 
effect  \cite{ABC} which has been  recently  observed  
by WASA collaboration at  
COSY \cite{abcPRL,Bashkanov}. 
The ABC effect refers to a pronounced low-mass enhancement in 
the $\pi\pi$ invariant mass spectrum in the reactions 
where  the  pion pair is produced together  with  the final nucleus in, e.g.,  
$pn\to d\pi\pi$ and $dd\to\alpha\pi\pi$ (these reactions 
are often named  double-pionic fusion reactions). 
Furthermore,  the data of Ref.\cite{abcPRL}  revealed that the 
ABC effect in  $p n\to d\pi\pi$  is uniquely correlated with
a resonance-like energy dependence in the total cross section.   
We stress that the  peak  in the  total  cross section  appears at  the energy 
  more than  200 ~MeV  above $NN\pi\pi$ threshold, 
which precludes  the use of  chiral EFT for understanding this phenomenon.
  On the other  hand,  the understanding of the  near-threshold   behaviour  of  $p n\to d\pi\pi$, 
  which provides the background  contribution,   is   possible  within chiral EFT.
  
\begin{figure}[htbp] \vspace{0.cm}
\begin{center} 
\includegraphics[height=6.41cm,keepaspectratio]{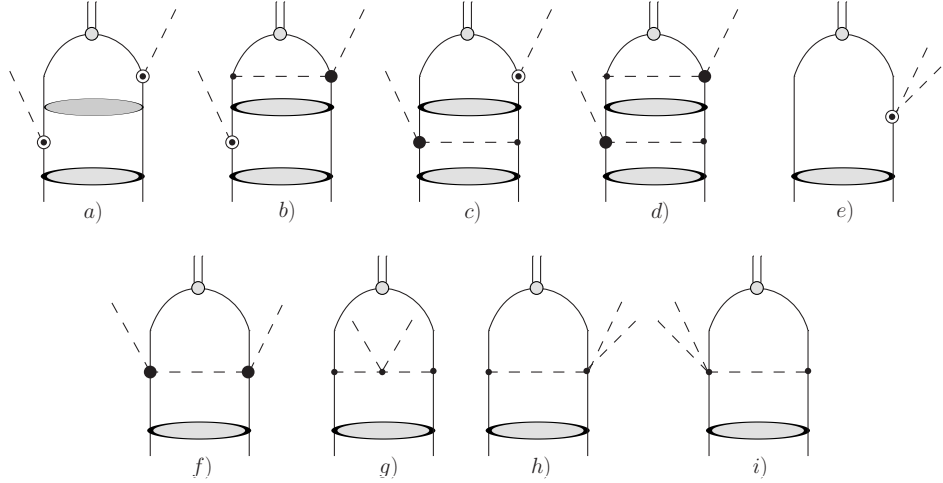}
 \caption{Leading order diagrams for the reaction $pn \rightarrow d \pi\pi$ at
   threshold.
Solid (dashed) lines denote nucleons (pions), filled
ellipses correspond to continuum $NN$ wave functions (including plane wave)
in the initial and intermediate state,  the outgoing double line
denotes the deuteron. 
Subleading vertices are marked as $\odot$, while  the blob  denotes   
the  leading $\pi N$ WT vertex  
from Eq.\eqref{eq:la0} and its recoil correction from Eq.\eqref{eq:la1}.
}
 \label{feyn_fig1}
\end{center}
\end{figure}

A complete leading order calculation for the
reaction $pn\to d\pi^0\pi^0$ at threshold within chiral EFT was presented in Ref.~\cite{liu}.   
The diagrams evaluated in Ref.~\cite{liu} are displayed in Fig.~\ref{feyn_fig1} and 
at LO there is no free parameter.  
Previous investigations of this reaction~\cite{ruso,ruso2} included 
only the diagrams in the second row whereas Ref.~\cite{liu} showed that 
a complete LO evaluation requires all diagrams in Fig.~\ref{feyn_fig1} to be considered.
The most important findings of Ref.~\cite{liu} are: 
\begin{enumerate}
\item all diagrams evaluated are of similar magnitude;        
\item there are                                                
important interferences among the                           
diagram contributions;
\item an accurate inclusion of the $NN$ interaction in both
intermediate as well as initial $NN$ states is very important.
\end{enumerate} 
In order to produce  two  pions  at  threshold,  the  
initial $NN$ momentum in their center of mass system, $p_{\tiny{\rm thr}}^{(2)}$,  
must be larger than that for the single pion production. 
Consequently, the expansion parameter  is 
$$
 \chi=\frac{p_{\tiny{\rm thr}}^{(2)}}{m_N}\approx 0.5 \ .
$$ 
An important future task is 
therefore  to calculate higher order contributions in order 
to check the rate of convergence.  As was pointed out, despite the proximity of the \Del mass to the $\pi\pi N$ threshold, the
potentially most important $N\Delta$ intermediate state is not allowed for
the reaction $pn\to d\pi^0\pi^0$ due to isospin conservation.
The reason is that  the $NN$ systems in the initial and 
final states must have isospin zero,  while the
$N\Delta$ system has   isospin one or two. 
Hence, to fulfil isospin conservation the isospin-1 $N\Delta$ state  can only
occur after  one-pion  emission and, thus, the role of  the 
$\Delta$ in $pn\to d\pi\pi$ is expected 
to be analogous to that in one-pion production, as  discussed in the previous sections.  
In particular, the $\Delta$ starts to contribute at NLO 
due to the fact that  the $\Delta$ propagator 
is suppressed by $1/p$ as compared to 1/$m_\pi$ in the nucleon case.

It is  known from phenomenological studies of $NN\to NN\pi\pi$~\cite{ruso,ruso2}
that the Roper resonance ($N^*$) can play a significant role already 
near threshold and that it
will become even more important when considering 
this reaction at higher energies.
This $N^*$ resonance is, however, not included explicitly 
in the LO evaluation of Ref.~\cite{liu}.
The reason is  that  in  chiral  EFT       
its contribution is absorbed into low energy constants that  start to  
matter  only at  N$^2$LO in the MCS. 
On the other hand,  an interesting  question to address would be 
to understand if its  contribution  is natural  or 
  it  is enhanced  compared to the  dimensional analysis estimate. 
The contribution of the Roper resonance to the $c_i$ parameters, that scale the
strength of the leading isoscalar $\pi N$ scattering, has been discussed, e.g.,
in Ref.~\cite{bernard}.
It seems that it plays only a minor role there. 
Similar
conclusions were drawn from systematic studies within ChPT of the double-pion
photoproduction process \cite{BKMSgamma,BKMgamma} and the reaction $\pi N\to
\pi\pi N$ \cite{BKM_pipiN95,Fettes_pipiN,Mobed} near threshold.  
In
particular, for the reaction $\gamma p\to \pi^0\pi^0 p$ the contribution of
the Roper was found to be rather moderate as compared to the large
contribution of chiral loops \cite{BKMgamma}.  
A  model calculation of
$\pi N\to \pi\pi N$~\cite{schneider} suggests also that
the Roper resonance plays a rather minor role.  
On the other hand, the Roper
might contribute significantly to the $(N  N)^2\pi\pi$ counterterms, which
enter at \NNLO{} in a chiral EFT calculation of $NN\to NN\pi\pi$.  
Based on the
discussions above, one can not expect that a leading order calculation of Ref.~\cite{liu} for 
$NN\to NN\pi\pi$ can
describe the experimental data well at higher energies and, hence,  
to answer  the  question about  the role  of  the $N^*$ resonance. 
However, it provides an estimate for the contribution of the non--resonant
background near threshold.   It therefore forms a basis for future studies and
is thus a precondition to extract reliable information on the Roper resonance
from future near-threshold experiments.

\section{The role of $NN\to NN\pi$ in  $\pi d$--atom observables}
\label{sec:pid-atom}

The isoscalar and  isovector pion-nucleon  scattering lengths  are fundamental
quantities of low--energy hadron physics.  
They serve as tests the QCD
symmetries and the  chiral symmetry breaking pattern. 
As stressed by
Weinberg a long time ago, chiral symmetry suppresses 
the isoscalar $\pi N$ scattering length $a^+$ compared to its isovector
counterpart $a^- \, $. 
These scattering lengths appear  as  subtraction constants in the 
Goldberger--Miyazawa--Oehme sum rule \cite{GMO,AMS,ELT},  
that is used to  determine  precisely the  pion-nucleon coupling constant $g_{\pi NN}$.  The scattering lengths 
are  also needed  in the dispersive analyses  of  
the pion-nucleon $\sigma$ term \cite{sigmaterm}. 
The most  natural  choice to determine these  quantities is  to measure 
the ground state  pionic hydrogen atom level shift  and  width 
due to strong  $\pi N$ interaction.  
It has been known   from pioneering  work by Deser 
 {\it et al.}~\cite{Deser} that the  hadronic level shift  is  proportional  
to  the real part of  the scattering length of  strongly interacting particles,  
 while  the  lifetime of  the atom        
is related to  the imaginary part of the scattering length.   Specifically,  in isospin limit the atomic level shift in $\pi^- p$      
reads  $\varepsilon_{1S}\propto a_{\pi^- p}=a^+ + a^-$,   
while  the width is 
$\Gamma_{1S}\propto a_{\pi^- p\to \pi^0 n}^2 \sim (a^-)^2$.  
A systematic inclusion of isospin-violating (IV) electromagnetic  
corrections  to  the  
relation between the  energy level  shifts and  width  and  
the pionic atom  scattering lengths  
is   performed in  Refs.~\cite{LR00,Gasser:2002am,zemp}, 
see also the review 
article \cite{hadatoms}.  
In addition,  taking into account    isospin violation in  the $\pi N$ scattering 
lengths \cite{Meissner:2005ne,HKM,HKMlong,Gasser:2002am}  is very  important.  
From   the   PSI measurement~\cite{Gottawidth} the   
shift   $\varepsilon_{1S}$ is known with unprecedented 
accuracy to be $\varepsilon_{1s}=(-7.120\pm 0.012)\,{\rm eV}$. 
Meanwhile,   the  width  is known  less accurately, 
$\Gamma_{1s} =(0.823\pm 0.019)\,{\rm eV}$\cite{Gottawidth},   
yielding  a  significant  uncertainty  in the determination of   $a^+$  
such that  even its  sign is unknown. 
Fortunately, the ground state level shift  of the pion-deuteron atom 
($\pi d$) is  experimentally   
known to a 2\% accuracy \cite{pidexp}   
and contains complementary information on 
$a^+ $ and $a^-$.  
Schematically,  the real part of  the pion-deuteron scattering  length  
can be  written as 
\be 
\mbox{Re}(a_{\pi d}) = 2a^+ + (\mbox{few--body corrections}) + (\mbox{IV corrections}) \ .
\ee
The first term $\sim a^+$
originates from the impulse approximation and is independent of the deuteron structure. 
Provided one can calculate the few--body corrections accurately the 
$\pi d$ scattering is ideally suited to extract $a^+$.  
However, already at threshold the $\pi d$ scattering length is a
complex-valued quantity and it is therefore crucial to obtain a precise
understanding of its imaginary part.

A  systematic  study  of  the  few-body  corrections  to  
the $\pi d$  scattering length has  been done 
within chiral  EFT\footnote{ 
The interested reader can find 
references to numerous phenomenological studies in, 
e.g., Ref.~\cite{ELT} }  
in the  last  two decades~\cite{swein1,
beane98,beane,Liebig,recoil,disp,
piddelta,Nogga,MRR,JOB,longJOB,MSS}.  
Unlike the pion-production reaction,
elastic pion-deuteron scattering is  a low-momentum transfer process, 
and  is essentially controlled
 by the standard ChPT expansion parameter  $\chi =\mpi/m_N$\footnote{ 
 The  characteristic distances in the deuteron are normally  estimated  by   
  taking the expectation   value of $1/r$  
between  the deuteron wave functions,  
  which gives  $\langle  1/r \rangle\simeq 0.5~\mbox{fm}^{-1}\sim \mpi$.  }.   
The other  scales,   $\sqrt{\mpi  m_N}$  
 due to   pion-absorption,    $\delta  \sim \sqrt{\mpi m_N}$ 
due to the explicit  $\Delta$-resonance,   the  deuteron  binding 
 momentum  $\gamma$   and     $\sqrt{\mpi/ m_N}\, \mpi$  
due to the 3-body  $\pi NN$  cut        
 have to be properly accounted for.  Ref.\cite{longJOB} demonstrated 
that the few-body corrections can be reliably calculated up to 
$\chi^{3/2}= \chi_{\rm MCS}^3$, 
which is smaller than 
the leading unknown $(NN)^2\pi\pi$ contact term  which enters at $\chi^2$.
 From a naive dimensional analysis the uncertainty anticipated 
due to the truncation of the higher-order terms is a few
percent.
 The  most  important  contributions   to the transition operator  
up to  $\chi^{3/2}$  
can be summarized  
as follows: 
\begin{itemize}
\item  the double-scattering diagram~\cite{swein1,beane98} is the    
most important contribution in the multiple-scattering series (MSS) 
as was recognized decades ago  

\item  the  triple-scattering diagram is the next correction in the MSS series~\cite{beane,Liebig} 

\item  nucleon recoil  corrections  to  the  double-scattering  process~\cite{recoil,recoil_BER}

\item  dispersive corrections due to $\pi d\to NN\to \pi d$ and $\pi d\to \gamma NN\to \pi d$~\cite{disp}

\item  explicit  $\Delta$-isobar contributions  \cite{piddelta}

\item  isospin violation  in the  3-body  sector:  few-body  corrections  
caused by   the pion mass  difference  and  virtual  photons \cite{JOB,longJOB}

 \end{itemize}
While  the detailed discussion of all these  effects  can be  found  in  
Ref.~\cite{longJOB},  
we will here focus  on the   role 
of pion production for  $a_{\pi d}$, since it  controls  
the strength of the  important dispersive corrections.

\begin{figure}[t!]
\begin{center} 
\includegraphics[height=2.1cm,keepaspectratio]{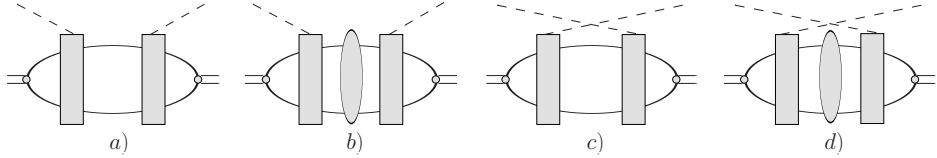}
\end{center}
\caption{Dispersive corrections to the $\pi d$ scattering length.  
The  notation of lines is the same  as  before.  
The rectangular blocks   refer to the pion-production (or absorption) 
operator  that consists 
of the direct and rescattering mechanisms including the recoil corrections,  
as  shown in the  first  row  of Fig.\ref{fig:allN2LO}.   }
\label{disp}
\end{figure}

The measurement of the $\pi d$ scattering length shows that its imaginary 
part is about 1/4 of its real part~\cite{pidexp}. 
The imaginary part can be expressed  in terms of the $\pi d$ 
total cross section through the
optical theorem, 
\begin{equation}
4\pi \mbox{Im}(a_{\pi d})=\lim_{q\to 0}q\left\{
\sigma(\pi d\to NN)+\sigma(\pi d\to \gamma NN)\right\} \ ,
\label{opttheo}
\end{equation}
where $q$ denotes the relative momentum of the initial $\pi d$ pair.
Furthermore, the ratio $R=\lim_{q\to 0}\left(\sigma(\pi d\to NN)/\sigma(\pi d\to
\gamma NN)\right)$ was measured to be $2.83\pm 0.04$
\cite{highland}. 
It is expected that diagrams which lead to a sizeable
imaginary part of $a_{\pi d}$,  will also contribute significantly via dispersive corrections  
 to the real part.
  In the mid of 70's the dispersive corrections were  calculated  within the  
Faddeev formalism~\cite{dispF1,dispF2}, showing  that   
the  dispersive and absorptive  contributions to the real and imaginary parts of the 
$\pi d$ scattering length are indeed of similar size.  
In Ref.\cite{disp} the   chiral EFT calculation of the 
dispersive corrections that emerge from $\pi d\to NN\to \pi d$ 
and $\pi d\to \gamma NN\to \pi d$ processes    was presented. 
The hadronic dispersive corrections were  defined as contributions from diagrams 
with an intermediate state that contains only nucleons as well as  crossed pion lines.   
Therefore, all the diagrams shown in Fig.~\ref{disp} were included in 
the evaluations in Ref. \cite{disp}.  
In other words, in addition 
to the diagrams  a)  and b)  in Fig.~\ref{disp}, that  are  present in the Faddeev calculations, 
 the contributions  with  crossed  pions  ( c)  and  d))  
were part of the  EFT  study   since  all diagrams  shown in Fig.~\ref{disp}
 are of the same  EFT  order.   
The rectangular blocks in the diagrams of Fig.~\ref{disp} refer
to the relevant transition operators for the reaction $NN\to NN\pi$ that consist 
of the direct and rescattering mechanisms including the recoil corrections,  
as  shown in the  first  row  of Fig.\ref{fig:allN2LO}.   
It was demonstrated in Sec.\ref{sec:obs}  that  the  
same production operator  resulted  in a  good  description of  the  
$pp\to d\pi^+$ total  cross section.
It is therefore not surprising  that  the calculation~\cite{disp} agreed nicely   
with the  imaginary part  of the $\pi d$-scattering length.
Using the CCF potential~\cite{CCF} for the $NN$ distortions 
Ref.~\cite{disp} found for the dispersive
correction from the purely hadronic transition
\begin{equation}
\delta a_{\pi d}^{\rm disp}=(-6.5+1.3+2.4-0.2)\times 10^{-3} \, m_\pi^{-1}
=-3.0\times 10^{-3} \, m_\pi^{-1} \ , 
\end{equation}
where the numbers in the first bracket are the  results from each 
diagram in Fig.~\ref{disp}, in order.  
Note that the diagrams with
intermediate $NN$ interactions and the crossed ones (diagram $(c)$ and $(d)$) 
give
significant contributions. 
None of of these diagrams were included in most of the previous calculations. 
The calculations were done with the four different phenomenological 
$NN$ potentials:  the CCF potential~\cite{CCF}, 
CD Bonn \cite{CDBonn}, Paris \cite{paris}, AV18 \cite{AV18}, and   
Refs.\cite{disp,piddelta}  obtained
\begin{equation}
\delta a_{\pi d}^{\rm disp} = (-2.9\pm 1.4) \times
\,10^{-3} \ m_\pi^{-1} \ ,
\label{result_disp}
\end{equation}
where the  number reflects the theoretical uncertainty of 
this calculation estimated conservatively --- 
see Ref.~\cite{disp,piddelta} for details.  
 In spite of the  cancellations  between  the diagrams  $(a)$ and $(b)$  
and  those with crossed  pions  ($(c)$ and $(d)$),   
  the  resulting  dispersive  correction in chiral EFT   constitutes  
an important  contribution to the  $\pi d$ scattering  length  
  of  the order of 10\%,  which  is  a  factor  of  
two larger  than  the  theoretical  uncertainty  estimate.

Including  the  few-body  and  isospin-violating  contributions  listed above,   
the  combined analysis  of the data on the  pionic  hydrogen shift 
and width and  deuterium shift  yielded~\cite{JOB,longJOB}
\beq
\tilde{a}^+=(1.9\pm 0.8)\cdot 10^{-3} \mpi^{-1},\quad 
a^-=(86.1\pm 0.9)\cdot 10^{-3}\mpi^{-1},
\label{atilde}
\eeq
where $\tilde a^+$ is defined as~\cite{Meissner:2005ne,Baru:2007ca}  
\beq
\tilde a^+ \equiv a^+ + \frac{1}{4\pi(1+\mpi/m_p)}
\bigg\{\frac{4\Delta_\pi}{\fpi^2}c_1-2e^2 f_1\bigg\}\label{atilde_def} \, , 
\eeq
and  $\Delta_\pi$  stands  for the mass difference  
between the charged and neutral pions,  $\Delta_\pi=\mpi^2-m^2_{\pi^0}$.
The isoscalar scattering length $a^+$ cannot be obtained independent of 
the LECs $c_1$ and $f_1$.
Using the rough estimate of $|f_1|\leq 1.4 \,{\rm GeV}^{-1}$ by 
Refs.~\cite{Gasser:2002am,Fettes:2000vm
}  and  
$c_1=(-1.0\pm 0.3)\,{\rm GeV}^{-1}$ used in Ref.\cite{longJOB},  
Eqs. \eqref{atilde} and \eqref{atilde_def} yield a non-zero $a^+$ 
at better than the $95\,\%$ confidence level
\beq
a^+=(7.6\pm 3.1)\cdot 10^{-3}\mpi^{-1}.
\label{apl}
\eeq
The results for the  scattering lengths were  used in Refs.~\cite{JOB,longJOB} 
to  infer the charged pion-nucleon coupling constant, $g_c$, from the GMO sum rule, 
with isospin-violating corrections to the $\pi N$ scattering
lengths fully under control for the first time. 
The result reads
\beq
\frac{g_c^2}{4\pi}=13.69\pm 0.12\pm0.15,\label{gc}
\eeq
where the first error gives the uncertainty due to the scattering 
lengths and the second that due to evaluation of  the cross section integral 
$J^-$, see  Refs.\cite{longJOB,AMS,ELT} for  
the definition and  further discussion. 
This value is in agreement with determinations from 
$NN$ ($g_c^2/4\pi=13.54\pm 0.05$~\cite{deSwart}) 
and $\pi N$ ($g_c^2/4\pi=13.75\pm 0.10$~\cite{FA02}, 
$g_c^2/4\pi=13.76\pm 0.01$~\cite{arndt06})          
scattering data where  the  errors  in Refs.~\cite{deSwart,FA02,arndt06} mainly 
reflect the statistical uncertainties.

\section{Implications for EFTs for decays of heavy quarkonia}
\label{heavyquarkonia}

In this review we have demonstrated that a consistent, 
convergent effective field theory can be designed
for systems controlled by two scales, one of which is a relatively large momentum.
This finding has direct implications for the design of  
effective field theories for heavy quarkonium systems where  one
needs to control
the effect of heavy meson loops on heavy quarkonium 
decays as stressed in Ref.~\cite{Zb}. 

For example the decays of regular quarkonia, like
$\psi(2S)\to J/\psi \pi^0$ and $\psi(2S)\to J/\psi \eta$,
can get sensitive to meson loops, since they
are 
 $SU(2)_F$ and $SU(3)_F$ violating, respectively~\cite{hml1,hml2,hml3,hml4}. 
The reason is that the mass differences of the open charm meson
loops induce enhanced violations of the mentioned symmetries, non--analytic
in the quark masses. 
\begin{figure}[t!]
\begin{center} 
\includegraphics[height=3.1cm,keepaspectratio]{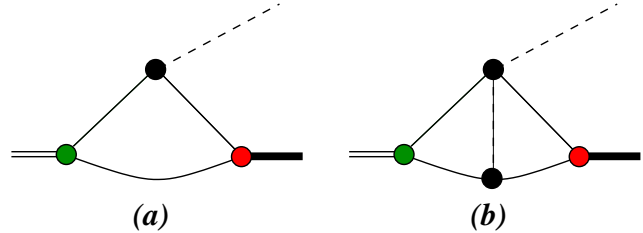}
\end{center}
\caption{ Loop contributions to transition matrix elements of heavy
quarkonia. The  double and bold lines denote two different heavy quarkonium states
and the red and green filled circles their coupling to heavy open heavy--flavor states,  denoted by the internal solid lines.
 The
dashed line denotes a light pseudo scalar that couples to the
heavy states
via the black solid dots.}
 \label{Qdec}
\end{figure} 
Furthermore, above the open charm thresholds a large number of states were recently 
found with properties in strong disagreement with well established quark model,
see Ref.~\cite{hqwg} for
a recent review. 
Several of these new states 
qualify as candidates for molecular states formed from heavy mesons. 
For those all transitions and decays are expected to be dominated by loops
of heavy mesons.

In an effective field theory formulation the evaluation of observables 
in the heavy quarkonium systems  depends on  the non-trivial interplay 
 of several different scales,   such as the pion  mass,  the pion 
 momentum and  the heavy meson     velocity  defined
as  a ratio of    the binding momentum $\sqrt{M_H E}$  
($E$ is the energy of the initial/final quarkonium with respect to the
$\bar HH$ threshold ($H= D, B, ...$))  to the  heavy meson mass $M_H$.
In order to establish which kind of transitions can be calculated in a controlled
way, a power counting was introduced in Refs.~\cite{Zb,Zb2} that
builds on the insights gained from $NN\to NN\pi$, also outlined in this review.
Especially, in order to compare the order estimate  for diagrams $(a)$ 
 and  $(b)$ of Fig.~\ref{Qdec} it was crucial to
use the findings of Ref.~\cite{lensky2}. 
Specifically,  similar to Eq. \eqref{eq:pipivert}  discussed  earlier in this review,  
the  $\pi \pi H H$  vertex  in diagram  (b)  was  decomposed 
in several terms in order to properly isolate the leading contribution of the diagram. 
For more details we refer to the cited references.

\section{Summary and outlook }
\label{sec:outlook}

Chiral perturbation theory  has been successfully applied in the past
decades to describe low-energy dynamics of pions and nucleons 
as well as electroweak processes. 
However, the application of this effective field theory to pion production in
nucleon-nucleon collisions turned out to be  considerably more challenging due to
the large three-momentum transfer involved in this reaction.  
This large momentum 
called for a modification of
the power counting. 
The resulting slower convergence of the chiral expansion for this reaction 
[the expansion  parameter $\chi_{\rm MCS} \sim \sqrt{m_\pi / m_N}$  defined
in Eq.~(\ref{expansionpapar})] 
provides a strong motivation for
extending the calculations to higher orders. 
In this review we demonstrated that   
the momentum counting scheme (MCS)  
 properly accounts for the additional scale
associated with the large momentum transfer. 
It is shown to be
 successful in  classify the various contributions to the
$NN\to NN\pi$ transition amplitudes according to their importance. 
%
Given the relatively large MCS expansion parameter $\chi_{\rm MCS}$, 
one may wonder whether 
the pion-production operator converges  sufficiently well to consider 
the theory predictive. 
The results of Refs.~\cite{lensky2,newpwave} 
provide 
strong indications for the  convergence of the 
s- and p-wave pion-production 
operators in the final two nucleon spin-triplet channels 
where no accidental cancellations emerge.  
Furthermore,  
good  convergence  of the  results  for spin observables  in  $pp\to pp \pi^0$, 
which probe  p-wave pion production in this channel,   was  observed  in Ref.\cite{ch3body}. 
On the other hand, s-wave pion production in $pp\to pp \pi^0$ 
looks exceptional since the experimental
cross section in this channel is suppressed by more than an order of 
magnitude as compared to the charged 
channels near threshold. 
The experimental evidence for the  
smallness of the s-wave operator in the neutral channel
is fully in line with the suppression of the 
Weinberg--Tomozawa (WT) operator for $pp\to pp\pi^0$, the almost complete cancellation  of the  pion  emission amplitude from 
the direct diagram at LO  and  the vanishing  of all loop contributions at NLO.  
Therefore, it is not a surprise that 
 the relative importance of N$^2$LO chiral loops  in this channel is significantly 
enhanced compared to the  case where the final $NN$ are in a spin-triplet state.  
As demonstrated in  Ref.~\cite{NNLOswave}, 
the nucleon-pion loops  are of a 
similar size for the spin  triplet  and  singlet  channels.  
In this sense  the experimentally measured  
$pp\to pp \pi^0$ reaction  is unique in that it directly probes 
the higher order MCS contributions which in the other reaction 
channels are masked by the dominant lower 
order WT term.  

In this review
the evaluation of all loop diagrams with pions, nucleons and $\Delta$ degrees 
of freedom up--to--and--including \NNLO{} has been summarized.
It was shown that all diagrams that contribute to NLO
cancel exactly --- a necessity from field theoretic consistency.
As presented, even 
among the \NNLO{} loop contributions 
there are significant cancellations as a results of chiral symmetry 
requirements and its symmetry breaking pattern. 
In particular, in Ref.~\cite{NNLOswave} it is shown that 
all $1/m_N$-corrections of the various diagrams cancel at \NNLO{}, 
and it is also shown that none of  the LECs $c_i$, $i=1\ldots  4$,
of ${\cal L}^{(2)}_{\pi\!N}$  contribute 
to the pion loops at this order. 

In Ref.~\cite{future} as well as in  Sec.~\ref{sec:Delta_N} 
it was demonstrated that
the amplitudes from the loops which include the $\Delta$, 
contribute at \NNLO{}  to  both  the $NN$ final state 
spin-triplet and  spin-singlet channels, and 
interfere destructively with the pure pion-nucleon 
ones.
The isoscalar amplitudes from the loops 
which include the $\Delta$             
are significantly smaller than the amplitudes from 
loops with only pions and nucleons,
while the isovector  amplitudes from the $\Delta$ loops  
are roughly half the size of the pure 
pion-nucleon ones.
The size of the isovector  contributions of  the $\Delta$  loops  is  in line  
with the  dimensional analysis estimate,  while  the  suppression  of  
the isoscalar  contributions originates  from specific
numerical factors (spin-isospin coefficients) in this channel~\cite{future}. 
What emerges from the resultant loop amplitudes looks  very promising in order 
to quantitatively describe the data for both 
$pp\to pp\pi^0$ and $pp\to d\pi^+$ for near threshold energies.  
While in the former reaction there persists a huge discrepancy
between data and the chiral EFT calculation to NLO, 
in the latter
 at NLO the description is already quite good~\cite{lensky2}, 
as  discussed in Sec.\ref{sec:obs}.  
Therefore,  the \NNLO\ chiral loop amplitude  
is expected  to be  much more  dominating   for 
the isoscalar  amplitudes  relevant for $pp\to pp \pi^0$ production
  than for the isovector ones.     
 In order to be able to compare the amplitudes presented here to experimental data
 and to make the qualitative arguments given more quantitative,
 a proper convolution with $NN$ wave functions is necessary.

The   
loop diagrams at \NNLO\ 
contributing to the pion-production amplitudes 
have  important phenomenological consequences, as 
was discussed in Refs.~\cite{NNLOswave,future}.   
Within various phenomenological meson-exchange approaches~\cite{eulogio,unsers,mosel},
the pion production is largely driven by tree-level pion rescattering off a nucleon
with the  $\pi N\to \pi N$ amplitude being far off shell.   The
scalar-isoscalar ($\sigma$-exchange) $\pi N$ interaction
is relevant for the isoscalar production amplitude 
while the isovector
($\rho$-exchange) $\pi N$ interaction contributes to the strength of  
the isovector one. 
Since the isoscalar  $\pi N$ scattering amplitude essentially 
vanishes on-shell, see Eq. ~\eqref{apl},
 the production mechanism proposed in Refs.~\cite{eulogio,unsers,mosel}
relies on the  significance  of the off-shell properties of the 
$\pi N$ scattering
amplitude.  
However,  the EFT consideration presented in this review puts this mechanism into question.
Pion rescattering via the  phenomenological  pion-nucleon transition amplitude
can in chiral EFT be mapped  onto  pion rescattering (at tree level)
via the low-energy constants $c_i$  
plus some contributions from pion loops.
The tree-level amplitude contributions  $\propto c_i$ are, even in the off-shell (pion
production) kinematics, far too small to explain the data for the
neutral pion production \cite{cohen,park}. 
As far as the loop contributions are concerned, only the diagrams IV
and the
mini-football may be regarded as an analog
of the corresponding phenomenological mechanism.  
However,     the remaining  N$^2$LO contribution of diagrams IV and $\Delta$IV and the mini-football diagram,  
see  Figs.~\ref{fig:allN2LO} and \ref{fig:allDLoops},  have a purely  isovector structure --- all isoscalar
contributions cancel exactly.
Therefore, none of the pion loop diagrams can be mapped 
into the particular phenomenological mechanism in the isoscalar case. 
Thus,  Refs.\cite{NNLOswave,future}  concluded  that the rescattering contribution with
the isoscalar $\pi N$ amplitude to  $pp\to pp\pi^0$ that could be modeled phenomenologically 
by a $\sigma$ exchange should be very small.

It is noted that none of the previous phenomenological
investigations take into  account  chiral  loops   which, as found, contribute significantly to the production amplitude. 
In particular, as presented in Sec.~\ref{sec:renorm1} the magnitude of the 
regularization-scheme independent  long-range contribution of the pion-nucleon 
loops to the isoscalar  production amplitude turns out to be 
comparable  to the magnitude of the
short-range amplitudes in phenomenological models of 
Refs.~\cite{Lee,HGM,Hpipl,jounicomment}.  
In some phenomenological models the short-range amplitudes originate
from heavy-meson z-diagrams which, in these studies, are advocated as
the necessary mechanism to describe experimental data. 
In chiral EFT these phenomenological heavy meson exchanges 
contribute via the contact interaction $(NN)^2\pi$ and 
normally would be used in order to estimate the magnitudes of 
the low-energy constants (LECs) of Eq.~(\ref{eq:4Npi}) 
as was done in Sec.~\ref{sec:renorm1}.     
Thus, the results of the loop diagrams
reviewed in Secs.~\ref{sec:swave} and \ref{sec:swaveD} raise doubts on the role  of the 
short-range physics in pion production as suggested by these
phenomenological studies. 
In order to draw more definite conclusions  the complete
\NNLO{} operator  for s-wave pion  production  
discussed in this review paper should be 
convoluted with the $NN$ wave functions and  
confronted with experimental data
in order to assess the magnitude of the LECs of Eq.~(\ref{eq:4Npi}).   
The work  in 
this  direction is currently in progress.

As  presented  in this review,  p-wave  pion production in  $NN\to NN\pi$  
opens  an appealing  possibility  to determine  the  
strength of  the  $(N N)^2 \pi$  contact operator,  the LEC $d$, 
which can be thought of as the 
two-nucleon analog of the axial constant $g_A$.  
In the introduction we outlined how this  LEC  contributes to several few-nucleon  processes,  
like  the  three--nucleon force,     the $pp$ fusion   
and the tritium $\beta$-decay, 
 the neutrino deuteron breakup and  the  muon capture on  the  deuteron,  
as well as reactions  involving   photons   
 $\pi  d \to \gamma NN$ and  $\gamma d\to \pi NN$.  
A   combined  analysis  of the  different  channels  of  
$NN\to NN\pi$  with chiral EFT by 
 Ref.~\cite{newpwave} and presented in Sec.~\ref{sec:pwave} 
revealed  that the reaction $pn\to pp\pi^-$ appears to be   the
most  interesting  channel   for  the extraction of the  LEC $d$.  
Unlike  s-wave pion production, the  p-wave pion-production operator   
up to  N$^2$LO~\cite{ch3body,newpwave}     
includes only the tree-level diagrams shown in Fig.~\ref{diag}.   
A large step towards 
an accurate extraction of the relevant  LEC $d$ from   data  was  provided   by the recent measurements  
of $\vec p p\to (pp)_s \pi^0$ and  $\vec p n\to (pp)_s \pi^-$    
at the COSY  accelerator in
J\"ulich\cite{ANKEpi0,ANKEpi-,ANKEAxx}.    As outlined in Sec.~\ref{COSY_data} the latest data from COSY   
allowed both s-wave and d-wave pion-production amplitudes 
to be  extracted  from the amplitude  analysis of  the data of $pp\to pp\pi^0$.   
It should be noted that in these recent  measurements   
at COSY \cite{ANKEpi0},   the final two nucleons were restricted to have a relative energy less than 3~MeV. 
This restricted phase space of the  final particles enhanced the importance of the pion d-wave.     
The  analysis of these COSY data for $pn\to pp\pi^-$,  
while revealing the dominant role  of  the  p-wave pion-production amplitudes,   
does not yield  the unique  solution  for these  amplitudes.  
In order to pin down the p-wave amplitudes and thereby the LEC $d$, 
will require  data from  the planned double polarisation measurement of 
$A_{xz}$  at COSY.   
As  advocated in Ref.~\cite{ANKEAxx},  this is  the  most  promising  
experiment which would  lift the ambiguity 
between  different solutions  and  ultimately allow an   extraction of  
the  p-wave amplitudes  from data. 

 In this review we also presented in some detail a few applications 
of the two scale expansion underlying the MCS:
\begin{itemize}
\item The extraction of the strong part of the proton--neutron mass difference from  charge symmetry breaking 
observables in $pn\to d\pi^0$ (Sec.~\ref{sec:CSB}) --- the value
extracted is consistent with other, completely different analyses.
\smallskip
\item The chiral expansion for $NN\to NN\pi\pi$ (Sec.~\ref{sec:twopiproduction}) 
which is an important
step towards a quantitative understanding of this class of reactions.
\smallskip
\item The calculation of dispersive corrections to $\pi d$ scattering at threshold. 
This evaluation allowed  a  high-precision  
extraction of  the isoscalar pion--nucleon scattering length $a^+$ from pionic atoms data, and   a  
reanalysis of  the  pion-nucleon coupling constant (Sec.~\ref{sec:pid-atom}).
\smallskip
\item The effective field theory outlined and confronted with data in this review 
is a model case for the design of an effective field theory to
investigate certain 
states in the heavy quarkonium sector (Sec.~\ref{heavyquarkonia}).
\end{itemize}
These examples show that an understanding
of reactions of the type $NN\to NN\pi$ within effective field theory is not only interesting in 
its own right, but also 
provides important new insights into the workings of the strong interaction in different energy regimes.

\vspace{5mm}

{\bf Acknowledgements}

\noindent
The hospitality of the University of Adelaide, where one of the authors (FM) 
spent his sabbatical while working on this review, is greatly appreciated. 
FM is supported in part by  the National Science Foundation (US), 
Grant No. PHY-1068305.  This work was supported in part by funds provided by 
the EU HadronPhysics3 project ``Study
of strongly interacting matter'', the European Research Council
(ERC-2010-StG 259218 NuclearEFT),  DFG-RFBR grant (436 RUS 113/991/0-1), 
and the DFG and NSFC funds to the Sino-German CRC 110 ``Symmetries  
and the Emergence of Structure in QCD''.
 %
%


\appendix 

\section{Definitions  for the various integrals}  
\label{sec:appbasicint}
 

In this subsection we give the explicit definitions of the 
common dimensionless loop integrals used in this work. The first integral 
$J_{\pi \Delta}= \mu^\epsilon J_0(-\delta )$ where the  integral $J_0(-\delta )$ is defined 
in Ref.~\cite{ulfbible}. 
\begin{eqnarray} 	\frac{1}{\delta}\jpid (\delta)
		   &=& \frac{\mu^\epsilon}{i \delta} \int \frac{d^{4- \epsilon}l}{(2\pi)^{4-\epsilon}}
	         \frac{1}{(l^2-\mpi^2+i0)( - v \cdot l - \delta +i0)}	
	         \nonumber 
\\
	       &=& 4 L + \frac{(-2)}{(4\pi)^2} \left[ -1 + \log \left( \frac{\mu^2}{\mpi^2} \right) \right] 
	       \nonumber 
\\
	       &&+ \frac{4}{(4\pi)^2} 
	           \left[ -1 + 
	                \frac{\sqrt{1-y-i0}}{\sqrt{y}} 
	                \left[ 
	                      -\frac{\pi}{2} + \arctan \left( \frac{\sqrt{y}}{\sqrt{1-y-i0}} \right)
	                \right] 
	           \right],
		\label{JpiD}
\\
	\ipipi (k_1^2)
		&=& \frac{\mu^\epsilon}{i} \int \frac{d^{4- \epsilon}l}{(2\pi)^{4-\epsilon}}
		\frac{1}{(l^2-\mpi^2+i0)((l+k_1)^2 - \mpi^2 +i0)} 
		\nonumber
\\
		&=& - 2 L - \frac{1}{(4 \pi)^2} \left [ \log \left( \frac{\mpi^2}{\mu^2} \right)
		- 1 + 2 F_1 \left(\frac{k_1^2}{\mpi^2}\right) \right],
		\label{Ipipi}
\end{eqnarray}
where  
\begin{eqnarray}
\newcommand{\myvar}{x}
F_1 (\myvar) = \frac{\sqrt{4-\myvar-i0}}{\sqrt{\myvar}}
\arctan \left( \frac{\sqrt{\myvar}}{\sqrt{4-\myvar-i0}} \right)\,,
\label{eq:F1} 
\end{eqnarray} 
\begin{equation}
	L = \frac{1}{(4\pi)^2} \left[ - \frac{1}{\epsilon} + \frac12 \left( \gamma_E -1 -\log (4\pi) \right)  \right],
\end{equation}
and the variables $x$, $y$ are defined via $x= k_1^2 / \mpi^2$, $y = \delta^2 / \mpi^2 $.

The integrals  with  two pion  and  one $\Delta$ propagator  and   in addition with   one nucleon propagator read
\begin{eqnarray}
	\hspace*{-0.64cm}\intT =\delta  \frac{\mu^\epsilon}{i} \int &&\hspace*{-0.51cm}\frac{d^{4- \epsilon}l}{(2\pi)^{4-\epsilon}}
         \frac{1}{(l^2-\mpi^2+i0)((l+k_1)^2 - \mpi^2 +i0)( - v \cdot l - \delta +i0)},
         \label{JpipiD}
\\
	\intL &=& k_1^2 \frac{\mu^\epsilon}{i}\int \frac{d^{4- \epsilon}l}{(2\pi)^{4-\epsilon}}
         \Big[ \frac{1}{(l^2-\mpi^2+i0)((l+k_1)^2 - \mpi^2 +i0)} 
         \nonumber
\\ 
         && \times \frac{1}{( - v \cdot l  +i0) ( - v \cdot l - \delta +i0)} \Big].
         \label{JpipiND}
\end{eqnarray}
The integrals in Eqs.~(\ref{JpipiD}) and (\ref{JpipiND}) 
can be reduced to simple 
one-dimensional integrals which  can be calculated numerically.

It is also  convenient  to define  finite, scale-independent parts of
$\jpid$ and $\ipipi$ in which  the  divergency  $L $ and the $\log
(\mpi/\mu)$   terms  are removed.   

\begin{eqnarray}
    \ipipi
        &=&  -2 L - \frac{1}{(4\pi)^2} \log \left( \frac{\mpi^2}{\mu^2} \right) + \ipipifsi,
\\
    \ipipifsi 
        &=&  
		\frac{1}{(4\pi)^2}
		\left( 
		1-  2 \frac{\sqrt{4-x-i0}}{\sqrt{x}} \arctan \left( \frac{\sqrt{x}}{\sqrt{4-x-i0}} \right)
		\right),
		\label{ipipifsi}
\\    
    \frac{1}{\delta}\jpid 
           &=& 4 L +
            \frac{2}{(4\pi)^2} 
           \log \left( \frac{\mpi^2}{\mu^2} \right)   + \frac{1}{\delta} \jpidfsi,
\\
    \frac{1}{\delta} \jpidfsi
           &=&  \frac{4}{(4\pi)^2} 
               \left[ -\frac{1}{2} + 
                    \frac{\sqrt{1-y-i0}}{\sqrt{y}} 
                    \left[ 
                          -\frac{\pi}{2} + \arctan \left( \frac{\sqrt{y}}{\sqrt{1-y-i0}} \right)
                    \right] 
               \right]. 
\end{eqnarray}

From the expressions above it is easy  to obtain  the  important
relation,  which  is  used  in the analysis of the  integral
combinations in Sec.\ref{sec:swaveD},  relevant for our study
\begin{eqnarray}
    \ipipi+\frac{1}{2\delta}\jpid = \ipipifsi + \frac{1}{2\delta}
    \jpidfsi.
\label{intfinite}
\end{eqnarray}

 
\end{document}